\documentclass[a4paper,onecolumn,11pt]{quantumarticle}
\pdfoutput=1
\usepackage[english]{babel}
\usepackage[T1]{fontenc}
\usepackage[utf8]{inputenc} 
\usepackage{microtype}
\usepackage[numbers]{natbib} 

\usepackage{amsmath,amssymb,amsthm}
\usepackage{bm}
\usepackage{bbm}       
\usepackage{nicefrac}

\usepackage{graphicx}
\usepackage{subcaption}
\usepackage{dcolumn}   
\usepackage{multirow}
\usepackage{ragged2e}

\usepackage{tikz}
\usepackage{quantikz}  

\usepackage{latexsym}
\usepackage{braket}
\usepackage{enumitem}
\usepackage{float}      
\usepackage{algorithm}
\usepackage{algpseudocode}
\usepackage{tabularray}

\usepackage[dvipsnames]{xcolor}
\usepackage{hyperref}
\hypersetup{
    colorlinks=true,
    pdfpagemode=FullScreen,
}
\usepackage[capitalise]{cleveref}
\usepackage[symbol]{footmisc}

\usepackage[font=small,labelfont=bf,justification=justified]{caption}
\newtheorem{theorem}{Theorem}
\newtheorem{lemma}[theorem]{Lemma}

\newtheorem{definition}[theorem]{Definition}

\newtheorem{assumption}[theorem]{Assumption}

\newcommand{\R}{\mathbb{R}}
\newcommand{\C}{\mathbb{C}}

\newcommand{\mc}[1]{\mathcal{#1}}

\begin{document}

\title{Quantum Algorithm for Low-Energy Effective Hamiltonians and Subspace Eigenvalue Problem}

\author[1,2,3]{Chun-Tse Li} 
\email{chuntsel@usc.edu}
\author[3]{T. Tzen Ong}
\email{tzen.ong@gmail.com}
\author[3,4]{Chih-Yun Lin}
\author[2]{Yu-Cheng Chen}
\author[3]{Hsin Lin}
\email{nilnish@gmail.com}
\author[2]{Min-Hsiu Hsieh}
\email{min-hsiu.hsieh@foxconn.com}

\affil[1]{Department of Electrical and Computer Engineering, University of Southern California, Los Angeles, California 90089, USA}
\affil[2]{Hon Hai (Foxconn) Quantum Computing Research Center, Taipei, 114, Taiwan}
\affil[3]{Institute of Physics, Academia Sinica, Taipei 115201, Taiwan}
\affil[4]{Department of Physics, National Taiwan University, Da'an District, Taipei 10617, Taiwan}


\begin{abstract}
Subspace eigenvalue problems arise ubiquitously in quantum chemistry and condensed-matter physics, where the relevant object is often a low-energy manifold rather than a single ground-state wavefunction. In this work, we propose a fault-tolerant quantum algorithm for this subspace-level task based on the Feshbach effective-Hamiltonian formalism. Given block-encoding access to the full Hamiltonian and a chosen $d$-dimensional reference subspace, the algorithm estimates eigenvalues of states with nonzero overlap with the reference subspace through a local secant fixed-point search. It then implements the associated wave operator and prepares an orthonormal basis whose span approximates the target invariant subspace. The construction combines projected block encodings with quantum singular value transformation (QSVT), which approximates the complementary-space resolvent and thereby provides both the self-energy used for eigenvalue estimation and the wave operator used for eigenstate reconstruction. For target accuracy $\varepsilon$, a single evaluation of the effective Hamiltonian has query complexity $\widetilde{\mathcal O}(d^3/(g^2\varepsilon))$, up to block-encoding normalization factors, where $g$ is the distance between the target eigenvalue and the nearest pole of the effective Hamiltonian. Under the stated local regularity conditions, the secant search requires only $\mathcal O(\log\log(1/\varepsilon))$ effective-Hamiltonian evaluations to reach the working precision. Classical numerical emulations for an open $4\times2$ Fermi--Hubbard cluster, all-electron LiH bond stretching, and $[\mathrm{Ru(bpy)}_{3}]^{2+}$ demonstrate the resolution and reconstruction of low-energy states and manifolds across spin-sector crossings, near-degeneracies, and dense excited-state spectra.
\end{abstract}

\maketitle

\section{Introduction}
Solving Hamiltonian eigenvalue problems and preparing eigenstates are among the most fundamental tasks in quantum many-body physics, quantum chemistry, and quantum information science. While several fault-tolerant quantum algorithms have been proposed for ground-state energy estimation and ground-state preparation
\cite{poulin_preparing_2009,ge_faster_2019,lin_nearoptimal_2020,gilyen_quantum_2019,martyn_grand_2021,ku_benchmarking_2025,ku_optimizing_2026,ren_hybrid_2026}
, these formulations are primarily designed for a single isolated eigenstate and do not directly address eigenvalue problems at the subspace level. In many practically relevant settings, the low-energy spectrum is degenerate or nearly degenerate, and the objective is not merely to prepare one eigenstate but to identify a low-energy subspace and determine its multiplicity.

Degenerate and nearly degenerate eigenvalue problems arise ubiquitously in condensed-matter physics and quantum chemistry. For instance, locating a quantum phase transition often requires tracking how the lowest few energy levels evolve as system parameters change
\cite{coleman_quantum_2005,sachdev_quantum_2011}. Near a transition, levels from different symmetry sectors may cross or nearly cross, and identifying the resulting low-energy manifold can help pinpoint the transition. More broadly, low-energy behavior is frequently governed by a small active manifold, as in Wannier downfolding or embedding constructions
\cite{martyn_grand_2021,souza_maximally_2001,georges_dynamical_1996}, where level splittings and projected matrix elements determine effective low-energy descriptions
\cite{anderson_antiferromagnetism_1950,zhang_effective_1988,kugel_jahnteller_1982}. In topologically ordered phases, the ground state may be degenerate on nontrivial compact manifolds, making the ground-state manifold rather than a single state the relevant target
\cite{kitaev_faulttolerant_2003,cincio_characterizing_2013}. In quantum chemistry, static correlation and near-degeneracy effects in open-shell molecules and catalytic complexes place multiple electronic configurations within chemical accuracy, motivating multireference active-space treatments
\cite{szalay_multiconfiguration_2012,helgaker_molecular_2000,abragam_electron_2013}. Similar phenomena occur near nonadiabatic seams, where conical intersections and avoided crossings couple several adiabatic surfaces and require a well-defined basis for evaluating forces, transition amplitudes, and nonadiabatic couplings
\cite{yarkony_diabolical_1996,levine_conical_2019,domcke_conical_2011}.

A canonical framework for addressing such subspace-level eigenvalue problems is the effective-Hamiltonian (downfolding) theory
\cite{feshbach_unified_1958,feshbach_unified_1962,lowdin_studies_1962,schrieffer_relation_1966,bravyi_schrieffer_2011}. The central idea is to map the full high-dimensional eigenvalue problem onto a chosen reference subspace, yielding a smaller effective problem in which the eliminated degrees of freedom enter through an energy-dependent self-energy. More explicitly, for a decomposition of the Hilbert space into a reference subspace $P\mathcal H$ and its complement $Q\mathcal H$, the Feshbach effective Hamiltonian takes the form
\[
    H_{\mathrm{eff}}(\lambda)
    =
    H_{PP}
    +
    \Sigma(\lambda),
    \qquad
    \Sigma(\lambda)
    =
    -H_{PQ}(H_{QQ}-\lambda I_Q)^{-1}H_{QP},
\]
where $H_{PP}=PHP$, $H_{PQ}=PHQ$, $H_{QP}=QHP$, and $H_{QQ}=QHQ$. This is the same convention as the Schur-complement formula below; equivalently, $-(H_{QQ}-\lambda I_Q)^{-1}=(\lambda I_Q-H_{QQ})^{-1}$. The self-energy $\Sigma(\lambda)$ accounts for virtual excursions from the reference subspace into the complementary $Q$ space and back. By solving the fixed-point (self-consistency) condition, the effective Hamiltonian reproduces the eigenvalues associated with eigenstates that have nonzero overlap with the reference subspace. This framework provides a natural way to identify low-energy manifolds, resolve degeneracies, and characterize effective low-energy physics.

However, exact downfolding on a classical computer is generally intractable for many-body systems. The main obstruction is the self-energy: evaluating $\Sigma(\lambda)$ requires applying or inverting the resolvent $(H_{QQ}-\lambda I_Q)^{-1}$ on the complement space $Q\mathcal H$, whose dimension typically grows exponentially with system size. Thus, an exact classical construction of $H_{\mathrm{eff}}(\lambda)$ requires manipulating exponentially large matrices or many-body operators. Consequently, practical classical implementations often rely on perturbative expansions, truncations, or approximate embeddings of the self-energy, which may become unreliable in strongly correlated or near-degenerate regimes.

In this work, we leverage the effective-Hamiltonian formalism to address Hamiltonian eigenvalue problems at the \emph{subspace} level. Rather than assuming that a single eigenstate is isolated by a resolvable spectral gap, we target an externally isolated low-energy invariant subspace and its spectrum to a prescribed working precision. This formulation naturally includes both exactly degenerate eigenspaces and nearly degenerate spectral clusters. When the internal splittings within a cluster are smaller than the working precision, the individual eigenstates are not uniquely meaningful from the energy information alone; the stable object is instead the spectral projector onto the corresponding low-energy manifold. Our goal is therefore to determine the resolved dimension of this manifold and to prepare an orthonormal basis spanning it, with guarantees stated in terms of subspace fidelity rather than basis-dependent state-by-state overlaps.

\subsection{Related Work and Positioning}

\emph{Fault-tolerant quantum algorithms for eigenstate preparation.} On quantum hardware, several rigorous algorithms for ground-state or low-energy eigenvalue problems have been developed. Canonical examples include quantum phase estimation (QPE)
\cite{poulin_preparing_2009,ge_faster_2019}, polynomial filters based on qubitization and QSVT
\cite{lin_nearoptimal_2020,gilyen_quantum_2019,martyn_grand_2021}, and dissipative preparation schemes that engineer Lindbladian dynamics to mix into the ground state
\cite{ding_singleancilla_2024,zhan_rapid_2026}. These methods are usually formulated around single-target state preparation, thus do not provide a systematic procedure to detect the (nearly) degenerate low-energy subspace and also prepare the corresponding orthonormal basis of the subspace.

Quantum filter diagonalization~\cite{cohn_quantum_2021}, multireference Krylov methods~\cite{stair_multireference_2020}, exact quantum Lanczos algorithms~\cite{kirby_exact_2023}, block-Lanczos methods~\cite{baker_block_2024}, and quantum block-Krylov projector construct~\cite{oliveira_quantum_2025} and diagonalize reduced subspaces~\cite{yu_quantumcentric_2025} containing several low-lying states.
In particular, block and multireference variants can recover degenerate eigenstates and their multiplicities, and therefore provide the closest comparison to the present work. Their reduced spaces are generated from one or more trial states by Hamiltonian polynomials, real- or imaginary-time evolution, or related basis-generation procedures. Their performance consequently depends on the coverage and conditioning of the generated subspace, as well as on basis size, propagation parameters, and measurement precision. Our method uses a different reduction: the reference space $P\mathcal H$ is prescribed directly, while the complimentary space $Q\mathcal H$ enters through the ``self-energy'' inside the effective Hamiltonian.

\vspace{0.2cm}

\noindent
\emph{Quantum algorithms for effective Hamiltonians.}
Previous quantum algorithms have constructed low-energy effective descriptions using Schrieffer--Wolff transformations or by optimizing over candidate effective theories
\cite{zhang_quantum_2022,yang_quantum_2023}.
These approaches seek a unitary or variational decoupling between low- and high-energy sectors, or identify an effective model that reproduces selected properties of the full theory. The present construction instead implements the energy-dependent Feshbach self-energy
$-H_{PQ}(H_{QQ}-\lambda I_Q)^{-1}H_{QP}$ and solves a nonlinear eigenvalue fixed-point problem. The same resolvent construction also yields a wave operator that maps reduced-space vectors back to the full Hilbert space. This combination of energy-dependent downfolding, local fixed-point estimation, and explicit full-space reconstruction is the main distinction from previous quantum effective-theory algorithms.

\vspace{0.2cm}

\noindent
\emph{Variational multistate methods.}
On noisy and early fault-tolerant devices, variational quantum deflation~\cite{higgott_variational_2019}, subspace-search VQE~\cite{nakanishi_subspacesearch_2019}, equation-of-motion methods~\cite{mcclean_hybrid_2017}, and related multistate algorithms~\cite{motta_determining_2020, parrish_quantum_2019} have been used to approximate several low-lying states.
These methods can be effective when compact ansatzes and stable optimizations are available, but their accuracy is generally governed by ansatz expressibility, optimization quality, and sampling error and do not have a clear and in-depth scaling analysis and complexity guarantee.

\vspace{0.3cm}

The distinguishing contribution of our approach is therefore not merely the estimation of several low-lying eigenvalues. Instead, we formulate the task as an energy-dependent effective-Hamiltonian problem on a prescribed reference subspace. The exact Feshbach--Schur complement incorporates the eliminated Hilbert space through a resolvent self-energy, while the associated wave operator coherently lifts reduced-space vectors to the full Hilbert space. Under the isolation, access, and conditioning assumptions stated below, the framework combines local spectral estimation with preparation of a basis approximating the target invariant subspace, and makes the relevant stability parameters explicit.

\subsection{Problem Setup and Main Results Overview}
We first make precise the resolution-dependent notion of degeneracy considered in this work. Let $H$ be a Hermitian operator with spectrum $\mathrm{spec}(H)$.

\begin{definition}[$\delta$-degenerate eigenvalue cluster]
Given a set of eigenvalues $\Lambda=\{\lambda_k\}\subset\mathrm{spec}(H)$, we say that $\Lambda$ is $\delta$-degenerate if
\begin{equation}
    \max_{\lambda_k,\lambda_{k'}\in\Lambda}
    |\lambda_k-\lambda_{k'}|
    \leq \delta .
\end{equation}
That is, the maximum splitting among eigenvalues in $\Lambda$ is at most $\delta$. The case $\delta=0$ recovers exact degeneracy.
\end{definition}

This definition is tied to the precision at which the eigenvalue problem is resolved. If a cluster $\Lambda$ is $\delta$-degenerate with $\delta\lesssim\varepsilon$, then the splittings among eigenvalues in $\Lambda$ are not resolved at accuracy $\varepsilon$. Resolving those splittings would require a higher target precision and, in phase-estimation-type procedures, longer coherent evolution times. Moreover, when the internal splitting is below the physical or algorithmic resolution, individual eigenvectors inside the cluster will not be the stable objects of interest: numerical perturbations comparable to the unresolved splitting can rotate the basis within the cluster substantially. The more robust target is the spectral projector onto the clustered subspace,
\begin{equation}
    \Pi_{\Lambda}
    =
    \sum_{\lambda_k\in\Lambda}
    |\Psi_k\rangle\langle\Psi_k| ,
\end{equation}
provided that the cluster is separated from the rest of the spectrum by an external gap
\begin{equation}
    \Delta
    =
    \mathrm{dist}\left(
        \Lambda,\mathrm{spec}(H)\setminus\Lambda
    \right)
\end{equation}
that is large compared with the working precision.

Accordingly, the goal of this work is not always to resolve every eigenvalue inside a low-energy cluster when the internal splitting is too small. Instead, at a prescribed precision $\varepsilon$, we aim to resolve the spectrum up to $\varepsilon$-degenerate low-energy manifolds. For an externally isolated cluster $\Lambda$ whose internal splitting is at most of order $\varepsilon$, our algorithm manages to:
\begin{enumerate}
    \item[(1)] return a representative eigenvalue $
    \lambda$ for the cluster $\Lambda$,
    \item[(2)] determine the resolved dimension $m=|\Lambda|$,
    \item[(3)] and prepare an orthonormal basis whose span approximates the invariant subspace $\mathrm{range}(\Pi_{\Lambda})$. 
\end{enumerate}
When the internal splittings are larger than $\varepsilon$, the same framework can be applied at higher resolution to distinguish the individual eigenvalues and eigenstates within the cluster.

\begin{figure}[t!]
    \centering
    \includegraphics[width=1.0\linewidth]{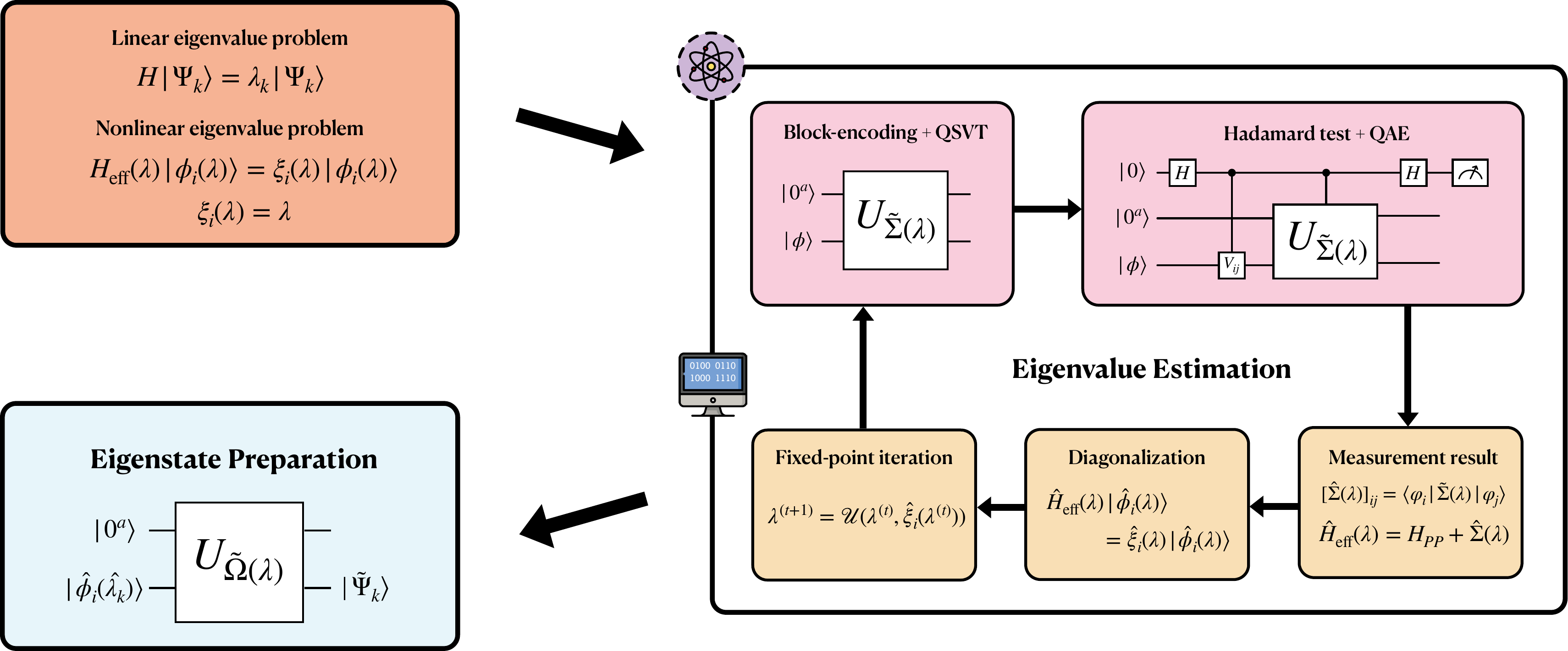}
    \caption{\justifying
    Conceptual and circuit-level workflow for the algorithm. The diagram separates the eigenvalue-estimation stage from the state-lifting and orthonormalization stage.
    \emph{Pink blocks:} Given block-encodings of $H$, QSVT implements a polynomial approximation to the resolvent
    $(H_{QQ}-\lambda I_Q)^{-1}$, yielding a block-encoding of the self-energy term $\widetilde \Sigma(\lambda)$.
    Generalized Hadamard tests (optionally with QAE) produce the $d\times d$ matrix
    $\hat H_{\mathrm{eff}}(\lambda)=H_{PP}+\hat\Sigma(\lambda)$. \emph{Yellow blocks:} the effective Hamiltonian $\hat H_{\mathrm{eff}}(\lambda)$ is diagonalized to obtain the
    eigenbranches $\hat\xi_i(\lambda)$. Then, a local secant iteration updates the spectral parameter $\lambda^{(t)}$ using consecutive evaluations of the residual $\hat\mu_i(\lambda)=\hat\xi_i(\lambda)-\lambda$. \emph{Blue block:} a
    block-encoded wave operator evaluated at $\lambda$ prepares the corresponding full-space
    eigenstates; Löwdin orthonormalization yields an orthonormal basis for the (nearly) degenerate manifold.
    }

    \label{fig:eigenvalue-estimation}
\end{figure}

Our main results are summarized by the following informal theorems.

\begin{theorem}[Informal -- Eigenvalue estimation]
\label{thm:eigen-est-informal}
Given block-encoding access to $H$ and a $d$-dimensional reference subspace $P\mathcal H$, the algorithm in \cref{fig:eigenvalue-estimation} returns an estimate $\hat\lambda_k$ of a reference-supported target eigenvalue $\lambda_k$ satisfying
\[
    |\hat\lambda_k-\lambda_k| =\mathcal O(p_k\varepsilon),
\]
where $p_k=\langle\Psi_k|P|\Psi_k\rangle$ is the $P$-space overlap and $\varepsilon$ denotes the effective-Hamiltonian approximation tolerance. Under the local secant-convergence conditions stated below, the number of effective-Hamiltonian evaluations satisfies
\[
    T_{\mathrm{fp}}
    =\mathcal O\left(
      \log\log\frac{\rho}{p_k\varepsilon}
    \right),
\]
where $\rho$ is a fixed local convergence radius. Hence, up to logarithmic factors, the query complexity is
\[
    \mathcal{O}\left(
      \frac{d^3}{g^2\,\varepsilon}
      \log\frac{1}{g\varepsilon}
      \log\log\frac{\rho}{p_k\varepsilon}
    \right),
\]
where $g=\mathrm{dist}(\lambda_k,\mathrm{spec}(H_{QQ}))$ is the distance from the target eigenvalue to the poles of the $Q$-space resolvent. A precise version, including normalization factors and oracle costs, is given in \cref{thm:eigen-estimation}.
\end{theorem}

\begin{theorem}[Informal -- Eigenstate and subspace preparation]
Given approximate eigenvalues $\hat\lambda_k\approx\lambda_k$ returned by the eigenvalue-estimation subroutine, the algorithm in \cref{fig:eigenvalue-estimation} prepares the approximate eigenstates with the following guarantees:
\\[0.2cm]
\emph{(i) Non-degenerate case.} If $\lambda_k$ is isolated from the rest of $\mathrm{spec}(H)$ by a gap $\Delta>0$, then the approximate eigenstate $|\widetilde\Psi_k\rangle$ satisfies
\[
1-F\big(\widetilde\Psi_k,\Psi_k\big)
=\mc O\left(
\frac{p_k\,\varepsilon^2}{\Delta^2}
\right),
\]
where $F(\widetilde\Psi_k,\Psi_k):=|\langle\widetilde\Psi_k|\Psi_k\rangle|^2$ is the pure-state fidelity.
\\[0.2cm]
\emph{(ii) Subspace case.} Let $\Lambda$ be a $\delta$-degenerate cluster ($\delta<\varepsilon$) of dimension $m$ separated from the rest of $\mathrm{spec}(H)$ by an external gap $\Delta>0$. Then the worst case infidelity satisfies
\[
1-F_{\min} = \mc O\left(\frac{\|G^{-1}\|\,p_{\max}\,\varepsilon^2}{\Delta^2}\right).
\]
where $p_{\max}:=\max_{\lambda_k\in\Lambda}p_k$ and $G$ is the Gram matrix of the approximate (non-orthogonal) eigenstates prepared by the algorithm.
\end{theorem}

These results highlight three central features of our guarantees.
First, the eigenvalue error scales linearly with the target tolerance $\varepsilon$, whereas the
state or subspace infidelity improves quadratically, scaling as $\varepsilon^{2}$ up to the Gram-conditioning factor in the subspace case; the overall query-complexity
dependence on $\varepsilon$ is $\widetilde{\mathcal O}(1/\varepsilon)$ up to logarithmic factors.
This reflects that eigenvalue accuracy depends directly on the precision of the effective-Hamiltonian
entries, while state preparation enjoys second-order robustness: small perturbations induce only
quadratic changes in fidelity.
Second, the bounds exhibit a linear dependence on the overlap
$p_k=\langle\Psi_k|P|\Psi_k\rangle$. At first glance this may look paradoxical, since it
suggests smaller error in the lower-overlap regime. The explanation is that the (normalized) eigenvector
$|\phi_k\rangle\in P\mathcal H$ of the effective Hamiltonian is lifted to the full eigenstate
$|\Psi_k\rangle\in\mathcal H$ by the wave operator $\Omega(\lambda_k)$ under the intermediate-normalization
convention $\langle\phi_k|\Psi_k\rangle=1$. Consequently, an error of order $\varepsilon$ introduced in
the subroutines is effectively shrunk by a factor $p_k$ once we enforce the full-state
normalization $\|\,|\Psi_k\rangle\|^2=1$. At the same time, small overlap often coincides with more
difficult oracle instances (see below), so it should not be interpreted as making the overall problem
easier.
Third, the complexity depends on a parameter $g>0$, defined as the minimal
distance between the target eigenvalue and the spectrum of $H_{QQ}=QHQ$. Intuitively, $g$ measures proximity to ``intruder states''---nearby out-of-subspace levels that
increase the condition number of the matrix we wish to invert. Hence, we require a higher-degree QSVT
polynomial to accurately approximate the matrix inverse, producing the characteristic $\frac{1}{g^2}$
dependence. This is the same obstruction long recognized in multireference/effective-Hamiltonian
formalisms, and classical quantum-chemistry techniques can be used to diagnose such intruders. In fact,
there are subtle relationships between $p$ and $g$. Loosely speaking, smaller overlap $p$
can imply smaller distance $g$ to intruders, while large $g$ often implies larger $p$; however,
the converse does not hold, so there is no nontrivial lower bound on $g$ in terms of $p$. We
elaborate on these relationships in a later section. From Theorem~\ref{thm:eigen-est-informal}, as long
as the spectral distance $g$ is $\mathcal O(1)$ (i.e., the instance is intruder-free), our
algorithm achieves a nearly optimal parameter dependence.

Taken together, our algorithm is advantageous when (i) the low-energy spectrum is nearly degenerate
and we want to prepare an orthonormal basis spanning the low-energy subspace; (ii) a single good trial
state is hard to identify (e.g., in open-shell problems) and the target is instead specified by a
reference subspace generated from multiple states; and (iii) we seek a set of low-lying
excited states beyond the ground state. Overall, the algorithm provides a unifying and systematic
framework for these settings, with rigorous guarantees and a clear physical interpretation.

\section{Theoretical Framework}
\label{sec:effective_hamiltonian}
This section provides the theoretical foundation for our projection-based approach to the eigenvalue problem. Throughout this paper, we denote the full $n$-qubit Hilbert space as $\mc H\cong (\C^{2})^{\otimes n}$ and we use the notation $N=2^n=\dim\mc H$ for the Hilbert space dimension, and we fix a $d$-dimensional reference subspace $P\mc H\subset\mc H$ (``$P$-space'') together with its orthogonal complement $Q\mc H$ (``$Q$-space''). For Hilbert spaces $\mc H_1,\mc H_2$, let $\mc L(\mc H_1,\mc H_2)$ denote the bounded linear maps $\mc H_1\to\mc H_2$, and write $\mc L(\mc H)\coloneqq\mc L(\mc H,\mc H)$. The symbol $\|\cdot\|$ denotes the $\ell_2$ norm for vectors in $\mc H$ and the operator (spectral) norm for elements of $\mc L(\mc H)$. We write $\mathrm{spec}(\cdot)$ for the spectrum, and use the notation $\mathrm{Block}(\cdot)$ to denote the matrix element of block-encoding by projecting out the ancilla qubits. For the notation convention, given $A\in\mc L(\mc H)$, the symbol $\widetilde A$ denotes a circuit-level (approximated) implementation of $A$ on a quantum computer, while $\hat A$ denotes a measured/estimated quantity (operator or scalar) derived from the estimation subroutine.

\subsection{Spectral Schur-complement and effective Hamiltonian.} Consider a Hamiltonian $H \in \mathcal{L}(\mathcal{H})$ acting on a
finite-dimensional Hilbert space $\mathcal{H} \cong \mathbb{C}^N$, and fix
an orthogonal projector $P$ of rank $d$ with complementary projector
$Q := I - P$. Relative to the direct sum decomposition
$\mathcal{H} = P\mathcal{H} \oplus Q\mathcal{H}$, any operator
$H \in \mathcal{L}(\mathcal{H})$ and state $|\Psi\rangle \in \mathcal{H}$
admit the block representations
\begin{align}
  H &= \begin{bmatrix} H_{PP} & H_{PQ} \\[2pt] H_{QP} & H_{QQ} \end{bmatrix},
  \qquad
  |\Psi\rangle = \begin{bmatrix} \Psi_P \\ \Psi_Q \end{bmatrix},
\end{align}
where $H_{PP} = PHP\big|_{P\mathcal{H}} \in \mathcal{L}(P\mathcal{H})$,
$H_{QQ} = QHQ\big|_{Q\mathcal{H}} \in \mathcal{L}(Q\mathcal{H})$, and
$H_{PQ} = PHQ\big|_{Q\mathcal{H}}^{\,P\mathcal{H}}
\in \mathcal{L}(Q\mathcal{H},P\mathcal{H})$ (for Hermitian $H$,
$H_{QP} = H_{PQ}^\dagger$).

In particular, the projector $P$ in quantum chemistry and condensed matter physics is not arbitrary, one can
further introduce a ``zeroth-order'' Hamiltonian $H_0$ that is easy to diagonalize and a residual interaction $V := H - H_0$.  Let
$H_0|\varphi_i\rangle = \varepsilon_i |\varphi_i\rangle$ define an
orthonormal eigenbasis $\{|\varphi_i\rangle\}_{i=1}^N$ of $H_0$, and choose
$P$ to project onto the span of $d$ low-energy eigenstates, e.g.,
\begin{align}
  P = \sum_{i=1}^d |\varphi_i\rangle\langle\varphi_i|, 
  \qquad 
  Q = I - P = \sum_{i=d+1}^{N} |\varphi_i\rangle\langle\varphi_i| ,
\end{align}
so that $[P,H_0] = [Q,H_0] = 0$.  In this common situation the off-diagonal
blocks of $H$ originate entirely from $V$:
\begin{align}
  H_{PQ} = PVQ\big|_{Q\mathcal{H}}^{\,P\mathcal{H}},
  \qquad
  H_{QP} = QVP\big|_{P\mathcal{H}}^{\,Q\mathcal{H}}.
\end{align}

In this block (projector) representation, the eigenvalue equation $(H-\lambda I)|\Psi\rangle=0$ reads
\begin{align}
  \begin{bmatrix}
    H_{PP}-\lambda I_P & H_{PQ} \\
    H_{QP} & H_{QQ}-\lambda I_Q
  \end{bmatrix}
  \begin{bmatrix}\Psi_P \\ \Psi_Q\end{bmatrix} = 0 .
\end{align}
When $\lambda\notin\mathrm{spec}(H_{QQ})$, the lower block is invertible and
\begin{align}
  \Psi_Q = -\,(H_{QQ}-\lambda I_Q)^{-1} H_{QP}\,\Psi_P. \label{eq:Psi2-solve}
\end{align}
Substituting into the upper block gives the \emph{spectral Schur complement} on $P\mc{H}$,
\begin{align}
  S(\lambda)\,\Psi_P 
  = \Bigl(H_{PP}-\lambda I_P + \Sigma(\lambda)\Bigr)\,\Psi_P = 0,
  \label{eq:Schur-null-equation}
\end{align}
where we denote $\Sigma(\lambda)=- H_{PQ}(H_{QQ}-\lambda I_Q)^{-1}H_{QP}$. The above equation can be viewed as the block-diagonalization via the (invertible, block-unitriangular) \emph{wave operator}
\begin{align}
  \Omega(\lambda) =
  \begin{bmatrix}
    I_P & 0 \\
    -\bigl(H_{QQ}-\lambda I_Q\bigr)^{-1} H_{QP} & I_Q
  \end{bmatrix}
\end{align}
where conjugation by the wave operator leads to block Gaussian elimination of the off-diagonal blocks
\begin{align}
  \Omega^\dagger(\lambda)\,(H-\lambda I)\,\Omega(\lambda)
  =
  \begin{bmatrix}
    S(\lambda) & 0 \\
    0 & H_{QQ}-\lambda I_Q
  \end{bmatrix}.
\end{align}
Because $\Omega(\lambda)$ is unitriangular, $\det\Omega(\lambda)=\det\Omega^\dagger(\lambda)=1$, and hence
\begin{align}
  &\det\begin{bmatrix} S(\lambda) & 0 \\ 0 & H_{QQ}-\lambda I_Q \end{bmatrix}
  = \det(H-\lambda I)
  \nonumber
  \\[0.5cm]
  &\qquad\implies
  \det S(\lambda) = \frac{\det(H-\lambda I)}{\det(H_{QQ}-\lambda I_Q)}
  = \frac{p_H(\lambda)}{p_{H_{QQ}}(\lambda)} .
\end{align}
Thus $\det S(\lambda)$ is a rational function whose zeros coincide with those of $p_H(\lambda)$ except at points where $p_H(\lambda)$ and $p_{H_{QQ}}(\lambda)$ vanish simultaneously. In particular, for $\lambda\notin\mathrm{spec}(H_{QQ})$, $S(\lambda)$ is singular if and only if $\lambda\in\mathrm{spec}(H)$. 

Having established that the roots of $\det S(\lambda)$ and $\det(H-\lambda I)$ coincide we can further show that the eigenstate we obtained from the reduced problem $S(\lambda)|\phi(\lambda)\rangle=0$ can be mapped to the eigenstate of the full problem $(H-\lambda I)|\Psi\rangle=0$. Equivalence with the full problem is transparent in block form: 
\begin{align}
   &\begin{bmatrix}
       S(\lambda) & 0 \\ 0 & H_{QQ}-\lambda I_Q
   \end{bmatrix}\begin{bmatrix}
       |\phi(\lambda)\rangle \\ 0
   \end{bmatrix}=0 
   \qquad  \Longleftrightarrow \qquad (H-\lambda I)\Omega(\lambda)|\phi(\lambda)\rangle = 0,
\end{align}
where the right hand side is obtained by multiplying $(\Omega^\dagger)^{-1}$ on the left hand side. That is, given any nonzero $|\phi(\lambda)\rangle\in P\mc{H}$ with $S(\lambda)|\phi(\lambda)\rangle=0$, the lifted vector
\begin{align}
  |\Psi\rangle = \Omega(\lambda)\,|\phi(\lambda)\rangle
  = \begin{bmatrix}
          |\phi(\lambda)\rangle \\[2pt]
          -\,(H_{QQ}-\lambda I_Q)^{-1}H_{QP}\,|\phi(\lambda)\rangle
        \end{bmatrix}
        \label{eq: Psi-eigstate}
\end{align}
satisfies $(H-\lambda I)|\Psi\rangle=0$. Hence we obtain a bijection
\[
  \mathrm{ker}\bigl(S(\lambda)\bigr)
  \xleftrightarrow[\ P\ ]{\ \Omega(\lambda)\ }
  \mathrm{ker}(H-\lambda I),
  \qquad \lambda\notin\mathrm{spec}(H_{QQ}),
\]
with inverses $P|\Psi\rangle=|\phi(\lambda)\rangle$ and $\Omega(\lambda)|\phi(\lambda)\rangle=|\Psi\rangle$. We summarize the results in the following theorem:

\begin{theorem}[Spectral Schur complement]\label{thm:spectral_schur_complement}
Let $P,Q$ be orthogonal projectors with $P+Q=I$ and $\mathrm{Tr}\,P=d$. Define
\[
  S(\lambda) = H_{PP} - \lambda I_P - H_{PQ}\bigl(H_{QQ}-\lambda I_Q\bigr)^{-1}H_{QP},
\]
where $S(\lambda)\in\mc L(P\mc{H})$. If $\lambda\notin\mathrm{spec}(H_{QQ})$, then:
\begin{enumerate}[leftmargin=*]
  \item $\lambda\in\mathrm{spec}(H)$ if and only if $S(\lambda)$ is singular.
  \item For any eigenpair $H|\Psi\rangle=\lambda|\Psi\rangle$ one has $|\phi(\lambda)\rangle\coloneqq P|\Psi\rangle\in\mathrm{ker}(S(\lambda))$ and $|\Psi\rangle=\Omega(\lambda)|\phi(\lambda)\rangle$; conversely, any nonzero $|\phi(\lambda)\rangle\in\mathrm{ker}(S(\lambda))$ lifts to $|\Psi\rangle=\Omega(\lambda)|\phi(\lambda)\rangle$ with $(H-\lambda I)|\Psi\rangle=0$.
  \item $\dim\mathrm{ker}(H-\lambda I) = \dim\mathrm{ker}\bigl(S(\lambda)\bigr)$.
\end{enumerate}
\end{theorem}

A complementary viewpoint, more familiar in physics and quantum chemistry,
recasts the $P/Q$-formalism in terms of an energy-dependent effective
Hamiltonian $H_{\mathrm{eff}}(\lambda)$ on $P\mathcal{H}$,
\begin{align}
  H_{\mathrm{eff}}(\lambda)
  := H_{PP} +\Sigma(\lambda)
  = S(\lambda) + \lambda I_P ,
\end{align}
well-defined for $\lambda \in \R\setminus\mathrm{spec}(H_{QQ})$. Solving $S(\lambda)|\phi(\lambda)\rangle=0$ is equivalent to the
self-consistent eigenproblem
\begin{align}
\label{eq:nonlinear-eigenproblem}
  H_{\mathrm{eff}}(\lambda)\,|\phi(\lambda)\rangle = \lambda\,|\phi(\lambda)\rangle,
\end{align}
which has the form of an eigenvalue equation except that the effective
Hamiltonian itself depends on the spectral parameter $\lambda$. In this
sense, the Feshbach formalism reduces the original $N$-dimensional linear
eigenproblem $H|\Psi\rangle=\lambda|\Psi\rangle$ to a $d$-dimensional
\emph{nonlinear} eigenproblem on $P\mathcal{H}$.

For any real $\lambda \notin \mathrm{spec}(H_{QQ})$, $H_{\mathrm{eff}}(\lambda)$
is Hermitian and admits an orthonormal eigenbasis
$\{|\phi_i(\lambda)\rangle\}_{i=1}^d$,
\begin{align}
  H_{\mathrm{eff}}(\lambda)\,|\phi_i(\lambda)\rangle = \xi_i(\lambda)\,|\phi_i(\lambda)\rangle,
\qquad 1\leq i \leq d,
\end{align}
We refer to the functions $\xi_i(\lambda)$ as
\emph{eigenbranches} to distinguish them from the true eigenvalues
$\lambda_k$ of $H$. Here an eigenbranch means a continuously tracked local eigenvalue branch on a pole-free interval. At an isolated crossing, the label $i$ is understood through this branch-continuation convention; for an exactly degenerate branch bundle, the corresponding eigenspace projector is the canonical object, and a smooth basis convention may be chosen inside the bundle. The nonlinear eigenproblem in Eq.~\eqref{eq:nonlinear-eigenproblem} is thus equivalent to finding the fixed point:
\[
    \xi_i(\lambda)=\lambda \quad \implies \quad \lambda\in\mathrm{spec}(H).
\]
We now state an important property of eigenbranch in the following Lemma:

\begin{lemma}[Monotone eigenbranches and $P/Q$ overlap]
\label{lem:monotone-branches}
Assume $\lambda\in\mathbb{R}\setminus\mathrm{spec}(H_{QQ})$.
Then, on any interval that avoids the poles of $H_{QQ}$, each chosen continuous eigenbranch $\xi_i(\lambda)$ is monotonically nonincreasing. Let $|\phi_i(\lambda)\rangle\in P\mathcal H$ be a differentiable normalized eigenvector along that branch, we have
\begin{align}
  \frac{d\xi_i(\lambda)}{d\lambda}
  = -\,\Big\|(H_{QQ}-\lambda I_Q)^{-1} H_{QP}\,|\phi_i(\lambda)\rangle\Big\|^2
  \le 0,
\end{align}
with equality at that value of $\lambda$ if and only if $H_{QP}|\phi_i(\lambda)\rangle=0$. Moreover, let $|\Psi(\lambda)\rangle$ denote the (unnormalized) reconstructed
eigenvector in $\mathcal{H}$ associated with $|\phi_i(\lambda)\rangle$ via the
wave-operator:
\begin{align}
    |\Psi(\lambda)\rangle
    =
    \begin{bmatrix}
      |\phi_i(\lambda)\rangle \\[2pt]
      -\,(H_{QQ}-\lambda I_Q)^{-1}H_{QP}\,|\phi_i(\lambda)\rangle
    \end{bmatrix}.
\end{align}
Then, with $\||\phi_i(\lambda)\rangle\|=1$, one has
\begin{align}
    \bigl\|\,|\Psi(\lambda)\rangle\bigr\|^2
    &= 1 - \frac{d\xi_i(\lambda)}{d\lambda}, 
\end{align}
and the $P$-space overlap 
\begin{align}
    \langle \Psi'(\lambda)|P|\Psi'(\lambda)\rangle
    &= \left(1-\frac{d\xi_i(\lambda)}{d\lambda}\right)^{-1}, 
    \label{eq:overlap-squared}
\end{align}
where $|\Psi'(\lambda)\rangle
= |\Psi(\lambda)\rangle / \bigl\|\,|\Psi(\lambda)\rangle\bigr\|$ denotes the normalized state.
Thus the magnitude of the negative slope, 
$-\tfrac{d\xi_i}{d\lambda}$, equals the squared $Q$-space norm of the
reconstructed eigenstate and directly controls the normalization overhead.
\end{lemma}

The lemma has two immediate consequences used later. First, between
successive poles of $H_{QQ}$, monotonicity implies that a tracked eigenbranch
can cross the line $\xi(\lambda)=\lambda$ at most once in each pole-free interval; this will later enable us to perform the
one-dimensional fixed-point iteration in a pole-free interval. Second, in the
decoupled case $H_{QP}|\phi_i(\lambda)\rangle=0$, the branch is locally flat and
the reconstructed state already lies entirely in $P\mathcal{H}$.
The detailed proof can be found in Appendix~\ref{app:branches}.

\section{Main result -- Quantum Algorithm}
We now present the quantum algorithm for the resolution-dependent degenerate eigenvalue problem. Our algorithm is built upon the effective Hamiltonian formalism introduced earlier and has two main stages:
\begin{enumerate}
    \item \emph{Eigenvalue estimation}: solve the nonlinear effective eigenproblem by locally searching fixed points of the eigenbranches of $H_{\mathrm{eff}}(\lambda)$.
    \item \emph{Eigenstate and subspace preparation}: lift the corresponding eigenvectors of effective Hamiltonian (in $P\mathcal H$) to full-space eigenstates (in $\mathcal H$) using the wave operator, followed by L\"owdin orthonormalization when the target is a ($\delta$-degenerate) subspace.
\end{enumerate}
The workflow is summarized in \cref{fig:eigenvalue-estimation}. The first stage of the algorithm can be further split into three parts
\begin{enumerate}
    \item \emph{Block-encoding construction}: Given an LCU block-encoding of $H$ and the residual interaction $V$, we can construct the block-encoding of the projected matrix elements $H_{QQ}-\lambda I_Q$ and $H_{QP}$ and therefore the approximate block-encoding of the self-energy $\Sigma(\lambda)$ and the wave operator $\Omega(\lambda)$.

    \item \emph{Matrix elements estimation}: After we construct the block-encoding of self-energy operator on the quantum computer. We can utilize the generalized Hadamard test to estimate and retrieve the matrix elements $\langle\varphi_i|\Sigma(\lambda)|\varphi_j\rangle$ efficiently.

    \item \emph{Classical diagonalization and local secant search}: After obtaining the (noisy) self-energy matrix, we diagonalize the effective Hamiltonian to obtain the (noisy) eigenbranches. Consecutive evaluations of the branch residual are then used in a secant update of the energy parameter $\lambda$.
\end{enumerate}

These three subroutines will keep continuing until the algorithm discovers a fixed-point $\xi_i(\lambda) \approx \lambda$. Then we can inspect if the fixed-point $\lambda$ has the $\delta$-degeneracy by looking at the spectrum of the effective Hamiltonian $H_{\mathrm{eff}}(\lambda)$. If several eigenvalues of $H_{\mathrm{eff}}(\lambda)$ satisfy $|\xi_i(\lambda)-\lambda|<\delta$ then the dimension of this (nearly) degenerate subspace can be identified.

After we get the eigenvalue estimate from the first stage, we can perform the eigenstate preparation by applying the block-encoded wave operator (with the energy parameter obtained from the first stage) on the eigenvectors of $H_{\mathrm{eff}}(\lambda)$. If we have a (nearly) degenerate subspace, and extra Lowdin orthonormalization procedure is required in order to prepare the orthonormal eigenbasis of the subspace.

The remaining parts of this section are organized as follows. \Cref{sec:block-encoding} gives the projected block-encodings used to implement the self-energy and wave operator. \Cref{sec:eigenvalue-estimation} then treats eigenvalue estimation as a one-dimensional fixed-point problem for the energy-dependent effective Hamiltonian $H_{\mathrm{eff}}(\lambda)$. Finally, \cref{sec:fidelity-bound} analyzes state and subspace preparation after the reduced-space solutions have been lifted by the wave operator. For clusters with dimension $m>1$, L\"owdin orthonormalization converts the lifted columns into a controlled orthonormal basis of the target subspace.

\subsection{Block-encoding of the Self-energy and Wave Operator}
\label{sec:block-encoding}

The central ingredient of our algorithm is the efficient construction of a block-encoding of the
effective Hamiltonian \(H_{\mathrm{eff}}(\lambda)\). In many practical settings, the \(P\)-space we
consider is spanned by classically tractable basis states, such as a mean-field Hartree--Fock state or
computational basis states. In this case, the matrix \(H_{PP}\) can be efficiently constructed on a
classical computer. Therefore, we do not need to construct the whole operator
\(H_{\mathrm{eff}}(\lambda)\) on the quantum computer; instead, we focus on the self-energy term:
\[
    \Sigma(\lambda) = -H_{PQ} (H_{QQ} - \lambda I_Q)^{-1} H_{QP}.
\]
We now state the projected-block-encoding access model used in our algorithm.
\begin{assumption}[Projected block-encoding access]\label{ass:projected-block-encodings}
    We assume access to \((\alpha_\lambda,a,0)\)- and \((\widetilde\alpha,\tilde a,0)\)-block-encoding unitaries
    \(U_{H_{QQ}-\lambda I_Q}\) and \(U_{H_{QP}}\) of the operators \(H_{QQ}-\lambda I_Q\) and \(H_{QP}\) such that:
    \begin{align*}
        (\langle 0^{a}| \otimes I)U_{H_{QQ}-\lambda I_Q}(|0^{a}\rangle \otimes I)&=-\frac{1}{\alpha_\lambda}(H_{QQ}-\lambda I_Q), 
        \\[0.2cm]
        (\langle 0^{\widetilde a}|\otimes Q)U_{H_{QP}}(|0^{\widetilde a}\rangle\otimes P)&=\frac{1}{\widetilde\alpha}H_{QP}.
    \end{align*}
\end{assumption}
This block-encoding access model is nontrivial: the projected operators \(H_{QQ}-\lambda I_Q\) and \(H_{QP}\) are matrices of size
\((N-d)\times (N-d)\) and \((N-d)\times d\), respectively, and may not admit a simple LCU decomposition
that leads to polynomial-overhead block-encodings. Fortunately, these two operators can be efficiently constructed as long as the basis of $P\mathcal H$ can be efficiently prepared by an unitary (see Appendix~\ref{app:BE-of-H22-H21} and~\ref{appendix:P-controlled-gate}). More specifically, assume we have a set of computational basis states $\{|s_i\rangle\}_{i=1}^d$ and a set of basis states $\{|\varphi_i\rangle\}_{i=1}^d$ that spans $P\mathcal H$. We want to construct an unitary that efficiently implements the mapping
\[
    U|s_i\rangle =|\varphi_i\rangle, \qquad \forall i=1,...,d.
\]
This allows us to construct a $P$-controlled NOT gate $P\otimes X+Q\otimes I$ through the multi-controlled NOT gate and $U$, and then project out the unwanted blocks of $H$ through some standard block-encoding techniques.

\begin{figure}[t!]
\centering
\begin{quantikz}
    \lstick{$|0\rangle_A$} &\gate[4]{U_{\widetilde{\Sigma}(\lambda)}}& \\
    \lstick{$|0\rangle_B$} && \\
    \lstick{$|0\rangle_C$} && \\
    \lstick{$|\psi\rangle$} && 
\end{quantikz}
$\quad \equiv \quad$
\begin{quantikz}
    \lstick{$|0^{\widetilde{a}}\rangle$}&&&\gate{U^\dagger_{H_{QP}}}\wire[d]{q}&\\
    \lstick{$|0^{\widetilde{a}}\rangle$}&\gate{U_{H_{QP}}}\wire[d]{q}&&\wire[d]{q}&\\
    \lstick{$|0^{a+1}\rangle$}&\wire[d]{q}&\gate[2]{U_{f(H_{QQ}-\lambda I_Q)}}&\wire[d]{q}&\\
    \lstick{$|\psi\rangle$}&\gate{U_{H_{QP}}}&&\gate{U^\dagger_{H_{QP}}}&
\end{quantikz}
\caption{Block-encoding circuit of the approximate self-energy $\widetilde{\Sigma}(\lambda)$.}
\label{fig:block-encoding-self-energy}
\end{figure}

\begin{figure}[t!]
\centering
\begin{quantikz}
    \lstick{$|0\rangle_{A'}$} &\gate[4]{U_{\widetilde{\Omega}(\lambda)}}& \\
    \lstick{$|0\rangle_{B'}$} && \\
    \lstick{$|0\rangle_{C'}$} && \\
    \lstick{$|\psi\rangle$} && 
\end{quantikz}
$\quad\equiv\quad$
\begin{quantikz}
    \lstick{$|0\rangle$} & \gate{H} & \ctrl{1} & \ctrl{1} & \gate{R_Y(-2\phi')} & \\
    \lstick{$|0^{\,\widetilde a}\rangle$} & & \gate{U_{H_{QP}}} \wire[d]{q} & \wire[d]{q} & & \\
    \lstick{$|0^{\,a+1}\rangle$}        & & \wire[d]{q} & \gate[2]{U_{f(H_{QQ}-\lambda I_Q)}} & & \\
    \lstick{$|\psi\rangle$}             & & \gate{U_{H_{QP}}} & & &
\end{quantikz}
\caption{\justifying
    Block-encoding circuit for the approximate wave operator $\widetilde\Omega(\lambda)$. 
    Postselecting all ancillas onto $\lvert 0\cdots 0\rangle$ implements $I - f(H_{QQ}-\lambda I_Q)H_{QP}$ up to a known scalar.
}
\label{fig:block-encoding-wave-operator}
\end{figure}

The next step is to implement the quantum singular value transformation $f(H_{QQ}-\lambda I_Q)$ with $f(x)\approx \frac1x$ up to the precision $\varepsilon_{f}$. The matrix-inverse block-encoding follows the standard QSVT construction, where we choose the polynomial degree high enough such that
\begin{equation}
\label{eq:poly-approx-f(x)-main}
    \left|f(x)-\frac{1}{x}\right| \le \varepsilon_{f}, \quad \text{for all } \,\, x\in [-\alpha_\lambda,\alpha_\lambda]\setminus\left(-\frac{g}{2},\frac{g}{2}\right).
\end{equation}
The detailed circuit construction and error analysis are given in Appendix~\ref{app:BE-of-Heff-Omega}; the two lemmas below state the resulting guarantees in the form used by the main algorithm:

\begin{figure}[t!]
    \centering
    \includegraphics[width=1.0\linewidth]{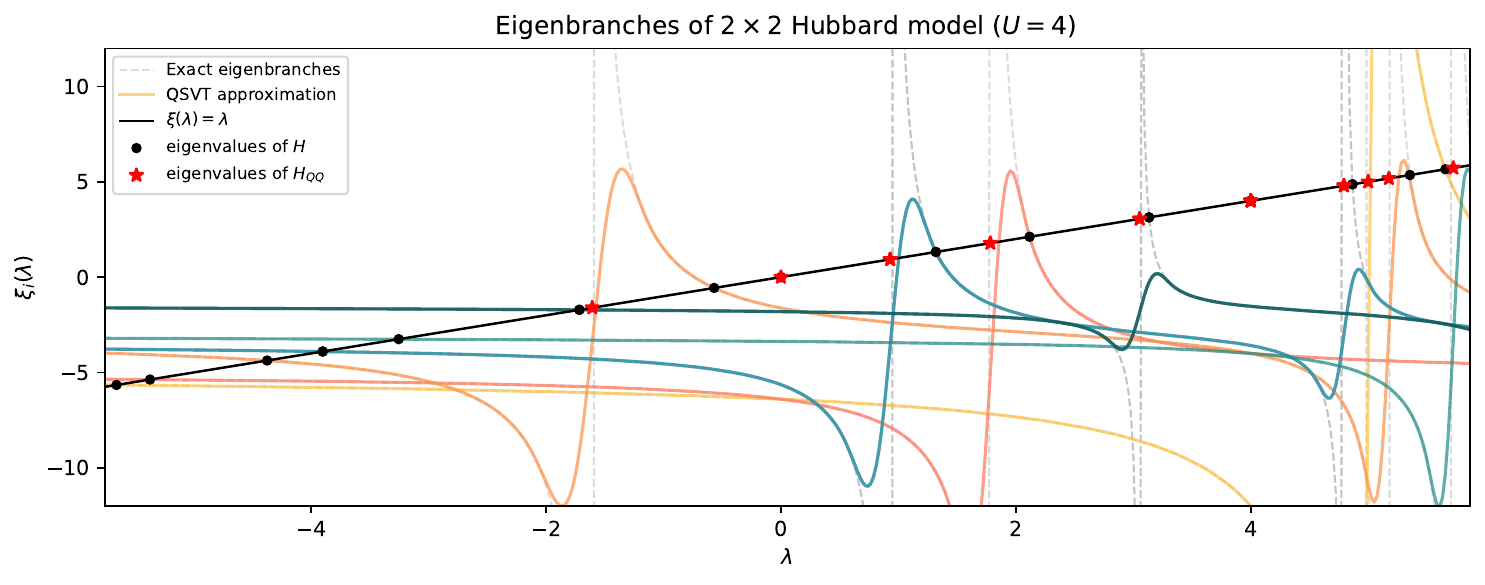}
    \caption{\justifying
    Eigenbranch plot for the $2\times 2$ Hubbard model. Solid colored curves are QSVT eigenbranches obtained by diagonalizing the approximate effective Hamiltonian $\widetilde H_{\mathrm{eff}}(\lambda)$, while dashed gray curves are the exact eigenbranches. Black dots indicate true eigenvalues of $H$, red stars indicate poles from $\mathrm{spec}(H_{QQ})$, and fixed points are obtained when an eigenbranch satisfies $\xi_i(\lambda)=\lambda$.}
    \label{fig:2x2-hubbard-eigenbranches}
\end{figure}

\begin{lemma}[Block-encoding of $\widetilde{\Sigma}(\lambda)$ and $\widetilde{\Omega}(\lambda)$] 
\label{lem:block-encoding-sigma-omega}
Let $\Lambda=\{\lambda_k\}_k$ be the target eigenvalues satisfying
\[
    \mathrm{dist}(\lambda,\mathrm{spec}(H_{QQ}))\ge g,
    \qquad \text{for all} \,\,\lambda\in \Lambda.
\]
One can construct the block-encoding $U_{\widetilde{\Sigma}(\lambda)}$ and $U_{\widetilde{\Omega}(\lambda)}$ with normalization factor $\Theta(\widetilde{\alpha}^{2}/g)$ and 
$\Theta(\sqrt{1+(\widetilde\alpha/g)^2})$
such that
\begin{align*}
\Big\| \widetilde{\Sigma}(\lambda) - \Sigma(\lambda) \Big\| \le \|H_{QP}\|^2\,\varepsilon_{f}, \qquad \Big\| \widetilde{\Omega}(\lambda)-\Omega(\lambda)\Big\|\leq \| H_{QP}\|\,\,\varepsilon_f,
\end{align*}
for all $\lambda\in \Lambda$.
The resource costs per invocation of $U_{\widetilde{\Sigma}(\lambda)}$ are:
\begin{enumerate}
    \item $\mathcal{O}\left(\frac{\alpha}{g}\log \frac{1}{g\varepsilon_{f}}\right)$ queries to $U_{H_{QQ}-\lambda I_Q}$ and $\mathcal{O}(1)$ queries to $U_{H_{QP}}$ and its adjoint.
    \item $a + 2\widetilde{a} + 1$ ancillary qubits.
    \item $\mathcal{O}\left(\frac{\alpha(a+2\widetilde{a}+n)}{g}\log \frac{1}{g\varepsilon_{f}}\right)$ additional one- and two-qubit gates.
\end{enumerate}
And the resource costs per invocation of $U_{\widetilde{\Omega}(\lambda)}$are:
\begin{enumerate}
    \item $\mathcal{O}\left(\frac{\alpha}{g}\log \frac{1}{g\varepsilon_{f}}\right)$ queries to $U_{H_{QQ}-\lambda I_Q}$ and $\mathcal{O}(1)$ queries to $U_{H_{QP}}$.
    \item $a + \widetilde{a} + 2$ ancillary qubits.
    \item $\mathcal{O}\left(\frac{\alpha(a+\widetilde{a}+n)}{g}\log \frac{1}{g\varepsilon_{f}}\right)$ additional one- and two-qubit gates.
\end{enumerate}
\end{lemma}

To get a sense on how $\widetilde\Sigma(\lambda)$ approximates the ideal $\Sigma(\lambda)$ and the effect on the eigenbranches $\xi_i(\lambda)$, \cref{fig:2x2-hubbard-eigenbranches} illustrates these eigenbranches for the $2\times2$ Hubbard model with $t=1$, $U=4$, and reference subspace dimension $d=10$. The dotted gray curves show the exact branches $\xi_i(\lambda)$, while the solid colored curves show $\widetilde{\xi}_i(\lambda)$ obtained from the QSVT approximation of $\Sigma(\lambda)$ (degeneracies reduce the number of visible curves).
The black dots mark the true eigenvalues $\lambda_k\in\mathrm{spec}(H)$, and the red stars mark the poles $\chi_\ell\in\mathrm{spec}(H_{QQ})$ of the self-energy. As $\lambda\to\chi_\ell$, the exact branches diverge to $\pm\infty$, whereas the QSVT eigenbranches remain finite and interpolate across $\chi_\ell$ due to the polynomial approximation to $\frac1x$. In the plotted instance, as $\lambda \to \chi_\ell$, only one branch exhibits divergence at each pole $\chi_\ell$, while the remaining branches extend continuously across $\lambda=\chi_\ell$; this \emph{eigenbranch continuity} phenomenon was proved in Ref.~\cite{kalantzis_spectral_2016} for simple poles, and we provide a mild generalization by considering the degeneracy of the pole $\chi_\ell\in\mathrm{spec}(H_{QQ})$ in Appendix~\ref{app:continuity-of-eigenbranches}.

\subsection{Matrix Elements Estimation}
\label{sec:eigenvalue-estimation}

\begin{figure}[t!]
\centering
\scalebox{0.95}{%
\begin{quantikz}
\lstick{$|+\rangle$ / $|+i\rangle$} & \ctrl{2} & \ctrl{1} & \gate{H} & \meter{} \\
\lstick{$|0^{a+2\widetilde{a}+1}\rangle$} & \qw & \gate[wires=2]{U_{\widetilde \Sigma(\lambda)}} & \qw & \qw \\
\lstick{$|\phi_i\rangle$} & \gate{V_{ij}} & \qw & \qw & \qw
\end{quantikz}
}
\caption{\justifying Generalized Hadamard tests using $U_{\widetilde \Sigma(\lambda)}$ and $V_{ij}$. Left: measures $\mathrm{Re}\langle\phi_i|\widetilde \Sigma(\lambda)|\phi_j\rangle$. Right: measures $\mathrm{Im}\langle\phi_i|\widetilde \Sigma(\lambda)|\phi_j\rangle$.}
\label{fig:generalized-HT}
\end{figure}

After we implement the block-encoding of $\widetilde{\Sigma}(\lambda)$ on quantum computer, the next step is to efficiently estimate the $d\times d$ matrix elements
\[
    \big[\widetilde\Sigma(\lambda)\big]_{ij}=\langle\varphi_i|\widetilde{\Sigma}(\lambda)|\varphi_j\rangle,
\]
which gives us a noisy self-energy matrix $\hat \Sigma(\lambda)$. These matrix elements can be estimated with the matrix norm error bounded by $\varepsilon_{\mathrm{est}}$ with success probability $1-\theta$ using generalized Hadamard test. The explicit circuit for the generalized Hadamard test is shown in \cref{fig:generalized-HT}, where the unitary $V_{ij}$ implements the following mapping:
\[
    V_{ij}|\phi_i\rangle = |\phi_j\rangle.
\]
When the ancilla state is $|+\rangle$, the probability of getting $0$ for the measurement is 
\[
    \mathrm{Pr}(0)=\frac{1}{2}\left(1+\frac{g}{2\beta\widetilde\alpha^2} \mathrm{Re}([\widetilde{\Sigma}(\lambda)]_{ij})\right),
\]
while as the initial state is $|+i\rangle$, we have:
\[
    \mathrm{Pr}(0)=\frac{1}{2}\left(1+\frac{g}{2\beta\widetilde\alpha^2} \mathrm{Im}([\widetilde{\Sigma}(\lambda)]_{ij})\right).
\]
After we obtain the estimated matrix elements, we can Hermitian symmetrize the matrix by $\hat\Sigma(\lambda)\leftarrow\frac{1}{2}\Big(\hat\Sigma(\lambda)+\hat\Sigma^\dagger(\lambda)\Big)$, and therefore, the generalized Hadamard test gives the estimates $\hat\Sigma(\lambda)$ such that 
\[
    \Big\| \hat\Sigma(\lambda) - \widetilde\Sigma(\lambda)\Big\| \le \varepsilon_{\mathrm{est}},
\]
with success probability $1-\theta$ with the total number of query 
\begin{equation}
    \left\{ 
    \begin{array}{ll}
    \displaystyle\mathcal{O}\left(\frac{\alpha\widetilde\alpha^4d^4}{g^3\varepsilon^2_{\mathrm{est}}}\log\frac{1}{g\varepsilon_f}\log\frac{d}{\theta}\right) &  \text{to } U_{H_{QQ}-\lambda I_Q}, \\[0.4cm]
    \displaystyle\mathcal{O}\left(\frac{\widetilde\alpha^4\,d^4}{g^2\,\varepsilon_{\mathrm{est}}^2}\,\log\frac{d}{\theta}
    \right) & \text{to } U_{H_{QP}}.
    \end{array}
    \right.
\end{equation}
While the dependence of $\varepsilon_{\mathrm{est}}$ can be quadratically improved via amplitude estimation (Appendix~\ref{app:eigenvalue_details}). This requires querying the block-encoding $U_{\widetilde{\Sigma}(\lambda)}$ for 
\[
\mathcal O\left(\frac{\widetilde\alpha^2 d^3}{g\varepsilon_{\mathrm{est}}}\log\Big(\frac{d}{\theta}\Big)\right)
\]
times. Therefore, we have the \emph{noisy oracle} for the effective Hamiltonian,

\begin{equation}
\label{eq:Heff-hat}
    \hat{H}_{\mathrm{eff}}(\lambda) \coloneqq H_{PP} + \hat{\Sigma}(\lambda),
\end{equation}
whose eigenbranches $\hat{\xi}_i(\lambda)$ approximate those of $\widetilde{H}_{\mathrm{eff}}(\lambda)\coloneqq H_{PP}+\widetilde{\Sigma}(\lambda)$ up to $\varepsilon_{\mathrm{est}}$. We can also argue that $\hat\xi_i(\lambda)$ is a good surrogate to the ideal eigenbranch $\xi_i(\lambda)$ when performing the secant fixed-point search. From the previous choice of polynomial in \cref{eq:poly-approx-f(x)-main}, $f(x)$ is a good approximation to $\frac1x$ whenever $x\notin(-g/2,g/2)$. This translates into the condition that 
\[
    \Big\|\widetilde\Sigma(\lambda)-\Sigma(\lambda)\Big\|\le \| H_{QP}\|^2\varepsilon_f, 
\]
whenever $\lambda\notin (\chi-g/2,\chi+g/2)$, for all $\chi\in\mathrm{spec}(H_{QQ})$. Therefore, as long as $\lambda$ is at least $g/2$ away from the pole of the self-energy, the $\hat\xi_i(\lambda)$ is close to $\xi_i(\lambda)$ with the error bound
\begin{align}
    \big|\hat\xi_i(\lambda)-\xi_i(\lambda)\big|
    &\le \big|\hat\xi_i(\lambda)-\widetilde\xi_i(\lambda)\big|+\big|\widetilde\xi_i(\lambda)-\xi_i(\lambda)\big|
    \nonumber \\[0.2cm]
    &\le \varepsilon_{\mathrm{est}}+ \| H_{QP}\|^2\varepsilon_f=:\varepsilon_{\mathrm{app}},
\end{align}
where we introduce the approximation error $\varepsilon_{\mathrm{app}}$ composed of the sum of the error introduced by QSVT and matrix elements estimation.

\subsection{Local secant fixed-point search}
\label{sec:fixed-point-search}

We now describe the local solver used to search for a fixed point near an initial guess of a target eigenvalue,
\[
    \xi_i(\lambda_k)=\lambda_k, \qquad \lambda_k\in\mathrm{spec}(H).
\]
Define the exact and noisy branch residuals
\[
    \mu_i(\lambda):=\xi_i(\lambda)-\lambda,
    \qquad
    \hat\mu_i(\lambda):=\hat\xi_i(\lambda)-\lambda.
\]
By Lemma~\ref{lem:monotone-branches}, on a differentiable pole-free branch,
\begin{equation}
    \mu_i'(\lambda)=\xi_i'(\lambda)-1
    =-\frac{1}{p_i(\lambda)}\le -1,
    \qquad
    p_i(\lambda):=\left(1-\xi_i'(\lambda)\right)^{-1}.
    \label{eq:residual-slope-secant}
\end{equation}
Hence the fixed point of a tracked branch is unique within a pole-free interval and is never a flat root. Whenever a queried energy is at least $g/2$ away from $\mathrm{spec}(H_{QQ})$, the effective-Hamiltonian approximation derived above gives
\begin{equation}
    |\hat\mu_i(\lambda)-\mu_i(\lambda)|
    =|\hat\xi_i(\lambda)-\xi_i(\lambda)|
    \le \varepsilon_{\mathrm{app}}.
    \label{eq:residual-oracle-error}
\end{equation}
This ensures that the approximate eigenbranches $\hat{\xi}_i(\lambda)$ have small deviation from the true ones $\xi_i(\lambda)$ near the target fixed-point.

\paragraph{Secant method.}
A Newton update would require direct access to the derivative of the residual. In the present setting, estimating this derivative quantumly would require a QSVT approximation related to $-f'(x)$, whose magnitude scales as $\mathcal O(1/g^2)$ near the excluded resolvent window, compared with $\mathcal O(1/g)$ for $f(x)$. To avoid this additional derivative-estimation overhead, we instead use the derivative-free secant method.

Given two consecutive energies $\lambda^{(t-1)}$ and $\lambda^{(t)}$, the local slope is approximated by
\begin{equation}
    s^{(t)}
    :=
    \frac{\hat\mu_i(\lambda^{(t)})-\hat\mu_i(\lambda^{(t-1)})}
         {\lambda^{(t)}-\lambda^{(t-1)}},
    \label{eq:secant-slope}
\end{equation}
and the next energy is
\begin{equation}
    \lambda^{(t+1)}
    =\lambda^{(t)}-
    \frac{\hat\mu_i(\lambda^{(t)})}{s^{(t)}}.
    \label{eq:secant-update}
\end{equation}
The local slope automatically adapts to steep eigenbranches without requiring a separate quantum estimation of $\mu_i'(\lambda)$. In our numerical experiments this gives stable and rapid convergence even when the target branch lies close to a resolvent pole.

We initialize the secant method with two nearby points around an initial estimate $\lambda_{\mathrm{init}}$, separated by a small step $h$, and stop once the noisy residual reaches the effective-Hamiltonian approximation scale. The procedure used in this work is summarized below.

\begin{algorithm}[H]
\caption{Secant fixed-point search}
\label{alg:secant}
\begin{algorithmic}[1]
\Require Initial estimate $\lambda_{\mathrm{init}}$, step size $h$, noisy residual $\hat\mu_i(\cdot)$, tolerance $\varepsilon_{\mathrm{app}}$
\Ensure Approximate fixed point $\hat\mu_i(\lambda)\approx 0$
\State $\lambda^{(0)}\leftarrow \lambda_{\mathrm{init}}-h/2$, \quad $\lambda^{(1)}\leftarrow \lambda_{\mathrm{init}}+h/2$
\For{$t=1,\ldots,\text{max-iter}-1$}
    \If{$|\hat\mu_i(\lambda^{(t)})|\le \varepsilon_{\mathrm{app}}$}
        \State \Return $\lambda^{(t)}$
    \EndIf
    \State $s^{(t)}\leftarrow\dfrac{\hat\mu_i(\lambda^{(t)})-\hat\mu_i(\lambda^{(t-1)})}{\lambda^{(t)}-\lambda^{(t-1)}}$
    \State $\lambda^{(t+1)}\leftarrow \lambda^{(t)}-\dfrac{\hat\mu_i(\lambda^{(t)})}{s^{(t)}}$
\EndFor
\If{$|\hat\mu_i(\lambda^{(\text{max-iter})})|\le \varepsilon_{\mathrm{app}}$}
    \State \Return $\lambda^{(\text{max-iter})}$
\Else
    \State \Return ``Failed to converge''
\EndIf
\end{algorithmic}
\end{algorithm}

\paragraph{Error and complexity analysis.}
The relation between residual error and eigenvalue error is independent of the particular local root-finding update. At the exact fixed point,
\[
    p_k=\langle\Psi_k|P|\Psi_k\rangle
    =\left(1-\frac{d\xi_i}{d\lambda}(\lambda_k)\right)^{-1}.
\]
For a nearby point $\lambda$, the mean-value theorem gives a $\lambda^*$ between $\lambda$ and $\lambda_k$ such that
\begin{equation}
    |\mu_i(\lambda)|
    =\frac{|\lambda-\lambda_k|}{p_i(\lambda^*)}.
    \label{eq:residual-to-eigenvalue}
\end{equation}
Therefore, once the search is sufficiently local that $p_i(\lambda^*)=\Theta(p_k)$,
\[
    |\lambda-\lambda_k|
    =\mathcal O\bigl(p_k|\mu_i(\lambda)|\bigr).
\]
If the secant search returns $\hat\lambda_k$ with $|\hat\mu_i(\hat\lambda_k)|\le\varepsilon_{\mathrm{app}}$, then \cref{eq:residual-oracle-error} implies
\[
    |\mu_i(\hat\lambda_k)|
    \le |\hat\mu_i(\hat\lambda_k)|+\varepsilon_{\mathrm{app}}
    \le 2\varepsilon_{\mathrm{app}},
\]
and hence
\begin{equation}
    |\hat\lambda_k-\lambda_k|
    =\mathcal O\left(p_k\varepsilon_{\mathrm{app}}\right).
    \label{eq:secant-eigenvalue-error}
\end{equation}

We next state the local iteration count of the secant solver. Let
$e_t:=\lambda^{(t)}-\lambda_k$ and suppose that the tracked residual is twice continuously differentiable on a neighborhood $J$ of $\lambda_k$. Define
\[
    m_\mu:=\inf_{\lambda\in J}|\mu_i'(\lambda)|,
    \qquad
    M_\mu:=\sup_{\lambda\in J}|\mu_i''(\lambda)|.
\]
By \cref{eq:residual-slope-secant}, $m_\mu\ge 1$. For exact residual evaluations, the standard secant error identity gives, once the iterates lie in $J$,
\begin{equation}
    |e_{t+1}|
    \le C_{\mathrm{sec}}|e_t|\,|e_{t-1}|,
    \qquad
    C_{\mathrm{sec}}:=\frac{M_\mu}{2m_\mu}.
    \label{eq:secant-error-recurrence}
\end{equation}
If the two initial errors satisfy $|e_0|,|e_1|\le\rho$ with
$q:=C_{\mathrm{sec}}\rho<1$, then
\begin{equation}
    |e_t|
    \le \frac{1}{C_{\mathrm{sec}}}
    q^{F_{t+1}},
    \label{eq:secant-fibonacci-bound}
\end{equation}
where $F_t$ is the Fibonacci sequence. Since
$F_t=\Theta(\varphi_{\mathrm{sec}}^t)$ with
$\varphi_{\mathrm{sec}}=(1+\sqrt5)/2$, the secant method has superlinear order
$\varphi_{\mathrm{sec}}$. Thus, to reach the eigenvalue scale
$|e_t|=\mathcal O(p_k\varepsilon_{\mathrm{app}})$ requires
\begin{equation}
    T_{\mathrm{fp}}
    =\mathcal O\!\left(
        \log\log\frac{\rho}{p_k\varepsilon_{\mathrm{app}}}
      \right)
    \label{eq:Tfp-secant}
\end{equation}
effective-Hamiltonian evaluations, up to constants determined by the local branch regularity and the initial secant pair. With the noisy residual oracle, the same local rate applies until the residual reaches the estimation floor, provided the oracle error remains subdominant to the secant differences before termination; the stopping criterion then prevents resolving below the scale $\varepsilon_{\mathrm{app}}$.

\begin{theorem}[Eigenvalue estimation]
\label{thm:eigen-estimation}
Consider a target eigenvalue $\lambda_k$ with normalized eigenvector $|\Psi_k\rangle$, overlap $p_k=\langle\Psi_k|P|\Psi_k\rangle>0$, and pole separation
\[
    \mathrm{dist}(\lambda_k,\mathrm{spec}(H_{QQ}))\ge g.
\]
Assume that there exists a neighborhood $J$ of $\lambda_k$ such that:
\begin{enumerate}[leftmargin=*]
    \item the target eigenbranch is continuously tracked and twice differentiable on $J$;
    \item the secant iterates generated before termination remain in $J$, and
    $\mathrm{dist}(\lambda,\mathrm{spec}(H_{QQ}))\ge g/2$ for all $\lambda\in J$;
    \item the local conditioning factor $p_i(\lambda)=\Big(1-\frac{d\xi_i}{d\lambda}\Big)^{-1}$ satisfies $p_i(\lambda)=\Theta(p_k)$ on $J$;
    \item the initial pair lies in the local secant-convergence region of \cref{eq:secant-error-recurrence}, and the residual-estimation error remains subdominant to the secant differences until the stopping scale $\varepsilon_{\mathrm{app}}$ is reached.
\end{enumerate}
Let $\varepsilon_{\mathrm{app}}:=\varepsilon_{\mathrm{est}}+\|H_{QP}\|^2\varepsilon_f$. If \textnormal{max-iter} is chosen at least as large as the local iteration bound below, then Algorithm~\ref{alg:secant} returns an estimate $\hat\lambda_k$ satisfying
\[
    |\hat\lambda_k-\lambda_k|
    =\mathcal O\left(p_k\varepsilon_{\mathrm{app}}\right),
\]
and the number of effective-Hamiltonian evaluations obeys
\[
    T_{\mathrm{fp}}
    =\mathcal O\!\left(
        \log\log\frac{\rho}{p_k\varepsilon_{\mathrm{app}}}
      \right).
\]
Allocating failure probability $\theta/T_{\mathrm{fp}}$ to each effective-Hamiltonian evaluation gives total success probability at least $1-\theta$. Using iterative QAE for matrix-element estimation, the resulting query complexities are
\begin{enumerate}[leftmargin=*]
    \item
    \(
    \mathcal O\left(
          \frac{\alpha\widetilde\alpha^2d^3}
          {g^2\,\varepsilon_{\mathrm{est}}}
          \log\!\Big(\frac{1}{g\varepsilon_f}\Big)
          \log\!\Big(\frac{dT_{\mathrm{fp}}}{\theta}\Big)
          \log\log\!\Big(\frac{\rho}{p_k\varepsilon_{\mathrm{app}}}\Big)
    \right)
    \)
    queries to $U_{H_{QQ}-\lambda I_Q}$;

    \item
    \(
    \mathcal O\left(
          \frac{\widetilde\alpha^2d^3}
          {g\,\varepsilon_{\mathrm{est}}}
          \log\!\Big(\frac{dT_{\mathrm{fp}}}{\theta}\Big)
          \log\log\!\Big(\frac{\rho}{p_k\varepsilon_{\mathrm{app}}}\Big)
    \right)
    \)
    queries to $U_{H_{QP}}$;

    \item $a+2\widetilde a+1$ ancilla qubits for the self-energy block-encoding.
\end{enumerate}
\end{theorem}

\paragraph{Branch tracking.} As $\lambda$ is updated across different iterations, the index of each eigenbranch $\xi_i(\lambda)$ can be reshuffled and therefore ambiguous. In order to track the index of eigenbranch consistently, we consider that each eigenbranch is defined continuously with the infinitesimal change of $\lambda$.
In practice, this can be resolved by measuring the overlap matrix 
\[
    O_{ij}=\big|\langle\phi_i(\lambda^{(t)})|\phi_j(\lambda^{(t-1)})\rangle\big|,
\]
and then determine a permutation $\pi^\star$ through the following definition
\[
    \pi^\star = \arg\max_{\pi\in S_d} \sum_j O_{\pi(j)j}.
\]
This allows us to update the index $i\leftarrow\pi^\star(i)$ which captures the continuity of eigenbranches with different $\lambda$.

\subsection{Orthonormalization of the (nearly) degenerate subspace basis}
\label{sec:ortho-subspace}

\begin{figure}[t]
\centering
\begin{quantikz}
    \lstick{$|0\rangle^{a+\widetilde a + 2}$}  &  \gate[2]{U_{\widetilde\Omega(\lambda)}}\wire[d]{q} &  \meter{} \\
    \lstick{$|\phi_{k}\rangle$}  & & 
\end{quantikz}
\hspace{1cm}
\begin{quantikz}
    \lstick{$|+\rangle$ / $|\!+\!i\rangle$} & \ctrl{1} & & \gate{H} & \meter{} \\
    \lstick{$|0\rangle^{a+\widetilde a + 2}$} & \wire[d]{q}\wire[u]{q} &  \gate[2]{U_{\widetilde\Omega(\lambda)}}\wire[d]{q} & & \meter{} \\
    \lstick{$|\phi_{k'}\rangle$} & \gate{V_{kk'}} & & &
\end{quantikz}
\caption{\justifying
Quantum circuit for Gram matrix estimation. Left: estimating diagonal elements. Right: estimating off-diagonal elements, the unitary $V_{kk'}$ implements $|\phi_{k'}\rangle \mapsto |\phi_k\rangle$.}
\label{fig:G-matrix-estimation}
\end{figure}

In the (nearly) degenerate case, we can have multiple eigenbranches that satisfy the condition $|\xi_i(\lambda)-\lambda|\le \varepsilon_{\mathrm{app}}$ in the secant fixed-point search. We then use a representative energy $\lambda$ returned from the algorithm as the input of wave operator $\widetilde\Omega(\lambda)$.

We now describe how to construct $m$ (approximately) orthonormal physical states spanning a (nearly) degenerate subspace.
We start from the $P$-space eigenvectors returned by diagonalizing the effective Hamiltonian,
\begin{align}
  \Phi := \big[\,\hat\phi_1,\ldots,\hat\phi_m\,\big]\in\C^{d\times m}.
\end{align}
We lift these eigenvectors in $P\mathcal H$ to the approximate eigenstate in the full Hilbert space $\mathcal H$ using the (approximate) wave operator
$\widetilde\Omega(\lambda)$, implemented via its block-encoding $U_{\widetilde\Omega(\lambda)}$ with known normalization factor $\kappa=\Theta(\sqrt{1+(\widetilde\alpha/g)^2})$:
\begin{align}
  \Psi :=
  \big[\,|\widetilde\Psi_1\rangle,\ldots,|\widetilde\Psi_m\rangle\,\big]
  = \frac{1}{\kappa}\,\widetilde\Omega(\lambda)\,\Phi
  \in \C^{N\times m}.
\end{align}
In general, the columns $|\widetilde\Psi_j\rangle$ are neither normalized nor mutually orthogonal. Their Gram matrix is
\begin{align}
\label{eq:G_Lowdin_def}
  G
  := \Psi^\dagger \Psi
  = \frac{1}{\kappa^2}\,\Phi^\dagger \widetilde\Omega^\dagger(\lambda)\widetilde\Omega(\lambda)\,\Phi
  \in \C^{m\times m},
\end{align}
and the corresponding matrix elements are
\begin{align}
  [G]_{kk'} := \frac{1}{\kappa^2}\langle\hat\phi_k|\widetilde\Omega^\dagger(\lambda)\widetilde\Omega(\lambda)|\hat\phi_{k'}\rangle.
\end{align}
The matrix elements can be estimated by a Hadamard test-like circuit in \cref{fig:G-matrix-estimation}. The diagonal elements can be estimated by measuring the postselection probability:
\begin{equation}
    \mathrm{Pr}(0^{a+\widetilde a + 2})=\frac{1}{\kappa^2} \langle \hat\phi_k|\widetilde\Omega^\dagger(\lambda)\widetilde\Omega(\lambda)|\hat\phi_k\rangle.
\end{equation}
While for the off-diagonal elements, consider the case that the top ancilla initialized in $|+\rangle$, the circuit implements the following mappings:
\begin{align}
    &\xmapsto{c-V_{kk'}}\frac{1}{\sqrt{2}}|0\rangle ||0^{a+\widetilde a+2}\rangle|\phi_{k'}\rangle + \frac{1}{\sqrt{2}}|1\rangle|0^{a+\widetilde a+2}\rangle|\phi_k\rangle
    \nonumber \\[0.2cm]
    &\xmapsto{U_{\widetilde\Omega(\lambda)}} \frac{1}{\sqrt{2}}|0\rangle||0^{a+\widetilde a+2}\rangle\frac{1}{\kappa}\widetilde\Omega(\lambda)|\phi_{k'}\rangle + \frac{1}{\sqrt{2}}|1\rangle||0^{a+\widetilde a+2}\rangle\frac{1}{\kappa}\widetilde\Omega(\lambda)|\phi_k\rangle+|\perp\rangle
    \nonumber \\[0.2cm]
    &\xmapsto{\,\,\,\, H\,\,\,\,} \frac{1}{2\kappa}|0\rangle|0^{a+\widetilde a+2}\rangle\widetilde\Omega(\lambda)\Big(|\phi_{k'}\rangle+|\phi_k\rangle\Big) 
    \nonumber \\[0.2cm]
    &\hspace{4cm} + \frac{1}{2\kappa}|1\rangle|0^{a+\widetilde a+2}\rangle\widetilde\Omega(\lambda)\Big(|\phi_{k'}\rangle-|\phi_k\rangle\Big)+|\perp'\rangle,
\end{align}
where $|\perp\rangle$ and $|\perp'\rangle$ are some unimportant states orthogonal to the states we are interested in. The corresponding difference of ancilla postselection probabilities is
\begin{align}
    \mathrm{Pr}(0, 0^{a+\widetilde a+2})-\mathrm{Pr}(1,0^{a+\widetilde a+2})=\frac{1}{\kappa^2}\mathrm{Re}(\langle\phi_k|\widetilde\Omega^\dagger(\lambda)\widetilde\Omega(\lambda)|\phi_{k'}\rangle).
\end{align}
The analysis of the case $|\!+i\rangle$ is similar and the circuit results in the state:
\begin{align}
    &\frac{1}{2\kappa}|0\rangle|0^{a+\widetilde a+2}\rangle\widetilde\Omega(\lambda)\Big(|\phi_{k'}\rangle+i|\phi_k\rangle\Big) 
    \nonumber \\[0.2cm]
    &\hspace{4cm} + \frac{1}{2\kappa}|1\rangle|0^{a+\widetilde a+2}\rangle\widetilde\Omega(\lambda)\Big(|\phi_{k'}\rangle-i|\phi_k\rangle\Big)+|\perp'\rangle,
\end{align}
and the corresponding difference of the postselection probabilities is
\begin{equation}
    \mathrm{Pr}(0, 0^{a+\widetilde a+2})-\mathrm{Pr}(1,0^{a+\widetilde a+2})=\frac{1}{\kappa^2}\mathrm{Im}(\langle\phi_k|\widetilde\Omega^\dagger(\lambda)\widetilde\Omega(\lambda)|\phi_{k'}\rangle).
\end{equation}
The estimation can be accelerated by quantum amplitude estimation similar to Appendix~\ref{app:matrix-element-estimation} compared to the Monte Carlo estimation.
Assuming success probability at least $1-\theta$ and the  additive error
$\varepsilon$ (in operator norm), the number of queries to $U_{\widetilde\Omega(\lambda)}$ required to estimate $G$ scales as
\begin{align}
  \mathcal O\left(\frac{\kappa^2 m^3}{\varepsilon}\log\frac{m}{\theta}\right)
  =
  \mathcal O\left(\frac{\widetilde \alpha^2 m^3}{g^2\,\varepsilon}\log\frac{m}{\theta}\right).
\end{align}

Now, given an estimate $\hat G$ of $G$, we compute the Löwdin orthonormalization matrix $\hat G^{-1/2}$ and form the orthonormalized columns
\begin{align}
  \Psi_0 := \widetilde\Psi\,\hat G^{-1/2}
  = \frac{1}{\kappa}\,\widetilde\Omega(\lambda)\,\Phi\,\hat G^{-1/2},
\end{align}
which satisfy
\begin{align}
  \Psi_0^\dagger \Psi_0 = \hat G^{-1/2} G\, \hat G^{-1/2} \approx I_m.
\end{align}
Equivalently, define the updated $P$-space vectors
\begin{align}
  \Phi_0 := \Phi\,\hat G^{-1/2}\Big(\mathrm{diag}(\hat G^{-1})\Big)^{-1/2},
  \qquad
  |\phi_{0,k}\rangle = \frac{\sum_{k=1}^m |\hat\phi_{k'}\rangle\, [\hat G^{-1/2}]_{k'k}}{\Big\| \sum_{k=1}^m |\hat\phi_{k'}\rangle\, [\hat G^{-1/2}]_{k'k}\Big\|},
\end{align}
where the extra $\Big(\mathrm{diag}(\hat G^{-1})\Big)^{-1/2}$ is to ensure each column of $\Phi_0$ is normalized and therefore each $|\phi_{0,k}\rangle$ is a normalized state.
Then the lifted states by this new basis is
\begin{align}
  |\Psi_{0,k}\rangle := \frac{1}{\kappa}\,\widetilde\Omega(\lambda)\,|\phi_{0,k}\rangle.
\end{align}
Up to these known normalization factors and the estimation error in $\hat G$, the states
$\{|\Psi_{0,k}\rangle\}_{k=1}^m$ form an (approximately) orthonormal basis spanning the (nearly) degenerate subspace.


\section{Fidelity Bounds and Performance Guarantees}
\label{sec:fidelity-bound}

After the eigenvalue-estimation stage returns a estimated eigenvalue
$\hat\lambda_k$ and a corresponding normalized eigenvector
$|\hat\phi_k\rangle\in P\mathcal H$ of
$\hat H_{\mathrm{eff}}(\hat\lambda_k)$, the approximate full-space
state is obtained by applying the block-encoded wave operator
$U_{\widetilde\Omega(\hat\lambda_k)}$. We first quantify the
additional cost of this lifting step and then state the resulting
state- and subspace-fidelity guarantees.

\subsection{Complexity of wave-operator state preparation}
\label{sec:wave-operator-preparation-complexity}

The wave-operator block encoding satisfies
\begin{equation}
\label{eq:wave-operator-block-action}
    U_{\widetilde\Omega(\hat\lambda_k)}
    |0^{a+\widetilde a+2}\rangle|\hat\phi_k\rangle
    =
    |0^{a+\widetilde a+2}\rangle
    \frac{
        \widetilde\Omega(\hat\lambda_k)|\hat\phi_k\rangle
    }{\kappa}
    +
    |\perp_k\rangle,
\end{equation}
where $|\perp_k\rangle$ has no support on the all-zero ancilla
subspace and
\begin{equation}
    \kappa
    =
    \Theta\!\left(
        \sqrt{
            1+\frac{\widetilde\alpha^2}{g^2}
        }
    \right).
\end{equation}
Define
\begin{equation}
\label{eq:eta-k-main}
    \eta_k
    \coloneqq
    \left\|
        \widetilde\Omega(\hat\lambda_k)
        |\hat\phi_k\rangle
    \right\|^{-2}.
\end{equation}
Postselecting the ancillas in the all-zero state prepares the
normalized lifted state
\begin{equation}
\label{eq:normalized-lifted-eigenstate}
    |\widetilde\Psi_k\rangle
    \coloneqq
    \sqrt{\eta_k}\,
    \widetilde\Omega(\hat\lambda_k)|\hat\phi_k\rangle
\end{equation}
with success probability
\begin{equation}
\label{eq:wave-operator-success-probability}
    \mathrm{Pr}(0^{a+\widetilde a+2}, k)
    =
    \frac{1}{\kappa^2\eta_k}.
\end{equation}
Under the norm-stability assumption
\begin{equation}
\label{eq:eta-overlap-stability}
    \eta_k
    =
    \mathcal O(p_k),
    \qquad
    p_k
    \coloneqq
    \langle\Psi_k|P|\Psi_k\rangle,
\end{equation}
amplitude amplification prepares
$|\widetilde\Psi_k\rangle$ with failure probability at most $\theta$
using
\begin{equation}
\label{eq:wave-operator-AA-calls}
    \mathcal O\!\left(
        \kappa\sqrt{p_k}
        \log\frac{1}{\theta}
    \right)
\end{equation}
applications of $U_{\widetilde\Omega(\hat\lambda_k)}$ and its
adjoint. We assume the coherent reduced-state access required by the
amplitude-amplification procedure and do not include its
implementation cost.

Combining Eq.~\eqref{eq:wave-operator-AA-calls} with the cost of one
wave-operator invocation from
Lemma~\ref{lem:block-encoding-sigma-omega}, the additional query
complexity after $\hat\lambda_k$ has been returned is
\begin{align}
\label{eq:eigenstate-preparation-QQ-cost}
    &\mathcal O\!\left(
        \kappa\sqrt{p_k}\,
        \frac{\alpha}{g}
        \log\frac{1}{g\varepsilon_f}
        \log\frac{1}{\theta}
    \right)
    &&\text{queries to }
    U_{H_{QQ}-\hat\lambda_k I_Q},
    \\[4pt]
\label{eq:eigenstate-preparation-QP-cost}
    &\mathcal O\!\left(
        \kappa\sqrt{p_k}
        \log\frac{1}{\theta}
    \right)
    &&\text{queries to }
    U_{H_{QP}}
    \text{ and }U_{H_{QP}}^\dagger.
\end{align}
The corresponding gate complexity is
\begin{equation}
\label{eq:eigenstate-preparation-gate-cost}
    \mathcal O\!\left(
        \kappa\sqrt{p_k}\,
        \frac{
            \alpha(a+\widetilde a+n)
        }{g}
        \log\frac{1}{g\varepsilon_f}
        \log\frac{1}{\theta}
    \right),
\end{equation}
with $a+\widetilde a+\mathcal O(1)$ ancillary qubits. These resource
counts are conditional on successful eigenvalue estimation and do
not include the cost of obtaining $\hat\lambda_k$.

For an $m$-dimensional target subspace, the procedure is applied to
the $m$ orthonormalized reduced-space inputs obtained from the
L\"owdin transformation. If their inverse squared lifted norms are
uniformly $\mathcal O(p_{\max})$, preparing all $m$ output basis
states with total failure probability at most $\theta$ requires
\begin{equation}
\label{eq:subspace-basis-wave-calls}
    \mathcal O\!\left(
        m\kappa\sqrt{p_{\max}}
        \log\frac{m}{\theta}
    \right)
\end{equation}
applications of the wave-operator block encoding and its adjoint, in
addition to the Gram-matrix estimation and classical L\"owdin
transformation described in Section~\ref{sec:ortho-subspace}.

\subsection{Residual and fidelity measures}

The fidelity analysis is based on the residual
\begin{equation}
\label{eq:state_residual_def}
    |r_k\rangle
    \coloneqq
    (H-\hat\lambda_k I)|\widetilde\Psi_k\rangle.
\end{equation}
A small residual, together with a gap separating the target spectral
subspace from the complementary spectrum, controls the corresponding
principal angles through a Davis--Kahan-type perturbation bound
\cite{davis_rotation_1970}. The residual estimates required below
are derived in Appendix~\ref{app:fidelity_proof}.

For a simple eigenvalue, we use the squared pure-state fidelity
\begin{equation}
    F(\Psi_k,\widetilde\Psi_k)
    \coloneqq
    |\langle\Psi_k|\widetilde\Psi_k\rangle|^2.
\end{equation}
For two $m$-dimensional subspaces with orthonormal basis matrices
$A,B\in\mathbb C^{N\times m}$, let
\begin{equation}
    A^\dagger B
    =
    U\,
    \operatorname{diag}
    (\cos\theta_1,\ldots,\cos\theta_m)
    V^\dagger
\end{equation}
define their principal angles. We use
\begin{equation}
    F_{\mathrm{avg}}
    \coloneqq
    \frac{1}{m}
    \sum_{j=1}^m\cos^2\theta_j,
    \qquad
    F_{\min}
    \coloneqq
    \cos^2\theta_{\max}.
\end{equation}

\begin{theorem}[Fidelity guarantees for eigenstate and subspace preparation]
\label{thm:prep-fidelity}
Define
\begin{equation}
    \varepsilon_{\mathrm{app}}
    \coloneqq
    \varepsilon_{\mathrm{est}}
    +
    \|H_{QP}\|^2\varepsilon_f,
    \qquad
    \varepsilon
    \coloneqq
    \varepsilon_{\mathrm{est}}
    +
    \varepsilon_{\mathrm{app}}
    +
    \alpha\|H_{QP}\|\varepsilon_f.
    \label{eq:total_error_budget}
\end{equation}
Assume that the fixed-point routine returns the approximate eigenpair $(\hat\lambda_k,|\hat\phi_k\rangle)$ and that
\begin{equation}
    \eta_k=\left\|
        \widetilde\Omega(\hat\lambda_k)
        |\hat\phi_k\rangle
    \right\|^{-2}
    =
    \mathcal O(p_k).
\end{equation}

\begin{enumerate}[leftmargin=*]

    \item \emph{Non-degenerate case.}
    Let $\lambda_k$ be simple and let
    \begin{equation}
        \Delta
        \coloneqq
        \operatorname{dist}\!\left(
            \hat\lambda_k,
            \operatorname{spec}(H)\setminus\{\lambda_k\}
        \right)
        >0.
    \end{equation}
    Then
    \begin{equation}
    \label{eq:fidelity_nondeg}
        1-F(\Psi_k,\widetilde\Psi_k)
        =
        \mathcal O\!\left(
            \frac{p_k\varepsilon^2}{\Delta^2}
        \right).
    \end{equation}

    \item \emph{Subspace case.}
    Let $\mathcal S$ be an $m$-dimensional target spectral subspace,
    and let
    \begin{equation}
        \Phi
        =
        \bigl[
            |\hat\phi_1\rangle,\ldots,|\hat\phi_m\rangle
        \bigr],
        \qquad
        \Phi^\dagger\Phi=I_m,
    \end{equation}
    contain the selected eigenvectors of
    $\hat H_{\mathrm{eff}}(\hat\lambda)$. Define
    $\hat\Xi$ by $\hat H_{\mathrm{eff}}(\hat\lambda)\Phi = \Phi\hat\Xi$,
    such that
    $\|\hat\Xi-\hat\lambda I_m\|\le\varepsilon_{\mathrm{app}}.$
    Let
    \begin{equation}
        \Delta
        =
        \mathrm{dist}\left(\hat\lambda,\mathrm{spec}(H)
        \setminus \mc S\right).
    \end{equation}
    If
    \begin{equation}
        \widetilde B
        =
        \bigl[
            |\widetilde\Psi_1\rangle,\ldots,
            |\widetilde\Psi_m\rangle
        \bigr],
        \qquad
        G
        \coloneqq
        \widetilde B^\dagger\widetilde B
        \succ0,
    \end{equation}
    then the L\"owdin-orthonormalized basis
    $B=\widetilde B G^{-1/2}$ satisfies
    \begin{align}
    \label{eq:fidelity_deg_min}
        1-F_{\min}
        &=
        \mathcal O\!\left(
            \frac{
                \|G^{-1}\|p_{\max}\varepsilon^2
            }{\Delta^2}
        \right),
        \\
    \label{eq:fidelity_deg_avg}
        1-F_{\mathrm{avg}}
        &=
        \mathcal O\!\left(
            \frac{
                \|G^{-1}\|p_{\mathrm{avg}}\varepsilon^2
            }{\Delta^2}
        \right),
    \end{align}
    where
    \begin{equation}
        p_{\max}
        \coloneqq
        \max_{1\le k\le m}p_k,
        \qquad
        p_{\mathrm{avg}}
        \coloneqq
        \frac{1}{m}\sum_{k=1}^m p_k.
    \end{equation}

\end{enumerate}
\end{theorem}

If $\Delta$ in the non-degenerate case is defined using the exact
eigenvalue $\lambda_k$, the same scaling holds whenever
$|\hat\lambda_k-\lambda_k|$ is at most a fixed fraction of that gap.
The amplitude-amplification failure probability is distinct from the
fidelity error: the former controls whether state preparation
succeeds, whereas Theorem~\ref{thm:prep-fidelity} controls the
accuracy of the normalized successful output. Allocating failure
probability $\theta/2$ to each of the eigenvalue-estimation and
state-preparation stages gives total success probability at least
$1-\theta$.

\section{Numerical Results}

In this section, we present numerical results for three representative systems with low-energy crossings or near-degeneracies. Section~\ref{sec:hubbard-model} studies an open $4\times2$ Fermi--Hubbard model with one hole away from half filling and resolves a finite-size Nagaoka-type crossing between the lowest $S=1/2$ and $S=3/2$ symmetry sectors. Section~\ref{sec:Molecular-benchmark-LiH} considers all-electron LiH along bond stretching, where the lowest singlet and triplet states approach one another toward dissociation. Section~\ref{sec:transition-metal-complex} examines $[\mathrm{Ru(bpy)}_3]^{2+}$ as a larger active-space molecular example. The system configuration can be found in Table~\ref{tab:benchmark-summary}.

Throughout this section, \emph{our method} refers to the complete numerical workflow consisting of the QSVT reciprocal approximation, QAE-based estimation of the effective-Hamiltonian matrix elements, and the secant fixed-point search. We use the convex-optimization-based implementation in QSPPACK to construct QSP phase sequences for a Chebyshev approximation of $p(x)=\frac{\delta}{\beta x}$~\cite{dong_efficient_2021}. For a fixed trial point $\lambda$, the QSVT polynomial is deterministic, so its finite-degree reciprocal approximation contributes a deterministic error. By contrast, the matrix-estimation stage is stochastic. We emulate ideal canonical quantum amplitude estimation (QAE)~\cite{brassard_quantum_2002} by sampling from its exact ideal outcome distribution whenever the secant solver requests a self-energy matrix element. Consequently, every reconstructed self-energy matrix, residual evaluation, and subsequent secant trajectory fluctuates from run to run. Each reported trial is an end-to-end run with fresh QAE samples at every matrix-element query. For Hubbard and LiH we perform ten independent trials at each parameter point and target sector, reporting the median together with the empirical $5$th--$95$th percentile interval. These error bars therefore quantify the stochastic QAE sampling propagated through the nonlinear secant search, while the QSVT approximation error is common to all trials.

\begin{table}[t]
\centering
\tiny
\begin{tblr}{
    width=\textwidth,
    colspec={X[1.35,l] X[0.8,c] X[1.0,c] X[1.85,l]},
    row{1}={font=\bfseries},
    cells={font=\small}
}
\hline
System & Hilbert dimension & $P$-space dimension & Target \\
\hline
Open $4\times2$ Hubbard
& $3{,}920$
& $4$ ($S=1/2$), $5$ ($S=3/2$)
& Lowest states in the two symmetry sectors across the Nagaoka-type crossing \\
\hline
LiH (6-31G)
& $3{,}025$
& $6$ singlet, $3$ triplet
& Lowest $A_1$ singlet and triplet along $R=0.8$--$6.0$~\AA \\
\hline
$[\mathrm{Ru(bpy)}_3]^{2+}$ CAS(12,9)
& $18{,}564$ (full CAS)
& $19$ singlet, $18$ triplet
& $13$ singlet roots, $18$ triplet roots, and the rank-$3$ $T_1$ multiplet \\
\hline
\end{tblr}
\caption{Summary of the numerical benchmarks and the corresponding target subspaces.}
\label{tab:benchmark-summary}
\end{table}

\subsection{Fermi--Hubbard Benchmark: Nagaoka-Type Spin-Sector Crossing}
\label{sec:hubbard-model}

\begin{figure}[t]
    \centering
    \includegraphics[width=0.98\linewidth]{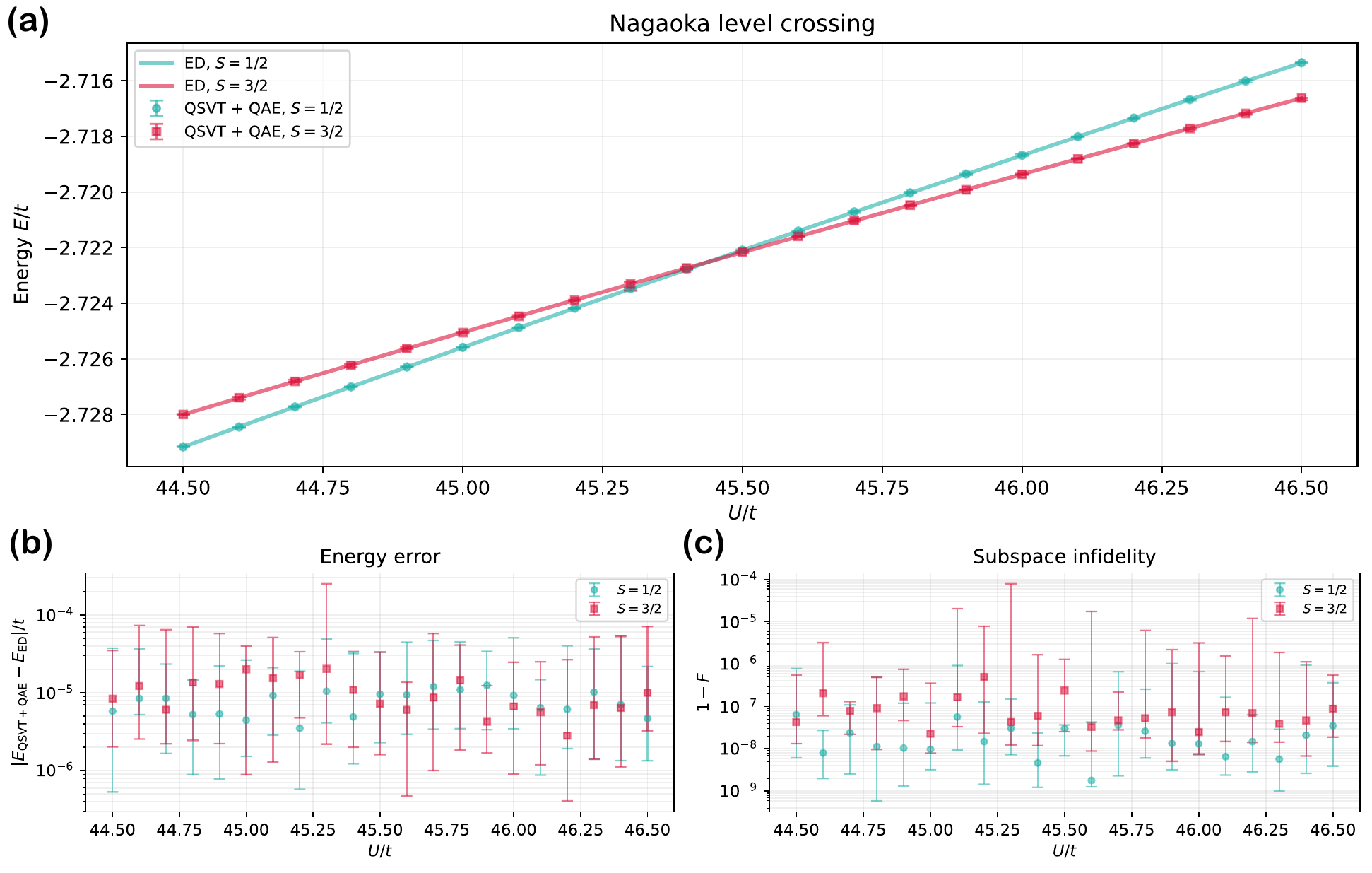}
    \caption{\justifying
    Open $4\times2$ Fermi--Hubbard benchmark with $(N_\uparrow,N_\downarrow)=(4,3)$ and $t=1$.
    (a) Lowest energies in the $S=1/2$ and $S=3/2$ symmetry sectors as functions of $U/t$. Solid curves are exact-diagonalization (ED) references, while markers show the median energies returned by our method over ten independent trials; the two sectors cross near $U/t\simeq45.44$.
    (b) Absolute energy error of our method, $|E_{\mathrm{method}}-E_{\mathrm{ED}}|/t$.
    (c) Infidelity $1-F$ of the wave-operator-reconstructed target state/subspace.
    For the stochastic quantities, markers denote medians and error bars denote the empirical $5$th--$95$th percentile interval over the ten end-to-end trials, with fresh ideal-QAE samples drawn for every self-energy matrix-element query during the secant search.
    }
    \label{fig:hubbard}
\end{figure}

As a many-body benchmark, we consider the repulsive Fermi--Hubbard model on an open $4\times2$ rectangular lattice,
\begin{equation}
    H = -t \sum_{\langle i,j \rangle,\sigma}
    \left(c_{i\sigma}^{\dagger}c_{j\sigma}+c_{j\sigma}^{\dagger}c_{i\sigma}\right)
    +U\sum_i n_{i\uparrow}n_{i\downarrow},
\end{equation}
with $t=1$. We work in the fixed-spin sector $(N_\uparrow,N_\downarrow)=(4,3)$, corresponding to seven electrons on eight sites, i.e., one hole away from half filling. The determinant-space dimension is
\begin{equation}
    \binom{8}{4}\binom{8}{3}=3920.
\end{equation}
The one-hole, strong-coupling Hubbard model is the canonical setting of Nagaoka ferromagnetism: in the $U\rightarrow\infty$ limit and under the appropriate connectivity and hopping-sign conditions, kinetic motion of the hole can stabilize a maximal-spin ground state~\cite{nagaoka_ferromagnetism_1966,tasaki_nagaoka_1998}. At finite $U$ and in finite clusters, the same competition between suppression of double occupancy and kinetic delocalization can appear through crossings or near-degeneracies among different total-spin sectors~\cite{yun_benchmark_2021,botzung_numerical_2024}. Our small open cluster is therefore used as a finite-size \emph{Nagaoka-type} benchmark rather than as a determination of the thermodynamic Nagaoka phase boundary. In the interaction window considered below, the lowest competing states are in the $S=1/2$ and $S=3/2$ sectors.

We resolve lattice-reflection parities in addition to total spin and focus on the two blocks participating in the lowest crossing. The $S=1/2$ block has $(p_x,p_y)=(-,+)$ and dimension $588$, while the $S=3/2$ block has $(p_x,p_y)=(-,-)$ and dimension $336$. For each block, the reference space is frozen from the strong-coupling limit: we first restrict to the no-doublon subspace of the $U\rightarrow\infty$ problem and diagonalize the kinetic Hamiltonian there, then retain the lowest four states for $S=1/2$ and the lowest five states for $S=3/2$. Thus the nonlinear effective-Hamiltonian problems have only $d=4$ and $d=5$ dimensions, respectively, despite the much larger symmetry blocks.

We scan $21$ equally spaced values of $U/t$ from $44.5$ to $46.5$. The secant solver uses residual tolerance $10^{-10}$, initial step $h=10^{-3}$, at most $100$ iterations, and an initial-energy offset of $0.2t$ when no previous root is available. For the QSVT reciprocal polynomial $p(x)=\frac{\delta}{\beta x}$ we use
\begin{equation}
    \delta=\frac{1}{200},\qquad
    \beta=2,
    \qquad \deg p = 2851,
\end{equation}
with target polynomial error $\varepsilon_{\mathrm{poly}}=10^{-6}$. For the ideal canonical-QAE sampling we use a $24$-qubit evaluation register, corresponding to $M_{\mathrm{QAE}}=2^{24}$ phase-estimation grid points and phase-grid spacing $2^{-24}$. This register size controls the resolution of each individual ideal-QAE sample; the resulting matrix and eigenvalue errors are determined empirically after propagation through the self-energy reconstruction and secant iterations. Ten independent end-to-end trials are performed for each $U/t$ and spin block.

Figure~\ref{fig:hubbard}(a) shows the finite-size spin-sector crossing. The ED curves intersect at $U/t\simeq45.44$, and the median estimates returned by our method follow both branches through the crossing on the scale of the plotted energies. Around the crossing, the identity of the lowest state among the two competing symmetry sectors changes even though the two energies become nearly degenerate. The benchmark therefore tests not only the accuracy of an isolated root, but also whether the complete effective-Hamiltonian workflow preserves the relative ordering of competing low-energy states.

The quantitative errors are shown in Fig.~\ref{fig:hubbard}(b). Across the full sweep, the median absolute energy errors of our method are predominantly at the $10^{-6}$--$10^{-5}t$ level. The percentile intervals show the run-to-run spread induced by QAE sampling and its nonlinear propagation through the secant search; they remain small compared with the energy scale on which the spin-sector crossing is resolved. Figure~\ref{fig:hubbard}(c) shows the corresponding wave-operator reconstruction. Median infidelities are generally between $10^{-9}$ and $10^{-7}$, with occasional values reaching a few $10^{-7}$ in the $S=3/2$ sector. The small reconstruction errors demonstrate that the low-dimensional $P$ spaces retain sufficient information to recover the full-space low-energy states even near the crossing.

The Hubbard Hamiltonian and symmetry-resolved determinant basis are constructed with OpenFermion~\cite{mcclean_openfermion_2020}.

\subsection{Molecular Benchmark: LiH Bond Stretching}
\label{sec:Molecular-benchmark-LiH}

\begin{figure}[t]
    \centering
    \includegraphics[width=0.98\linewidth]{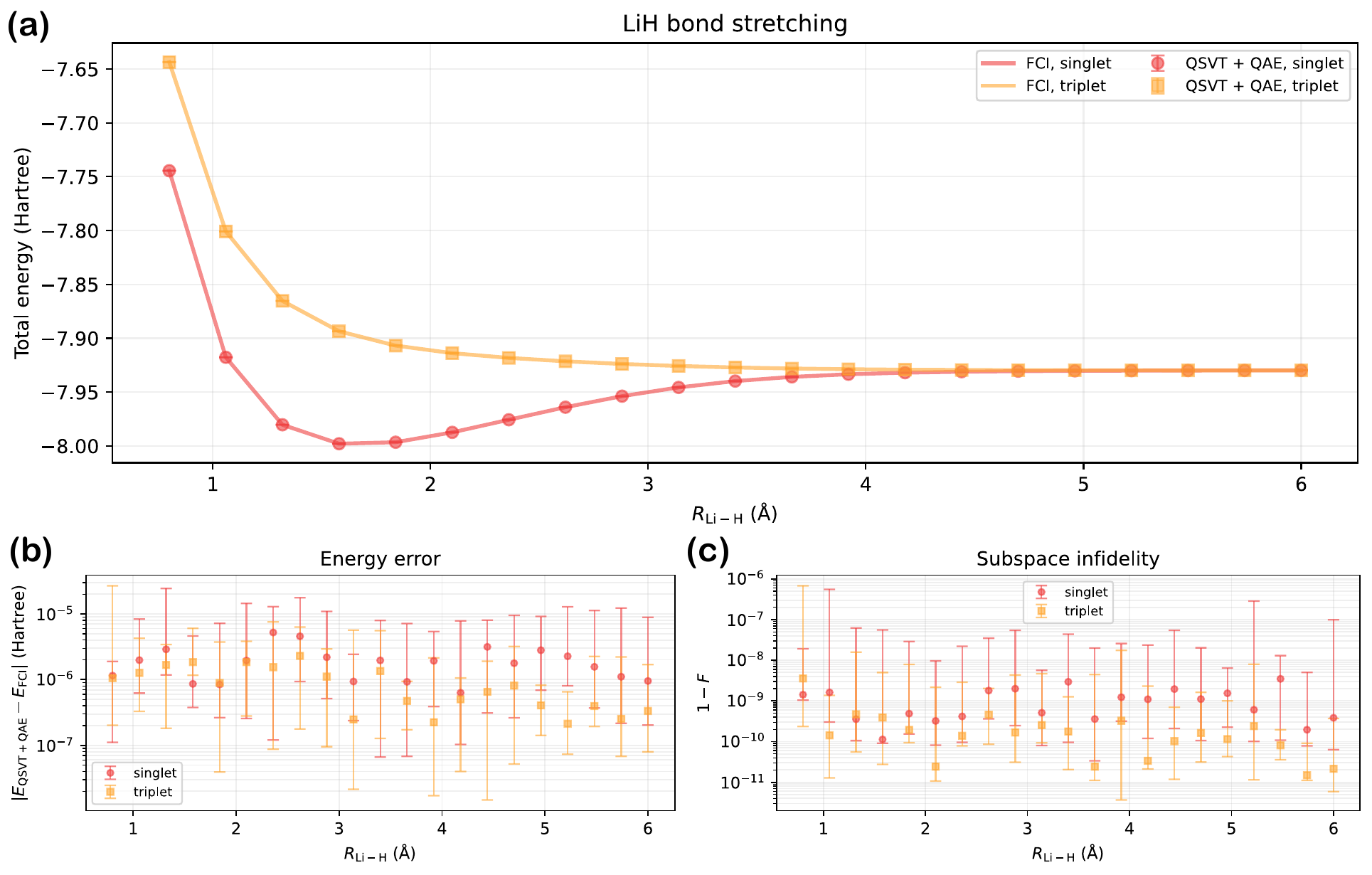}
    \caption{\justifying
    All-electron LiH bond stretching in the 6-31G basis.
    (a) Lowest $A_1$ singlet and triplet total energies from full configuration interaction (FCI, solid curves) and from our method (markers) over $R_{\mathrm{Li-H}}=0.8$--$6.0$~\AA.
    (b) Absolute energy error of our method, $|E_{\mathrm{method}}-E_{\mathrm{FCI}}|$.
    (c) Infidelity $1-F$ of the wave-operator-reconstructed target state/subspace.
    Markers denote medians over ten independent end-to-end trials and error bars denote the empirical $5$th--$95$th percentile intervals. In each trial, every self-energy matrix-element query during the secant search is independently sampled from the ideal canonical-QAE distribution.
    }
    \label{fig:LiH_energy_spectrum}
\end{figure}

\begin{figure}[t]
    \centering
    \includegraphics[width=\linewidth]{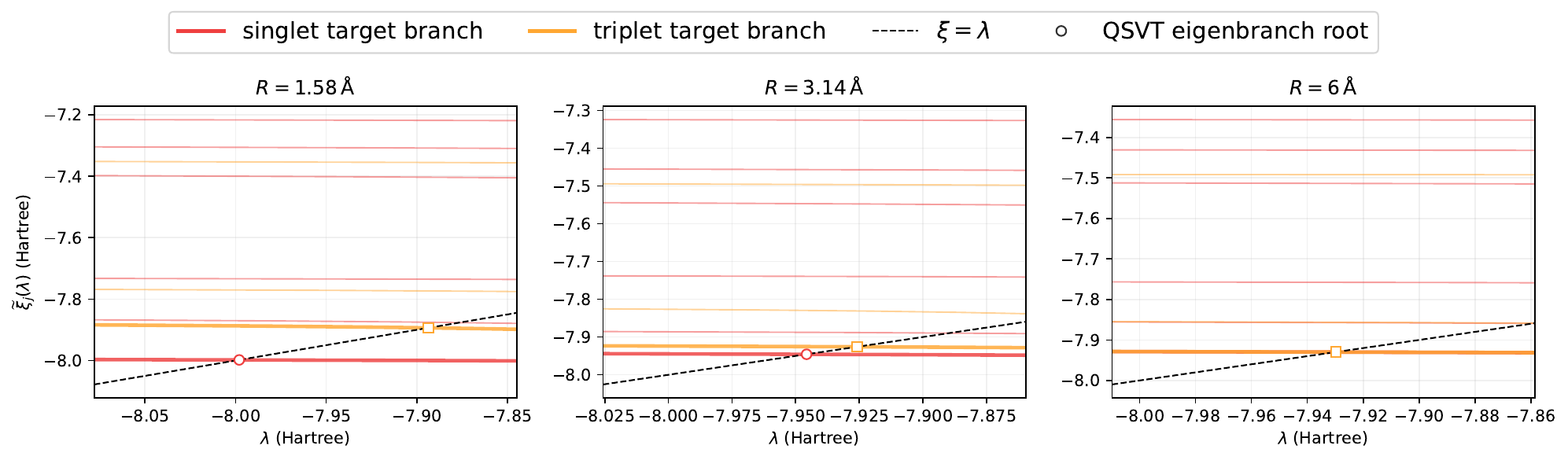}
    \caption{\justifying
    Effective-Hamiltonian eigenbranches $\widetilde\xi_j(\lambda)$ for LiH at representative bond lengths $R=1.58$, $3.14$, and $6.0$~\AA. The smooth curves are obtained from the deterministic QSVT reciprocal approximation: the thick red and orange curves are the target singlet and triplet branches, respectively, and thinner curves are the remaining eigenbranches in the corresponding spin-adapted $P$ spaces. The dashed line is $\xi=\lambda$, so its intersections with the target branches are fixed points of the nonlinear eigenvalue problem. Open markers indicate the median fixed-point estimates returned by our method at the corresponding geometries. The singlet--triplet separation decreases markedly on stretching and is nearly closed at $R=6.0$~\AA.
    }
    \label{fig:LiH_eigenbranches}
\end{figure}

We next consider LiH as a molecular benchmark in which the character of the low-energy wavefunction changes continuously from the equilibrium region toward dissociation. The calculation is all-electron in the 6-31G basis with four electrons and $(N_\alpha,N_\beta)=(2,2)$. The resulting determinant space contains $3025$ configurations. At each bond length we perform a restricted Hartree--Fock calculation with convergence tolerance $10^{-10}$ and construct the electronic Hamiltonian in the resulting molecular-orbital basis using PySCF~\cite{sun_scf_2018}. The Li $1s$-like lowest $A_1$ orbital is kept doubly occupied, while the next three $A_1$ ($\sigma$) orbitals define a CAS(2,3) reference space. With one active $\alpha$ and one active $\beta$ electron, this gives nine reference determinants. Exact spin adaptation separates them into a six-dimensional singlet block and a three-dimensional triplet block; the complementary $Q$ space therefore has dimension $3016$. The targets are the lowest $A_1$ singlet and triplet FCI states of the full $3025$-dimensional determinant problem.

We scan $21$ equally spaced bond lengths over $R_{\mathrm{Li-H}}=0.8$--$6.0$~\AA. The secant solver again uses residual tolerance $10^{-10}$, step $h=10^{-3}$, and at most $100$ iterations, with an initial-energy offset of $0.02$ Hartree. For LiH the reciprocal polynomial $p(x)=\frac{\delta}{\beta x}$ is chosen with
\begin{equation}
    \delta=\frac{1}{100},\qquad
    \beta=2,
    \qquad \deg p = 1451,
\end{equation}
with target polynomial error $\varepsilon_{\mathrm{poly}}=10^{-6}$. The ideal canonical-QAE sampling uses a $20$-qubit evaluation register, i.e., $M_{\mathrm{QAE}}=2^{20}$ phase-estimation grid points with phase-grid spacing $2^{-20}$. As for Hubbard, each geometry and spin sector is evaluated with ten independent end-to-end trials using fresh QAE samples at every matrix-element query.

Figure~\ref{fig:LiH_energy_spectrum}(a) shows the bond-stretching curves. Near the equilibrium region the singlet is well separated from the triplet; for example, at $R=1.58$~\AA{} the FCI reference energies are
\begin{equation}
    E_{S=0}=-7.9979911951~\mathrm{Ha},\qquad
    E_{S=1}=-7.8935556629~\mathrm{Ha},
\end{equation}
corresponding to a singlet--triplet gap of approximately $0.10444$ Hartree. Upon stretching, the singlet rises while the triplet approaches the same dissociation energy, so the two states become nearly degenerate at large $R$. The median energies returned by our method are visually indistinguishable from the FCI curves on the scale of panel (a), including the strongly correlated stretched-bond regime.

The absolute errors in Fig.~\ref{fig:LiH_energy_spectrum}(b) remain far below chemical accuracy throughout the scan. The median energy errors of our method lie between roughly $10^{-7}$ and a few $10^{-6}$ Hartree, while the empirical $5$th--$95$th percentile intervals are typically below a few $10^{-5}$ Hartree. Figure~\ref{fig:LiH_energy_spectrum}(c) shows that the wave operator also reconstructs the full-space states accurately: median infidelities are generally in the $10^{-11}$--$10^{-9}$ range, and even the upper percentile remains below the $10^{-6}$ level over the displayed geometries. The percentile spread here reflects stochastic QAE matrix-element sampling propagated through the complete root-search and reconstruction workflow. These results are particularly notable near dissociation, where a compact single-determinant description deteriorates but the spin-adapted CAS(2,3) reference space continues to support both target states.

The fixed-point geometry is illustrated directly in Fig.~\ref{fig:LiH_eigenbranches}. At $R=1.58$~\AA, the target singlet and triplet branches intersect $\xi=\lambda$ at well-separated energies. At $R=3.14$~\AA{} the separation has narrowed substantially, while at $R=6.0$~\AA{} the two fixed points are nearly coincident. In all three cases the target roots are identified using small spin-adapted effective Hamiltonians of dimensions six and three. The figure therefore complements the bond-stretching accuracy data by showing explicitly how the fixed-point construction continues to identify the relevant branches as the molecule evolves from a well-separated spectrum to a near-degenerate singlet--triplet pair.

\subsection{Transition Metal Complex Benchmark: [Ru(bpy)$_3$]$^{2+}$}
\label{sec:transition-metal-complex}

As a larger molecular benchmark, we consider tris(bipyridine)ruthenium(II), $[\mathrm{Ru(bpy)}_3]^{2+}$, whose low-energy spectrum contains dense metal-to-ligand charge-transfer (MLCT) and ligand-to-ligand charge-transfer (LLCT) manifolds. We use the geometry and active-space construction of Yang et al.~\cite{yang_theoretical_2022}; the active orbitals were selected from orbital-composition and symmetry analysis, with additional inspection using Multiwfn~\cite{lu_multiwfn_2012}. The underlying molecular calculation uses the def2-SVP basis and the corresponding relativistic effective-core potential for Ru, with electronic-structure quantities generated in PySCF~\cite{sun_scf_2018}. A state-averaged CASSCF(12,9) calculation targeting 13 singlet roots provides a frozen active-space Hamiltonian and reference states for the present benchmark.

The nine active spatial orbitals consist of three predominantly Ru $d$ orbitals, three occupied ligand $\pi$ orbitals, and three ligand $\pi^*$ orbitals. With 12 active electrons, the unrestricted fixed-particle-number CAS space has dimension
\begin{equation}
    \binom{18}{12}=18{,}564.
\end{equation}
Rather than treating this full space as one block, we exploit spin projection. The $M_S=0$ determinant sector has dimension $\binom{9}{6}^2=7056$, whereas the $M_S=\pm1$ triplet sectors have dimension $\binom{9}{7}\binom{9}{5}=4536$. The frozen singlet reference data contain 13 validated CASCI roots: the ground state $S_0$, nine low-lying MLCT states, and three LLCT states. The spin-free Hamiltonian is also used to obtain the triplet manifold independently.

For the present benchmark, the $P$ spaces are chemically structured and spin adapted. For the singlet calculation, $P$ has dimension $d=19$ and contains the closed-shell reference together with nine spin-adapted MLCT single excitations and nine spin-adapted LLCT single excitations. For the triplet calculation, each fixed-$M_S$ sector uses an $18$-dimensional MLCT/LLCT reference space; in the $M_S=0$ sector these are the antisymmetric triplet-adapted combinations orthogonal to the singlet reference configurations. In the classical emulation, $Q$-eigenmodes whose coupling to $P$ is below $10^{-8}$ are discarded as a numerical compression; this leaves 2501 coupled $Q$ modes in the singlet problem and 3384 in each triplet problem. The corresponding Feshbach identities are verified numerically before applying the approximate resolvent.

For the QSVT reciprocal we use
\begin{equation}
    \delta=\frac{1}{300},\qquad
    \beta=2,\qquad
    \deg p=3937,
\end{equation}
with target polynomial error $10^{-6}$. Matrix elements are sampled from ideal canonical QAE using a $24$-qubit evaluation register, corresponding to $M_{\mathrm{QAE}}=2^{24}$ phase-grid points, and ten independent end-to-end trajectories are generated for every target root. The deterministic secant search uses an initial step $h=10^{-3}$, residual tolerance $10^{-10}$, and at most $100$ iterations; the QAE-sampled searches terminate at the corresponding QAE resolution floor. For this multiroot validation, target labels are assigned by maximal overlap between the computed eigenbranches and the corresponding CASCI $P$-space components. This branch association is used only to identify the reference state against which the root and reconstructed subspace are benchmarked.

\begin{figure}[t!]
    \centering
    \includegraphics[width=0.98\linewidth]{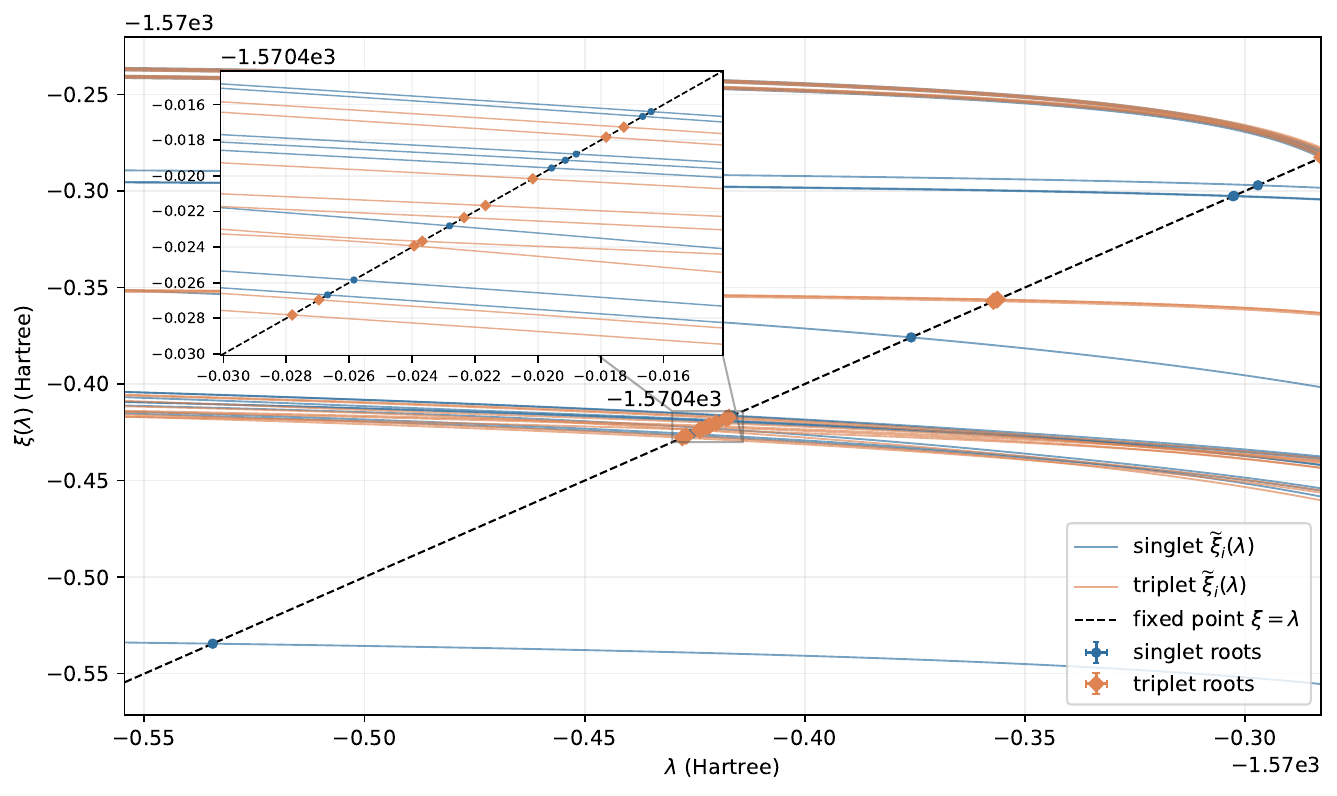}
    \caption{\justifying
    Effective-Hamiltonian eigenbranches for the $[\mathrm{Ru(bpy)}_3]^{2+}$ CAS(12,9) benchmark. Blue and orange curves show the singlet ($d=19$) and representative $M_S=+1$ triplet ($d=18$) QSVT eigenbranches, respectively, while the dashed line denotes $\xi=\lambda$. Their intersections are the nonlinear fixed points. Markers show median roots returned by our method over ten ideal-QAE trajectories, with horizontal error bars spanning the observed minimum and maximum. The inset enlarges the dense low-energy excited-state region. All 13 singlet roots and all 18 roots of the representative triplet sector are searched.
    }
    \label{fig:Ru_eigenbranches}
\end{figure}

Figure~\ref{fig:Ru_eigenbranches} illustrates the multiroot fixed-point structure. The ground-state singlet appears well separated at
\begin{equation}
    E_{S_0}=-1570.5344708432~\mathrm{Ha},
\end{equation}
whereas the first triplet and first excited singlet lie in the dense charge-transfer manifold,
\begin{equation}
    E_{T_1}=-1570.4278121373~\mathrm{Ha},\qquad
    E_{S_1}=-1570.4266870288~\mathrm{Ha}.
\end{equation}
The corresponding vertical excitation energies are approximately $2.902$~eV for $T_1$ and $2.933$~eV for $S_1$, and the $S_1$--$T_1$ separation is only about $1.13$~mHartree ($30.6$~meV). The inset shows that several additional singlet and triplet fixed points occur within a narrow energy window, making this a substantially denser multistate test than the Hubbard and LiH examples.

\begin{figure}[ht!]
    \centering
    \begin{subfigure}[t]{0.9\linewidth}
        \centering
        \caption{}
        \includegraphics[width=\linewidth]{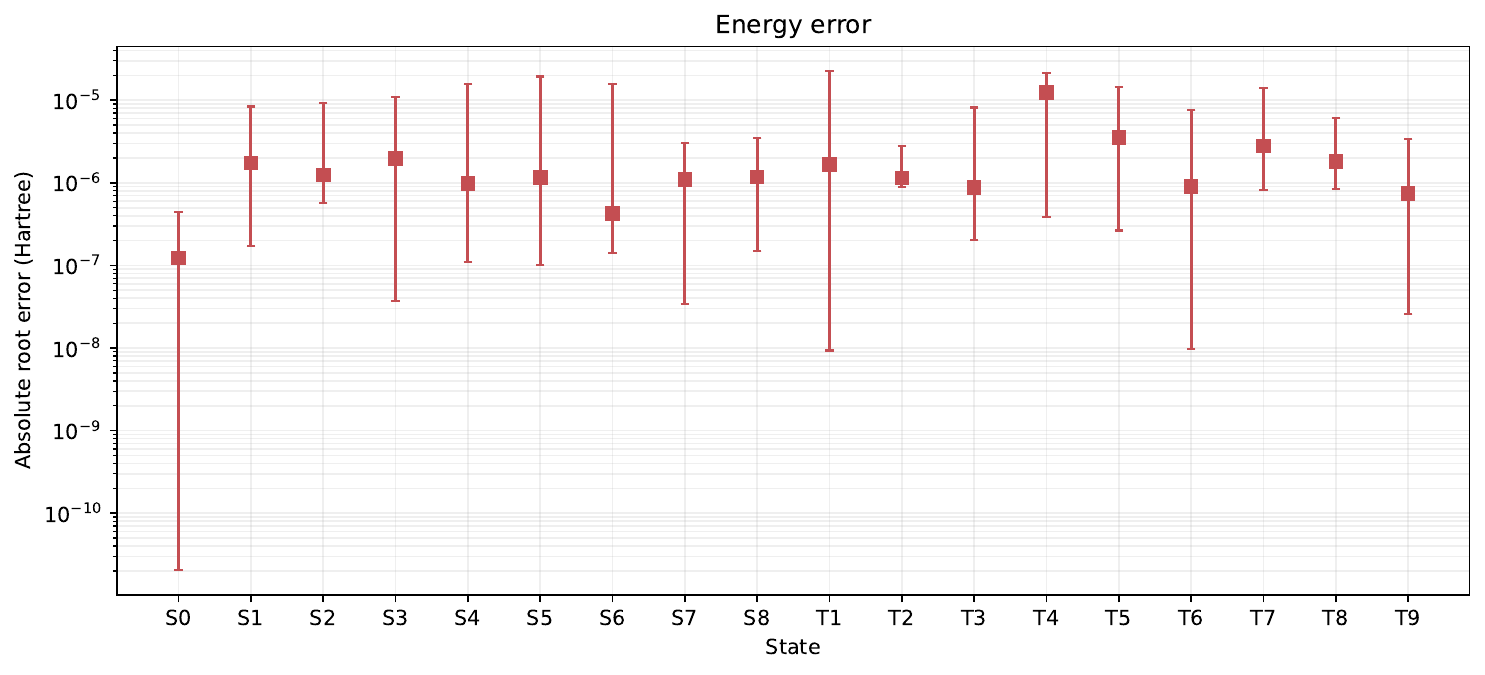}
        \label{fig:Ru_energy_error}
    \end{subfigure}\hfill
    \begin{subfigure}[t]{0.9\linewidth}
        \centering
        \caption{}
        \includegraphics[width=\linewidth]{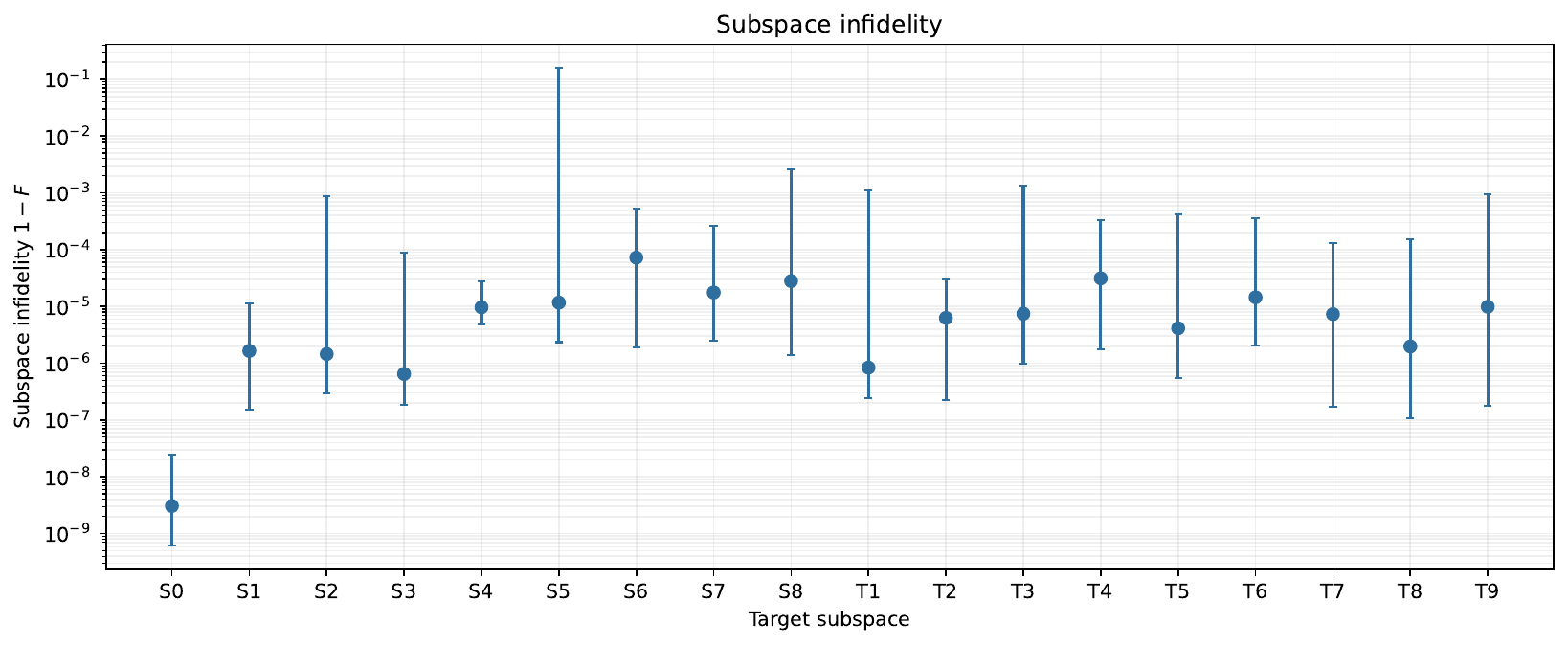}
        \label{fig:Ru_fidelity}
    \end{subfigure}
    \caption{\justifying
    Numerical performance for representative $[\mathrm{Ru(bpy)}_3]^{2+}$ singlet and triplet states. Panel~(a) shows absolute root errors and panel~(b) shows wave-operator reconstruction infidelity. The displayed states are $S_0$--$S_8$ and $T_1$--$T_9$. In both panels, markers show medians over ten independent ideal-QAE trajectories and error bars span the observed minimum and maximum. The larger stochastic ranges for some excited states reflect the sensitivity of individual eigenvectors inside the dense excited-state manifold.
    }
    \label{fig:Ru_error_fidelity}
\end{figure}

The energy accuracy is summarized in Fig.~\ref{fig:Ru_error_fidelity}(a). For most displayed roots, the median absolute error is between $10^{-7}$ and a few $10^{-6}$ Hartree. The ground state is especially accurate, with median error $1.23\times10^{-7}$ Hartree. The largest median error among the displayed roots occurs for $T_4$, approximately $1.24\times10^{-5}$ Hartree, while all ten trajectories converge for the singlet and representative-triplet roots. The spread of the error bars reflects the stochastic QAE sampling propagated through repeated self-energy estimation and the nonlinear secant search, whereas the finite-degree QSVT approximation is common to all trials.

The reconstructed states are shown in Fig.~\ref{fig:Ru_error_fidelity}(b). The $S_0$ state has median infidelity of approximately $3\times10^{-9}$, while the displayed excited states typically have median infidelities from $10^{-6}$ to a few $10^{-5}$. Some individual QAE trajectories exhibit substantially larger infidelity for closely spaced excited states, producing the long upper error bars in the figure.

The triplet calculation provides a direct test of the subspace viewpoint. Since the frozen CAS Hamiltonian is spin free, the three $M_S=-1,0,+1$ components of $T_1$ should be exactly degenerate. Their CASCI energies agree within $1.89\times10^{-11}$ Hartree, and each component satisfies $\langle S^2\rangle=2$ to numerical precision. Treating the three components together as a rank-$3$ target manifold, our method gives median average subspace fidelity
\begin{equation}
    F_{T_1}^{(3)}=0.999997078,
\end{equation}
with the ten observed trials ranging from $0.999977336$ to $0.999999643$. Thus the physically relevant conclusion is not a resolved spin splitting, where none is present in the spin-free Hamiltonian, but that the method preserves and reconstructs the degenerate $T_1$ multiplet with high subspace fidelity. Together with the 13 singlet and 18 representative-triplet roots, this benchmark demonstrates the use of compact, chemically motivated effective Hamiltonians to resolve and reconstruct a dense multistate molecular spectrum.

\section{Discussion}
\label{sec:discussion}
In this section we clarify the roles of two parameters that govern our guarantees: the
\(P\)-space overlap \(p_k=\langle\Psi_k|P|\Psi_k\rangle\) of the target eigenstate and the pole distance \(g:=\mathrm{dist}(\lambda_k,\mathrm{spec}(H_{QQ}))\).
Our eigenvalue and fidelity bounds carry a prefactor \(p_k\), indicating that for a fixed
oracle error budget, the \emph{final} error in the normalized output can be suppressed when
\(p_k\) is smaller. At first glance this may look paradoxical. The resolution is that
\(p_k\) primarily controls \emph{error propagation} from reduced-space quantities to the final
normalized full-space output (conditioning of the fixed-point and lifting step), whereas \(g\)
controls \emph{oracle difficulty}, i.e., the cost of approximating the resolvent
\((H_{QQ}-\lambda I_Q)^{-1}\) and hence the self-energy and wave operator.

These two parameters are not independent. Using the wave-operator representation
\(\frac{1}{p_k}=1+\big\|(H_{QQ}-\lambda_k I)^{-1}H_{QP}|\phi(\lambda_k)\rangle\big\|^2\) we can have the following bound:
\[
\frac{1}{p_k}\leq 
1+\frac{\|H_{QP}\|^2}{g^2}, \qquad \implies \qquad 
p_k \ge \frac{g^2}{g^2+\|H_{QP}\|^2}.
\]
In particular, for bounded \(\|H_{QP}\|\), the regime \(p_k\to 0\) can only occur when \(g\to 0\).
The converse does not hold: a small \(g\) does not necessarily imply a small \(p_k\), because the
nearest pole of \((H_{QQ}-\lambda I_Q)^{-1}\) might not have contribution to a given eigenbranch, so that the branch \(\xi_i(\lambda)\) remains
continuous across that point (see \cref{fig:2x2-hubbard-eigenbranches} and Appendix~\ref{app:continuity-of-eigenbranches}). In this sense, \(g\) is a conservative (worst case) pole distance
parameter, while a branch-dependent pole distance can be larger.

This one-sided relation helps explain the practical tradeoff. In low-overlap regimes, the same
\(p_k\) factor that shrinks the eigenvalue and fidelity errors is often
accompanied by a small \(g\), meaning that the target energy lies close to a pole of the self-energy.
Near poles, the QSVT approximation to \(\frac{1}{x}\) requires a higher polynomial degree (and typically a
larger normalization overhead), and matrix-element estimation becomes more expensive.
Moreover, the relation above implies that the pole-separation factor cannot remain bounded as
\(p_k\to 0\): rearranging gives
\[
\frac{1}{g^2}\ \geq \ \frac{1-p_k}{p_k}\cdot \frac{1}{\|H_{QP}\|^2},
\]
so the \(1/g^2\) term appearing in the complexity bounds typically grows rapidly in the small-overlap
limit. Finally, proximity to poles can complicate the local fixed-point iteration: an estimate
\(\hat\xi_i(\lambda)\) can land across a nearby pole and move the iterate outside the intended
pole-free interval, requiring additional checks to maintain a valid search region.

From a practical standpoint, this suggests choosing the reference space \(P\mathcal H\) to maintain
non-vanishing overlap with the low-energy sector while avoiding near-resonances with
\(\mathrm{spec}(H_{QQ})\), so that \(g\) remains sufficiently large to keep the resolvent
approximation tractable. Compared to ground-state preparation methods whose cost is governed primarily by overlap and target precision, our algorithm exhibits a different resource profile: the attainable accuracy is controlled by the oracle error budget and the subsequent renormalization $p_k$, whereas the runtime is dominated by pole
separation $g$ and the target precision $\varepsilon$.

\section{Conclusion}
\label{sec:conclusion}

In this work, we developed a fault-tolerant quantum algorithm to address the subspace eigenvalue problems based on the Feshbach effective Hamiltonians formalism. Rather than treating the target low-energy sector as a collection of independently prepared eigenstates, the proposed framework identifies and reconstructs the corresponding spectral manifold. By combining the Feshbach effective Hamiltonian with projected block encodings and QSVT, the algorithm estimates the spectrum of the energy-dependent effective Hamiltonian and locates the target energies through the fixed-point condition $\xi_i(\lambda)=\lambda$. Once an eigenbranch satisfies
\begin{equation}
    \bigl|\hat{\xi}_i(\hat{\lambda})-\hat{\lambda}\bigr|
    <
    \varepsilon_{\mathrm{app}},
\end{equation}
the full residual spectrum of the effective Hamiltonian is examined at the same energy. The number of eigenbranches satisfying the same condition determines the dimension of the manifold that remains unresolved at the working precision. The corresponding effective eigenvectors are then lifted to the full Hilbert space through the block-encoded wave operator and orthogonalized to reconstruct an approximation to the target spectral subspace. In this way, the algorithm provides not only eigenvalue estimates, but also information about the multiplicity and structure of the associated low-energy manifold.

Our analysis establishes explicit guarantees for the eigenvalue error, state fidelity, subspace fidelity, and query complexity of the proposed procedure. The numerical studies demonstrate the applicability of the framework to several qualitatively different problems. For the open $4\times2$ one-hole Hubbard cluster, the method resolves the Nagaoka-type crossing between the lowest $S=1/2$ and $S=3/2$ symmetry sectors while reconstructing the corresponding states with high fidelity. For all-electron LiH in the 6-31G basis, it follows the lowest singlet and triplet states from the equilibrium region toward their near-degeneracy at dissociation using only a spin-adapted CAS(2,3) reference space. For $[\mathrm{Ru(bpy)}_3]^{2+}$, chemically structured $d=19$ singlet and $d=18$ triplet reference spaces recover a dense set of singlet and triplet CAS(12,9) roots and reconstruct the exactly degenerate rank-$3$ $T_1$ multiplet with high subspace fidelity. These examples indicate that the framework is applicable to both lattice Hamiltonians and realistic molecular electronic-structure problems, including systems with strong correlation, excited states, and near-degenerate multiplets.

The proposed approach also provides a natural bridge between classical multireference quantum chemistry and fault-tolerant quantum computation. Classical multireference methods can be used to construct a compact and physically motivated reference space containing the dominant configurations or approximate states of interest. The quantum algorithm can then incorporate the contribution from the complementary $Q$ space through the self-energy, refine the corresponding energies, and reconstruct the associated states in the full Hilbert space. Consequently, existing classical procedures for selecting active spaces, generating multiconfigurational reference states, and identifying chemically relevant low-energy manifolds can serve directly as the starting point of the quantum calculation. This hybrid perspective substantially broadens the potential applications of the method, since the reference space can be adapted to the physical structure of the problem rather than being restricted to a single determinant or a particular ansatz.

Several challenges remain before the framework can be applied at larger scales. The dense estimation of a $d\times d$ effective Hamiltonian leads to a $d^3$-type matrix-estimation cost in the present implementation, motivating the development of strategies that exploit sparsity, symmetry, and additional structure in the reference space. Scalable procedures for coherently preparing general multiconfigurational reference states and implementing the associated $P$-controlled operations will also be important. Further improvements in secant initialization, pole diagnostics, and the numerical stability of subspace orthogonalization could make the method more robust in challenging strongly correlated regimes. Finally, end-to-end logical resource estimates for concrete Hamiltonian representations are required to determine the system sizes and reference-space dimensions for which effective-Hamiltonian reduction provides a practical quantum advantage. Overall, the proposed framework offers a general route for combining physically informed classical reference calculations with quantum effective-Hamiltonian techniques to obtain low-energy eigenvalues, degeneracy information, and spectral-subspace reconstructions for a broad range of quantum many-body problems.

\section*{Data availability}
The codes that support this study are available on GitHub via \url{https://github.com/chuntse0514/Quantum-Algorithm-Subspace-Eigenvalue-Problem}.


\appendix

\section{Proof of eigenbranch monotonicity and overlap identities}
\label{app:branches}

In this appendix we collect proofs of the basic properties of the
eigenbranches of $H_{\mathrm{eff}}(\lambda)$ stated in
Lemma~\ref{lem:monotone-branches}.

\begin{proof}[Proof of Lemma~\ref{lem:monotone-branches}]
Assume $\lambda\in\mathbb{R}\setminus\mathrm{spec}(H_{QQ})$ so that
$(H_{QQ}-\lambda I_Q)^{-1}$ is well-defined and Hermitian. Differentiating
the resolvent yields
\begin{align}
    \frac{d}{d\lambda}(H_{QQ}-\lambda I_Q)^{-1}
    = (H_{QQ}-\lambda I_Q)^{-1} I (H_{QQ}-\lambda I_Q)^{-1}
    = (H_{QQ}-\lambda I_Q)^{-2} ,
\end{align}
and hence
\begin{align}
    \frac{dH_{\mathrm{eff}}(\lambda)}{d\lambda}
    = -\,H_{PQ}\,(H_{QQ}-\lambda I_Q)^{-2}\,H_{QP}.
\end{align}
Because $(H_{QQ}-\lambda I_Q)^{-2}\succ 0$ and
$H_{PQ}(\cdot)H_{QP}\succeq 0$ on $P\mathcal{H}$, we obtain
\begin{align}
    \Big\langle \varphi \Big|\,\frac{dH_{\mathrm{eff}}}{d\lambda}\,\Big|\varphi\Big\rangle
    &= -\,\bigl\|(H_{QQ}-\lambda I_Q)^{-1}H_{QP}|\varphi\rangle\bigr\|^2 \le 0,
\end{align}
for all $|\varphi\rangle\in P\mathcal{H}$, so
$\tfrac{dH_{\mathrm{eff}}}{d\lambda}\preceq 0$.

Along any differentiable normalized eigenpair
$(\xi_i(\lambda),|\phi_i(\lambda)\rangle)$ belonging to a chosen continuous eigenbranch of
$H_{\mathrm{eff}}(\lambda)$, the Hellmann--Feynman theorem gives
\begin{align}
    \frac{d\xi_i(\lambda)}{d\lambda}
    =
    \Big\langle \phi_i(\lambda)\Big|
    \frac{dH_{\mathrm{eff}}(\lambda)}{d\lambda}
    \Big|\phi_i(\lambda)\Big\rangle
    \nonumber =
    -\,\bigl\|(H_{QQ}-\lambda I_Q)^{-1}H_{QP}
    |\phi_i(\lambda)\rangle\bigr\|^2
    \le 0,
\end{align}
where the equality holds if and only if
$H_{QP}|\phi_i(\lambda)\rangle=0$.
Thus each continuously tracked eigenbranch is monotonically nonincreasing on any pole-free interval, and strictly decreasing whenever there is nontrivial $P$--$Q$ mixing.
At an isolated degeneracy, the formula is understood branchwise along the chosen continuous eigenbranch, rather than for an arbitrary vector in the degenerate eigenspace.

For the overlap interpretation, use the wave-operator representation
(cf. Eq.~\eqref{eq: Psi-eigstate}) to write the unnormalized eigenvector
in $\mathcal{H}$ associated with $|\phi_i(\lambda)\rangle$ as
\begin{align}
    |\Psi(\lambda)\rangle
    =
    \begin{bmatrix}
      |\phi_i(\lambda)\rangle\\[2pt]
      -\,(H_{QQ}-\lambda I_Q)^{-1}H_{QP}\,|\phi_i(\lambda)\rangle
    \end{bmatrix}.
\end{align}
With $\||\phi_i(\lambda)\rangle\|=1$, its squared norm is
\begin{align}
    \bigl\|\,|\Psi(\lambda)\rangle\bigr\|^2
    = 1 + \Big\|(H_{QQ}-\lambda I_Q)^{-1}H_{QP}\,|\phi_i(\lambda)\rangle\Big\|^2
    = 1 - \frac{d\xi_i(\lambda)}{d\lambda}.
\end{align}
Let $|\Psi'(\lambda)\rangle
= |\Psi(\lambda)\rangle/\bigl\|\,|\Psi(\lambda)\rangle\bigr\|$ denote the
normalized state. Since
$\{|\phi_j(\lambda)\rangle\}_{j=1}^d$ forms an orthonormal basis of $P\mathcal{H}$,
we have
\begin{align}
  P = \sum_{j=1}^d|\phi_j(\lambda)\rangle\langle\phi_j(\lambda)|,
\end{align}
and therefore
\begin{align}
    \langle \Psi'(\lambda)|P|\Psi'(\lambda)\rangle
    = \sum_{j=1}^d \bigl|\langle \phi_j(\lambda)|\Psi'(\lambda)\rangle\bigr|^2
    = \left(\,1-\frac{d\xi_i(\lambda)}{d\lambda}\right)^{-1},
\end{align}
which identifies the magnitude of the negative slope
$-\tfrac{d\xi_i}{d\lambda}$ with the squared $Q$-space norm of the
reconstructed eigenstate and the associated normalization overhead.
\end{proof}

\section{Projected block-encodings of $H_{QQ}$ and $H_{QP}$}
\label{app:BE-of-H22-H21}

In this appendix we explain how to obtain efficient block-encodings of the sub-blocks
\[
    H_{QQ} = Q H Q, \qquad
    H_{QP} = Q H P,
\]
in the regime where the $P$-space is spanned by classically tractable many-body basis states. We first assume that the Hamiltonian has already been mapped to a sum of Pauli operators
\begin{equation}
    H = \sum_{\ell} w_\ell U_\ell,
    \qquad U_\ell \in \mc P_n / \{\pm1, \pm i\},
\end{equation}
where $\mc P_n$ denotes the $n$-qubit Pauli group and $w_\ell \in \mathbb C$ are scalar coefficients.
This assumption is standard and always valid: any fermionic Hamiltonian in second-quantized form can be mapped to a spin Hamiltonian by a fermion-to-qubit encoding (such as the Jordan--Wigner or Bravyi--Kitaev transformations).

The $P$-space is assumed to be spanned by a finite set of Fock states (for example, a small set of low-energy mean field states), and we require that membership in $P$ is classically tractable.
Concretely, we assume the existence of an efficiently computable predicate
\begin{equation}
    \chi_P(\mathbf s) =
    \begin{cases}
        1, & \ket{\mathbf s} \in P\mc H,\\
        0, & \ket{\mathbf s} \in Q\mc H,
    \end{cases}
\end{equation}
which can be realized by a reversible circuit on a quantum computer.
We then define the projectors
\begin{equation}
    P = \sum_{\ket{\mathbf s} \in P\mc H} \ket{\mathbf s}\bra{\mathbf s}, \qquad
    Q = I - P.
\end{equation}

From the Pauli expansion we can construct a standard LCU block-encoding of $H$.
Let $\alpha_H := \sum_{\ell} |w_\ell|$,
and define a ``PREP'' oracle and a ``SELECT'' unitary $\mathrm{SEL}$ acting on an index register as
\begin{align}
    \mathrm{PREP}\ket{0}
    = \sum_{\ell} \sqrt{\frac{|w_\ell|}{\alpha_H}} \ket{\ell}, \qquad 
    \mathrm{SEL}\ket{\ell}\ket{\psi}
    = \ket{\ell}\, (\mathrm{sgn}(w_\ell) U_\ell)\ket{\psi}.
\end{align}
Then the standard construction yields a unitary $U_H$ such that
\begin{equation}
    U_H=\Big(\mathrm{PREP^\dagger}\otimes I\Big)\mathrm{SEL}\Big(\mathrm{PREP}\otimes I\Big), \qquad  (\bra{0}\otimes I)\, U_H \,(\ket{0}\otimes I)
        = \frac{H}{\alpha_H},
\end{equation}
i.e.\ $U_H$ is an $\alpha_H$-block-encoding of $H$.

\begin{figure}[t!]
    \centering
    \scalebox{0.95}{%
    \begin{quantikz}
        \lstick{$|0\rangle$} &\gate[2]{U_{H-\lambda I}}& \\
        \lstick{$|\psi\rangle$} &&
    \end{quantikz}
    $\equiv$
    \begin{quantikz}
        \lstick{$|0\rangle$} &\gate{H}&\ctrl{1}&\gate{R_Y(2\phi)}& \\
        \lstick{$|0\rangle$}&&\gate[2]{U_{H}}&& \\
        \lstick{$|\psi\rangle$}&&&& \\
    \end{quantikz}
    }
    \caption{\justifying
    Quantum circuit gadget $U_{H-\lambda I}$ implementing an adjustable block-encoding of $(H-\lambda I)$, with angle $\phi = \cot^{-1}(\lambda/\alpha)$.
    }
    \label{fig:adjustable_H22}
\end{figure}

\subsection{Block-encoding of $H_{QQ}-\lambda I_Q$}
\begin{figure}[t]
\centering
\scalebox{0.9}{
\begin{quantikz}
    \lstick{$|0^{a}\rangle$} &\gate[2]{U_{H_{QQ-\lambda I_Q}}}& \\
    \lstick{$|\psi\rangle$} &&
\end{quantikz}
$\equiv$
\begin{quantikz}
    \lstick{$|0\rangle$} &&&\targ{}\wire[d]{q}& \\
    \lstick{$|0\rangle$} &\targ{}\wire[d]{q}&&\wire[d]{q}& \\
    \lstick{$|0\rangle$} &\wire[d]{q}&\gate[2]{U_{H-\lambda I}}&\wire[d]{q}& \\
    \lstick{$|\psi\rangle$} &\gate{P}&&\gate{P}&
\end{quantikz}
}
\caption{Quantum circuits for constructing the block encodings of $H_{QQ}-\lambda I_Q$.}
\label{fig:BE-1}
\end{figure}

The block-encoding of $H_{QQ}-\lambda I_{Q}$ can be realized by first realizing the block-encoding of $H-\lambda I$ by a single ancilla rotation, and then apply the technique of composition of block-encoding~\cite{gilyen_quantum_2019}.

We first introduce a single-ancilla-controlled version of $U_{H}$ so that the same block-encoding of $H-\lambda I$ can be reused for different values of the parameter $\lambda$. The $\lambda$-dependence is implemented via a rotation on the ancillary qubit, and the resulting circuit for the parameterized block-encoding is shown in \cref{fig:adjustable_H22}. It is easy to verify that the circuit gadget $U_{H_{QQ}-\lambda I}$ implements the block-encoding:
\begin{align}
\label{eq:adjustable-BE}
    \mathrm{Block}(U_{H-\lambda I}) = \frac{\sin(\phi)\big(\cot(\phi)I - H/\alpha_H\big)}{\sqrt{2}}
    \nonumber = -\frac{1}{\alpha_\lambda}\,(H-\lambda I),
\end{align}
where in the last equality we choose $\alpha\cot(\phi)=\lambda$, with the corresponding normalization factor $\alpha_\lambda=\sqrt{2(\alpha^2+\lambda^2)}\in\Theta(\alpha)$. This yields an $(\alpha_\lambda,a+1,0)$ block-encoding of $-(H-\lambda I)$.

Second, we apply the idea to sandwich the block encoding circuit $U_{H-\lambda I}$ by the $P$-controlled NOT gate $P\otimes X+Q\otimes I$ to get the projected matrix element $H_{QQ}-\lambda I_Q$, the circuit can be found in \cref{fig:BE-1}. The circuit gadget essentially implement the mapping:
\begin{align*}
    |00\rangle|0^a\rangle|\psi\rangle \xmapsto{U_{H_{QQ}-\lambda I_Q}} \,\, &|00\rangle (I\otimes Q)U_{H-\lambda I}(I\otimes Q)|0^a\rangle|\psi\rangle
    \\[0.2cm]
    +&|01\rangle (I\otimes Q)U_{H-\lambda I}(I\otimes P)|0^a\rangle |\psi\rangle 
    \\[0.2cm]
    +&|10\rangle (I\otimes P)U_{H-\lambda I}(I\otimes Q)|0^a\rangle|\psi\rangle 
    \\[0.2cm]
    +&|11\rangle (I\otimes P)U_{H-\lambda I}(I\otimes P)|0^a\rangle|\psi\rangle
\end{align*}
projecting the ancilla qubits to $|00\rangle |0^a\rangle$ then we have:
\[
    \mathrm{Block}(U_{H_{QQ}-\lambda I_Q})=-\frac{1}{\alpha_\lambda}Q(H-\lambda I)Q = -\frac{1}{\alpha_\lambda}(H_{QQ}-\lambda I_Q)
\]
which confirms the circuit gadget implement the $H_{QQ}-\lambda I_Q$ block-encoding. This construction is always valid when the $P$-controlled NOT gate is available, where we give the detail construction of this gate in Appendix~\ref{appendix:P-controlled-gate}.

\subsection{Block-encoding of $H_{QP}$}

We now construct the off-diagonal block $H_{QP}=QHP$. Unlike the $H_{QQ}$ construction,
here the algorithm only requires agreement on the projected block from $P\mc H$ to
$Q\mc H$. This is because the input quantum states $|\phi_i\rangle$ always lie in $P\mathcal H$, and the block-encoding $(H_{QQ}-\lambda I_Q)^{-1}$ only has non-zero matrix element on $Q\mathcal H$. This observation allows a possible reduction of the LCU normalization.

The term-wise criterion is whether a Pauli string has a nonzero $P\to Q$ block.
Define the index set for the Pauli string satisfying this criterion:
\begin{equation}
\mc I_{QP}
:=
\{\ell \quad \mathrm{s.t.} \quad QU_\ell P\neq 0\}.
\end{equation}
Note that by Hermitiancy of Pauli strings, the condition $QU_\ell P\neq 0$ automatically implies $P U_\ell Q\neq 0$. Equivalently, a Pauli string can be omitted from the $H_{QP}$ LCU if and only if $QU_\ell P=0.$
This condition means that $U_\ell$ maps every vector in $P\mc H$ back into $P\mc H$:
\begin{equation}
QU_\ell P=0
\qquad\Longleftrightarrow\qquad
U_\ell(P\mc H)\subseteq P\mc H.
\end{equation}
Such terms do not contribute to $QHP$ and can be safely removed from the LCU used for
the off-diagonal block.

When $P\mc H$ is spanned by a polynomial number of computational or Fock basis states,
this condition is often efficiently checkable. Since a Pauli string maps a computational
basis state to another computational basis state up to a phase, we can test whether
\begin{equation}
U_\ell\ket{\mathbf s}
=
e^{i\theta_{\ell,\mathbf s}}\ket{\mathbf s'}
\end{equation}
lies in $P\mc H$ or $Q\mc H$ for each reference basis state $\ket{\mathbf s}\in P\mc H$.
Then $\ell\in\mc I_{QP}$ precisely when there exists at least one reference basis state
$\ket{\mathbf s}\in P\mc H$ such that $U_\ell\ket{\mathbf s}\in Q\mc H.$
Typical terms that can be removed include diagonal $Z$-type Pauli strings in a computational-basis reference space, since they only add phases and never leave $P\mc H$, and Pauli strings that merely permute the selected reference configurations inside $P\mc H$. By contrast, a term must be kept if it maps even one reference basis state into $Q\mc H$.

Define the simplified Hamiltonian
\begin{equation}
\overline H
:=
\sum_{\ell\in\mc I_{QP}}w_\ell U_\ell,
\qquad
\widetilde\alpha
:=
\sum_{\ell\in\mc I_{QP}}|w_\ell|.
\end{equation}
Since every omitted term satisfies $QU_\ell P=0$, we have the exact projected identity
\begin{align}
Q\overline H P
=\sum_{\ell\in\mc I_{QP}}w_\ell QU_\ell P
=\sum_{\ell}w_\ell QU_\ell P
=QHP=H_{QP}.
\end{align}
Thus $\overline H$ need not agree with $H$ on all blocks; it only needs to agree after the left and right projections $Q$ and $P$ are applied.

We now define the restricted LCU oracles
\begin{align}
\widetilde{\mathrm{PREP}}|0\rangle
=
\sum_{\ell\in\mc I_{QP}}
\sqrt{\frac{|w_\ell|}{\widetilde\alpha}}\ket{\ell},
\qquad
\widetilde{\mathrm{SEL}}|\ell\rangle|\psi\rangle
=
|\ell\rangle
\bigl(\mathrm{sgn}(w_\ell)U_\ell\bigr)|\psi\rangle,
\quad
\ell\in\mc I_{QP}.
\end{align}
The corresponding LCU unitary
\begin{equation}
U_{\overline H}
:=
\bigl(\widetilde{\mathrm{PREP}}^{\dagger}\otimes I\bigr)
\widetilde{\mathrm{SEL}}
\bigl(\widetilde{\mathrm{PREP}}\otimes I\bigr)
\end{equation}
satisfies
\begin{equation}
(\bra{0^{\widetilde a}}\otimes Q)
U_{\overline H}
(\ket{0^{\widetilde a}}\otimes P)
=
\frac{H_{QP}}{\widetilde\alpha}, \qquad (\bra{0^{\widetilde a}}\otimes P)
U_{\overline H}
(\ket{0^{\widetilde a}}\otimes Q)
=
\frac{H_{PQ}}{\widetilde\alpha}.
\end{equation}

In the worst case, one may choose $\mc I_{QP}$ to be the full Pauli index set, giving
\begin{equation}
\widetilde\alpha=\alpha_H,
\end{equation}
so the construction is exact but gives no LCU normalization saving. When many Pauli terms
preserve the reference space and therefore satisfy $QU_\ell P=0$, resulting in the smaller
set $\mc I_{QP}$ with
\begin{equation}
\widetilde\alpha
=
\sum_{\ell\in\mc I_{QP}}|w_\ell|
\leq
\alpha_H,
\end{equation}
which reduces the cost of each use of the $H_{QP}$ block in the self-energy and
wave-operator constructions.

\section{$P$-controlled Gate Construction}
\label{appendix:P-controlled-gate}

This appendix gives an explicit implementation of the system-controlled NOT
\begin{equation}
    \mathsf{C}_{P}X
    := P\otimes X_a+(I-P)\otimes I_a,
    \label{eq:CPX-definition}
\end{equation}
where $a$ is a target qubit. Let $P\mc H$ be a $d$-dimensional reference
subspace with orthonormal basis
$\{\ket{\varphi_j}\}_{j=1}^d$. Choose distinct computational-basis states
$\{\ket{s_j}\}_{j=1}^d$ and define
\begin{equation}
    \Pi:=\sum_{j=1}^d \ket{s_j}\!\bra{s_j}.
\end{equation}
If a unitary $U$ satisfies
$U\ket{s_j}=\ket{\varphi_j}$ for all $j$, then
$P=U\Pi U^\dagger$.

\begin{lemma}[Conjugation identity]
\label{lem:conjugation}
With
\begin{equation}
    \mathsf{C}_{\Pi}X
    :=
    \Pi\otimes X_a+(I-\Pi)\otimes I_a,
\end{equation}
one has
\begin{equation}
    \mathsf{C}_{P}X
    =
    (U\otimes I_a)\,
    \mathsf{C}_{\Pi}X\,
    (U^\dagger\otimes I_a).
    \label{eq:CPX-conjugation}
\end{equation}
\end{lemma}

\begin{proof}
Substituting $P=U\Pi U^\dagger$ into
\cref{eq:CPX-definition} gives
\cref{eq:CPX-conjugation} directly.
\end{proof}

Thus, the nontrivial step is a reversible membership test for the
computational subspace $\Pi\mc H$. For a bit string
$x\in\{0,1\}^n$, define
\begin{equation}
    \chi_\Pi(x):=
    \begin{cases}
        1, & x\in\{s_1,\ldots,s_d\},\\
        0, & \text{otherwise}.
    \end{cases}
\end{equation}
For any pattern $C$, let $\chi_C(x)=1$ when $x$ matches $C$ and
$\chi_C(x)=0$ otherwise. The corresponding membership gadget
$\mathsf{M}_{C}$ acts on the system register and a clean flag qubit
$f$ as
\begin{equation}
    \mathsf{M}_{C}:\quad
    \ket{x}\ket{b}_f
    \longmapsto
    \ket{x}\ket{b\oplus\chi_C(x)}_f.
    \label{eq:membership-gadget}
\end{equation}
For an exact bit string $C=s_j$, $\mathsf{M}_{C}$ is an
$n$-controlled $X$ on $f$, with open controls on positions where
$(s_j)_i=0$. Because the strings $s_j$ are distinct, their indicators
are mutually exclusive, and hence
\begin{equation}
    \chi_\Pi(x)
    =
    \bigoplus_{j=1}^d \chi_{s_j}(x).
\end{equation}
Starting from $\ket{0}_f$, one may therefore compute $\chi_\Pi(x)$
into the flag, apply a CNOT from $f$ to the target qubit, and then
uncompute the flag. For $t\in\{0,1\}$, the resulting action is
\begin{equation}
    \ket{x}\ket{0}_f\ket{t}_a
    \longmapsto
    \ket{x}\ket{0}_f
    \ket{t\oplus\chi_\Pi(x)}_a,
\end{equation}
which implements $\mathsf{C}_{\Pi}X$ and, after conjugation by $U$,
implements $\mathsf{C}_{P}X$.

\begin{figure}[t!]
\centering
\begin{quantikz}[row sep=0.35cm, column sep=0.30cm]
\lstick{$\ket{\psi}_S$}
& \qwbundle{n}
& \gate{U^\dagger}
& \gate[2]{\mathsf{M}_{C_1}}
& \cdots
& \gate[2]{\mathsf{M}_{C_K}}
& \qw
& \gate[2]{\mathsf{M}_{C_K}}
& \cdots
& \gate[2]{\mathsf{M}_{C_1}}
& \gate{U}
& \qw \\
\lstick{$\ket{0}_f$}
& \qw
& \qw
& \ghost{\mathsf{M}_{C_1}}
& \qw
& \ghost{\mathsf{M}_{C_K}}
& \ctrl{1}
& \ghost{\mathsf{M}_{C_K}}
& \qw
& \ghost{\mathsf{M}_{C_1}}
& \qw
& \qw \\
\lstick{$\ket{t}_a$}
& \qw
& \qw
& \qw
& \qw
& \qw
& \targ{}
& \qw
& \qw
& \qw
& \qw
& \qw
\end{quantikz}
\caption{\justifying
Implementation of $\mathsf{C}_{P}X$.
The unitary $U^\dagger$ rotates the physical reference space to the
computational subspace $\Pi\mc H$. The membership gadgets for the
disjoint cubes $\mathsf{M}_{C_k}$ compute the indicator $\chi_\Pi$
into the clean flag $f$; the flag controls the target qubit $a$ and
is then uncomputed. Since every $\mathsf{M}_{C_k}$ is a
controlled-$X$, the same gates in reverse order implement the inverse.
}
\label{fig:optionB-single-flag-compact}
\end{figure}

\subsection{Disjoint-cube compression}

The exact-pattern construction uses $d$ membership gadgets. It can be
compressed when the selected bit strings admit an exact disjoint-cube
representation. A cube $C\in\{0,1,-\}^n$ denotes all bit strings that
agree with its fixed entries; the symbol ``$-$'' is a wildcard.
Suppose
\begin{equation}
    \{s_1,\ldots,s_d\}
    =
    \bigsqcup_{k=1}^K \mathcal B(C_k),
    \label{eq:disjoint-cube-cover}
\end{equation}
where $\mathcal B(C_k)$ is the set represented by $C_k$ and the union
is disjoint. Then
\begin{equation}
    \chi_\Pi(x)
    =
    \bigoplus_{k=1}^K\chi_{C_k}(x),
\end{equation}
and $\mathsf{M}_{C_k}$ needs controls only on the fixed entries of
$C_k$. Closed controls enforce a fixed $1$, open controls enforce a
fixed $0$, and wildcard positions require no control.

A simple greedy preprocessing repeatedly merges two cubes having the
same wildcard positions and differing in exactly one fixed bit.
Replacing such a pair by the cube with a wildcard in that position
preserves both the represented set and pairwise disjointness. The
result is a valid, although not necessarily minimum, disjoint cover.
For example,
\begin{equation}
    \{000,001,110\}
    =
    \mathcal B(00-)\sqcup\mathcal B(110),
\end{equation}
so three exact-pattern tests are replaced by one two-control test and
one three-control test.

Let $w_k$ be the number of fixed entries of $C_k$, and let
$C_{\mathrm{MCX}}(w)$ and $D_{\mathrm{MCX}}(w)$ denote the gate cost
and depth, in a fixed cost model, of the chosen synthesis of a
$w$-controlled $X$. Ignoring the single-qubit $X$ masks, the
membership part of the circuit requires
\begin{align}
    C(\mathsf{C}_{\Pi}X)
    &\leq
    2\sum_{k=1}^K C_{\mathrm{MCX}}(w_k)+O(1),\\
    D(\mathsf{C}_{\Pi}X)
    &\leq
    2\sum_{k=1}^K D_{\mathrm{MCX}}(w_k)+O(1),
    \label{eq:Cpix-resources}
\end{align}
because the gadgets share the same flag and are applied serially.
The worst case is $K=d$ and $w_k=n$. The circuit uses one clean flag
qubit in addition to any work qubits required by the selected
multi-controlled-$X$ decomposition.

\subsection{Efficient basis transformation}

Equation~\eqref{eq:CPX-conjugation} is useful only when $U$ and
$U^\dagger$ are efficient. This is an access assumption for a general
reference subspace. A common explicit realization occurs when
$P\mc H$ is spanned by selected Slater determinants in one orbital
basis. If $u\in U(n)$ is the corresponding single-particle orbital
rotation, its second-quantized unitary satisfies
\begin{equation}
    U a_p^\dagger U^\dagger
    =
    \sum_{q=1}^n u_{pq}a_q^\dagger.
\end{equation}
Standard Givens-rotation decompositions implement this fermionic
Gaussian unitary using $O(n^2)$ two-qubit rotations and $O(n)$ depth
on a linear architecture with an appropriate swap network
\cite{kivlichan_quantum_2018,wecker_solving_2015}. The complete cost
is therefore
\begin{align}
    C(\mathsf{C}_{P}X)
    &\leq
    2C(U)
    +
    2\sum_{k=1}^K C_{\mathrm{MCX}}(w_k)
    +O(1),\\
    D(\mathsf{C}_{P}X)
    &\leq
    2D(U)
    +
    2\sum_{k=1}^K D_{\mathrm{MCX}}(w_k)
    +O(1).
\end{align}
For reference vectors that are not generated from computational-basis
determinants by a common efficiently implementable $U$, an efficient
implementation of $U$ must be supplied separately as part of the
reference-subspace access model.

\section{Self-energy and wave-operator block-encodings}
\label{app:BE-of-Heff-Omega}

In this appendix, we detail the construction of the block-encoding circuit for the self-energy operator and the corresponding wave operator stated in Lemma~\ref{lem:block-encoding-sigma-omega}. 
Our construction relies on the direct availability of block-encodings for the specific sub-blocks of the Hamiltonian, as discussed in Appendix~\ref{app:BE-of-H22-H21}. Where we assume we already have these block-encoding unitaries:
\begin{align}
\label{eq:BE-H_QQ-H_QP}
    (&\langle 0^{a}|\otimes I)\,U_{H_{QQ}-\lambda I_Q}\,(|0^{a}\rangle\otimes I) = -(H_{QQ}-\lambda I_Q)/\alpha_\lambda, 
    \\[0.2cm]
    &(\langle 0^{\widetilde{a}}|\otimes Q)\,U_{H_{QP}}\,(|0^{\widetilde{a}}\rangle\otimes P) = H_{QP}/\widetilde{\alpha}. 
\end{align}
Note that the block-encoding of $H_{PQ}=H_{QP}^\dagger$ is obtained simply by taking the adjoint $U_{H_{QP}}^\dagger$. 

\subsection{Block-encoding construction}
With these primitives in place, we now construct the polynomial approximation for the resolvent $(H_{QQ} - \lambda I_Q)^{-1}$ using quantum singular value transformation (QSVT). The necessary polynomial approximation results from~\cite{gilyen_quantum_2019} are summarized below.
\begin{lemma}[Polynomial approximation of $1/x$]
\label{lem:poly-approx}
    Let $\delta,\varepsilon\in(0,\frac{1}{2}]$, there exists an odd polynomial $p(\cdot \ ;\delta,\varepsilon)\in \R[x]$ of degree $\mc{O}(\frac{1}{\delta}\log(\frac{1}{\varepsilon}))$ such that 
    \begin{enumerate}[leftmargin=*]
        \item $\vert p(x;\delta,\varepsilon)\vert \leq 1$ for all $x\in[-1, 1]$
        \item $\vert p(x;\delta,\varepsilon)-\frac{\delta}{\beta x}\vert\leq \varepsilon $ for all $|x|\leq 1,x\notin(-\delta,\delta)$ and $\beta>1$.
    \end{enumerate}
\end{lemma}

\noindent
In the following discussion, we assume that the scaling factor $\beta = \Theta(1)$ and suppress the explicit dependency in the asymptotic notation.

Now, it is convenient to introduce a rescaled polynomial $f(x)$ that directly approximates $\frac{1}{x}$ without the extra normalization:
\begin{align}
\label{eq:f-poly-approx}
    f(x)\coloneqq \frac{\beta}{\alpha_\lambda \delta}\,
    p\left(\frac{x}{\alpha_\lambda};\,\delta,\varepsilon_{\mathrm{poly}}\right),
\end{align}
where $p(\cdot)$ is the polynomial from Lemma~\ref{lem:poly-approx} that approximates $\frac{\delta}{\beta x}$. Constructing the block-encoding of $p\left(\frac{H_{QQ}-\lambda I_Q}{\alpha_\lambda}\right)$ via QSVT requires one additional ancilla qubit for the signal processing rotations. We denote the resulting unitary as $U_{f(H_{QQ}-\lambda I_Q)}$ satisfying:
\[
\mathrm{Block}\Big(U_{f(H_{QQ}-\lambda I_Q)}\Big) = -p\left(\frac{H_{QQ}-\lambda I}{\alpha_\lambda}; \delta,\varepsilon_{\mathrm{poly}}\right) = -\frac{\alpha_\lambda\delta}{\beta} f(H_{QQ}-\lambda I_Q),
\]
which is a $(\frac{\beta}{\alpha_\lambda\delta}, a+1, 0)$ block-encoding of $f(H_{QQ}-\lambda I_Q)$. And the polynomial $f(x)$ has the following error bound
\begin{equation}
\label{eq:poly-approx-f(x)}
    \left|f(x) - \frac1x\right| \le \varepsilon_{f} = \frac{\beta \varepsilon_{\mathrm{poly}}}{\alpha_\lambda \delta}, \qquad \text{for all} \quad |x|\leq \alpha_\lambda, \,\, x\notin (-\alpha_\lambda\delta,\alpha_\lambda\delta)
\end{equation}
whenever the input $x$ lies outside the region $(-\alpha_\lambda \delta, \alpha_\lambda \delta)$. 

The block-encoding of the self-energy 
\[
    \widetilde{\Sigma}(\lambda) = -H_{PQ}f(H_{QQ}-\lambda I_Q)H_{QP},
\]
is therefore realized by the product of three block-encoded matrices (see \cref{fig:block-encoding-self-energy}): $H_{PQ}$, the polynomial approximation $f(H_{QQ}-\lambda I_Q)$, and $H_{QP}$. We can easily verify that the composed unitary $U_{\widetilde{\Sigma}(\lambda)}$ in \cref{fig:block-encoding-self-energy} gives the block-encoding:
\begin{align}
    \mathrm{Block}\Big(U_{\widetilde{\Sigma}(\lambda)}\Big) 
    &= \left(\frac{H_{PQ}}{\widetilde\alpha}\right)\left(-\frac{\alpha_\lambda\delta}{\beta}f(H_{QQ}-\lambda I_Q)\right)\left(\frac{H_{QP}}{\widetilde\alpha}\right) =\frac{\alpha_\lambda \delta}{\beta \widetilde{\alpha}^2} \widetilde\Sigma(\lambda)
\end{align}
This constitutes a block-encoding of $\widetilde{\Sigma}(\lambda)$ with normalization factor $\frac{\beta \widetilde{\alpha}^2}{\alpha_\lambda \delta}$ and the total ancilla count $a + 2\widetilde{a} + 1$.

The block-encoding of the wave operator
\[
    \widetilde{\Omega}(\lambda) = I - f(H_{QQ}-\lambda I_Q)\,H_{QP},
\]
can be constructed by product of block-encoding unitaries followed by a single ancilla rotation (see \cref{fig:block-encoding-wave-operator}). The middle product of two block-encoding unitaries $U_{H_{QP}}$ and $U_{f(H_{QQ}-\lambda I_Q)}$ gives the block-encoding of the product
\[
    \mathrm{Block}\Big(U_{f(H_{QQ}-\lambda I_Q)}U_{H_{QP}}\Big)=-\frac{\alpha_\lambda\delta}{\beta\widetilde\alpha}f(H_{QQ}-\lambda I_Q)H_{QP}
\]
and then the single ancilla controlled rotation gives the overall block-encoding:
\begin{align}
    \mathrm{Block}\Big(U_{\widetilde{\Omega}(\lambda)}\Big)=\frac{\cos(\phi')\left(I - \tan(\phi') \left[ \frac{\alpha_\lambda \delta}{\beta \widetilde{\alpha}} f(H_{QQ}-\lambda I_Q)H_{QP} \right]\right)}{\sqrt{2}}.
\end{align}
To obtain the target form $I - f(H_{QQ}-\lambda I_Q)H_{QP}$, we define the parameter $\eta \coloneqq \frac{\beta \widetilde{\alpha}}{\alpha_\lambda \delta}$ and select the rotation angle $\phi'$ such that:
\begin{align}
    \tan(\phi') = \eta \implies \sin(\phi') = \frac{\eta}{\sqrt{1+\eta^2}}, \quad \cos(\phi') = \frac{1}{\sqrt{1+\eta^2}}.
\end{align}
Substituting these trigonometric values, the operator simplifies to:
\begin{align}
    \mathrm{Block}\left(U_{\widetilde{\Omega}(\lambda)}\right) &= \frac{1}{\sqrt{2(1+\eta^2)}} \Big( I -  f(H_{QQ}-\lambda I_Q)H_{QP} \Big) = \frac{1}{\kappa} \widetilde{\Omega}(\lambda),
\end{align}
with the normalization factor $\kappa = \sqrt{2(1+\eta^2)} = \sqrt{2 + 2(\frac{\beta \widetilde{\alpha}}{\alpha_\lambda \delta})^2}$. The total ancilla count for the block-encoding of wave operator is $a+\widetilde a + 2$.

\subsection{Choice of QSVT parameter and error bound}
Next, we address the choice of the parameter $\delta$ to ensure the polynomial approximation $|f(x)-\frac{1}{x}|\leq\varepsilon_{f}$ holds effectively. Crucially, we do not require this error bound to hold for all $\lambda \in \mathbb{R}$; rather, it must be satisfied whenever $\lambda$ is near the target eigenvalues we intend to resolve. Let $\Lambda=\{\lambda_k\}_k\subset \mathrm{spec}(H)$ denote this set of target eigenvalues, we define the spectral distance $g$ as the minimum distance between these targets and the spectrum of $H_{QQ}$:
\[
g := \min_{\lambda_k\in\Lambda} \mathrm{dist}(\lambda_k, \mathrm{spec}(H_{QQ})).
\]
We then conveniently choose the parameter $\delta$ such that 
\[
\alpha_\lambda\delta=\frac{g}{2}.
\]
Then the scalar condition in \cref{eq:poly-approx-f(x)} translates into 
\begin{equation}
    \Big\| f(H_{QQ}-\lambda I_Q)-(H_{QQ}-\lambda I_Q)^{-1} \Big\| \leq \varepsilon_f
\end{equation}
whenever
\begin{equation}
    \lambda \notin \left(\chi-\frac{g}{2}, \chi+\frac{g}{2}\right) \quad \forall \chi\in \mathrm{spec}(H_{QQ}),
\end{equation}
and since the target eigenvalues $\Lambda=\{\lambda_k\}$ has the minimal distance $g$ to the spectrum of $H_{QQ}$, then the error bound is satisfied near the target eigenvalues $\Lambda$.

With the choice of the parameter $\delta$, we can now bound the total error of the approximate self-energy operator 
\begin{align}
    \|\widetilde{\Sigma}(\lambda)-\Sigma(\lambda)\| &= \Big\| H_{PQ}f(H_{QQ}-\lambda I_Q)H_{QP}-H_{PQ}(H_{QQ}-\lambda I_Q)^{-1}H_{QP}\Big\|
    \nonumber \\[0.2cm]
    &\le \|H_{QP} \|^2 \cdot \Big\| f(H_{QQ}-\lambda I_Q)-(H_{QQ}-\lambda I_Q)^{-1} \Big\|
    \nonumber \\[0.2cm]
    &\le \|H_{QP}\|^2\,\varepsilon_{f},
\end{align}
and the approximate wave operator
\begin{align}
    \|\widetilde\Omega(\lambda) - \Omega(\lambda) \| &= \Big\| \big(I-f(H_{QQ}-\lambda I_Q)H_{QP}\big)-(I-(H_{QQ}-\lambda I_Q)^{-1}H_{QP})\Big\| 
    \nonumber \\[0.2cm]
    &\le \| H_{QP} \| \cdot \Big\| f(H_{QQ}-\lambda I_Q)-(H_{QQ}-\lambda I_Q)^{-1} \Big\|
    \nonumber \\[0.2cm]
    &\le \| H_{QP}\| \,\,\varepsilon_f.
\end{align}
This establishes the error bound stated in Lemma~\ref{lem:block-encoding-sigma-omega}.

\subsection{Resource Counts}
For both the block-encoding of self energy and wave operator, the dominant costs of them are the QSVT part to obtain the polynomial approximation $f(H_{QQ}-\lambda I_Q)$. From the previous analysis in \cref{eq:poly-approx-f(x)}, the approximation $f(x)$ of $\frac{1}{x}$ such that 
\[
    \Big|f(x)-\frac{1}{x}\Big|\leq \varepsilon_f, \qquad \text{for all} \quad x \in [-\alpha_\lambda, \alpha_\lambda] \setminus \left(-\frac{g}{2},\frac{g}{2}\right)
\]
is equivalent to
\[
    \Big| p(x;\delta,\varepsilon_{\mathrm{poly}})-\frac{\delta}{\beta x}\Big|\leq \varepsilon_{\mathrm{poly}}, \qquad \text{for all} \quad x\notin (-\delta,\delta), 
\]
with $\delta=\frac{g}{2\alpha_\lambda}, \, \varepsilon_{\mathrm{poly}}=\frac{g\varepsilon_{f}}{2\beta}$. Therefore, by Lemma~\ref{lem:poly-approx}, the total query to the block-encoding $U_{H_{QQ}-\lambda I_Q}$ is
\begin{equation}
    \mathcal O\left(\frac{1}{\delta}\log\Big(\frac{1}{\varepsilon_{\mathrm{poly}}}\Big)\right)=\mathcal{O}\left(\frac{\alpha}{g}\log\Big(\frac{1}{g\varepsilon_f}\Big)\right)
\end{equation}
while the total query to $U_{H_{QP}}$ is $\mathcal O(1)$.

\section{Matrix-element estimation details}
\label{app:eigenvalue_details}
\label{app:matrix-element-estimation}

This appendix derives the operator-norm error and query-complexity
bounds for the matrix-element-estimation procedure used in
Section~\ref{sec:eigenvalue-estimation}. Fix a spectral parameter
$\lambda$ in the safe region. From
Appendix~\ref{app:BE-of-Heff-Omega}, the unitary
$U_{\widetilde{\Sigma}(\lambda)}$ satisfies
\begin{equation}
    (\langle 0^{a_\Sigma}|\otimes I)\,
    U_{\widetilde{\Sigma}(\lambda)}\,
    (|0^{a_\Sigma}\rangle\otimes I)
    =
    \frac{\widetilde{\Sigma}(\lambda)}{\mathcal N_\lambda},
    \qquad
    \mathcal N_\lambda
    =
    \frac{2\beta\widetilde{\alpha}^2}{g},
    \label{eq:self-energy-block-appendix}
\end{equation}
where $a_\Sigma=a+2\widetilde{a}+1$. For the reference-space basis
states $|\varphi_i\rangle,|\varphi_j\rangle\in P\mathcal H$, assume
access to a unitary $V_{ij}$ satisfying
\begin{equation}
    V_{ij}|\varphi_i\rangle=|\varphi_j\rangle.
\end{equation}
The generalized Hadamard-test circuits are shown in
Fig.~\ref{fig:generalized-HT}. When the control qubit is initialized
in $|+\rangle$, the probability of measuring $0$ is
\begin{align}
    \Pr(0)
    &=
    \frac12\left(
        1+
        \operatorname{Re}\!\left[
        \langle 0^{a_\Sigma}|\langle\varphi_i|
        U_{\widetilde{\Sigma}(\lambda)}
        (I\otimes V_{ij})
        |0^{a_\Sigma}\rangle|\varphi_i\rangle
        \right]
    \right)
    \nonumber\\[0.15cm]
    &=
    \frac12\left(
        1+
        \frac{1}{\mathcal N_\lambda}
        \operatorname{Re}
        \bigl([\widetilde{\Sigma}(\lambda)]_{ij}\bigr)
    \right).
    \label{eq:HT-real-probability-appendix}
\end{align}
When the control qubit is initialized in $|+i\rangle$, the
corresponding probability is
\begin{equation}
    \Pr(0)
    =
    \frac12\left(
        1+
        \frac{1}{\mathcal N_\lambda}
        \operatorname{Im}
        \bigl([\widetilde{\Sigma}(\lambda)]_{ij}\bigr)
    \right).
    \label{eq:HT-imag-probability-appendix}
\end{equation}

Because $\widetilde{\Sigma}(\lambda)$ is Hermitian, it is sufficient
to estimate its $d$ diagonal entries and the real and imaginary parts
of the upper-triangular entries. This gives
\begin{equation}
    d+2\binom{d}{2}=d^2
\end{equation}
independent real quantities. The lower-triangular entries are then
assigned by Hermitian conjugation, so $\hat{\Sigma}(\lambda)$ is
Hermitian by construction. If each independent real quantity is
estimated to additive error at most
$\varepsilon_{\mathrm{est}}/(2d)$, then every matrix entry satisfies
\begin{equation}
    \bigl|
    [\hat{\Sigma}(\lambda)-\widetilde{\Sigma}(\lambda)]_{ij}
    \bigr|
    \le
    \frac{\varepsilon_{\mathrm{est}}}{d}.
\end{equation}
Consequently,
\begin{align}
    \|
    \hat{H}_{\mathrm{eff}}(\lambda)
    -
    \widetilde{H}_{\mathrm{eff}}(\lambda)
    \|
    &=
    \|
    \hat{\Sigma}(\lambda)
    -
    \widetilde{\Sigma}(\lambda)
    \|
    \nonumber\\[0.15cm]
    &\le
    \|
    \hat{\Sigma}(\lambda)
    -
    \widetilde{\Sigma}(\lambda)
    \|_F
    \le
    \varepsilon_{\mathrm{est}}.
    \label{eq:matrix-estimation-operator-bound}
\end{align}
Assigning failure probability $\mathcal O(\theta/d^2)$ to each
independent real estimate and applying a union bound ensures that
Eq.~\eqref{eq:matrix-estimation-operator-bound} holds with
probability at least $1-\theta$.

To obtain additive error $\varepsilon_{\mathrm{est}}/(2d)$ in a
matrix element, Eqs.~\eqref{eq:HT-real-probability-appendix} and
\eqref{eq:HT-imag-probability-appendix} require the relevant
probability to be estimated to accuracy
$\mathcal O(\varepsilon_{\mathrm{est}}/(d\mathcal N_\lambda))$.
Substituting
$\mathcal N_\lambda=2\beta\widetilde{\alpha}^2/g$ and suppressing
the constant $\beta=\Theta(1)$, the total numbers of calls to
$U_{\widetilde{\Sigma}(\lambda)}$ are
\begin{align}
    M_{\mathrm{MC}}
    &=
    \mathcal O\left(
        \frac{\widetilde{\alpha}^4d^4}
        {g^2\varepsilon_{\mathrm{est}}^2}
        \log\frac{d}{\theta}
    \right),
    \label{eq:ME-MC-calls-appendix}
    \\
    M_{\mathrm{IQAE}}
    &=
    \mathcal O\left(
        \frac{\widetilde{\alpha}^2d^3}
        {g\varepsilon_{\mathrm{est}}}
        \log\frac{d}{\theta}
    \right),
    \label{eq:ME-IQAE-calls-appendix}
\end{align}
for standard Monte Carlo sampling and iterative quantum amplitude
estimation, respectively
\cite{grinko_iterative_2021,suzuki_amplitude_2020}.

\begin{lemma}[Resource summary for matrix-element estimation]
\label{lem:ME-summary}
For any $\lambda$ in the safe region, one can construct a Hermitian
estimate
\begin{equation}
    \hat{H}_{\mathrm{eff}}(\lambda)
    :=
    H_{PP}+\hat{\Sigma}(\lambda)
\end{equation}
satisfying
\begin{equation}
    \|
    \hat{H}_{\mathrm{eff}}(\lambda)
    -
    \widetilde{H}_{\mathrm{eff}}(\lambda)
    \|
    \le
    \varepsilon_{\mathrm{est}}
\end{equation}
with probability at least $1-\theta$. The required resources are as
follows:
\begin{itemize}[leftmargin=*]
    \item \emph{Monte Carlo sampling:}
    \begin{align*}
        &\mathcal O\left(
            \frac{\alpha\widetilde{\alpha}^4d^4}
            {g^3\varepsilon_{\mathrm{est}}^2}
            \log\frac{1}{g\varepsilon_f}
            \log\frac{d}{\theta}
        \right)
        &&\text{queries to }U_{H_{QQ}-\lambda I_Q},
        \\
        &\mathcal O\left(
            \frac{\widetilde{\alpha}^4d^4}
            {g^2\varepsilon_{\mathrm{est}}^2}
            \log\frac{d}{\theta}
        \right)
        &&\text{queries to }U_{H_{QP}}.
    \end{align*}

    \item \emph{Iterative QAE
    \cite{grinko_iterative_2021,suzuki_amplitude_2020}:}
    \begin{align*}
        &\mathcal O\left(
            \frac{\alpha\widetilde{\alpha}^2d^3}
            {g^2\varepsilon_{\mathrm{est}}}
            \log\frac{1}{g\varepsilon_f}
            \log\frac{d}{\theta}
        \right)
        &&\text{queries to }U_{H_{QQ}-\lambda I_Q},
        \\
        &\mathcal O\left(
            \frac{\widetilde{\alpha}^2d^3}
            {g\varepsilon_{\mathrm{est}}}
            \log\frac{d}{\theta}
        \right)
        &&\text{queries to }U_{H_{QP}}.
    \end{align*}

    \item \emph{Ancillas:}
    $a+2\widetilde{a}+2$ qubits, including the Hadamard-test
    control qubit.
\end{itemize}
\end{lemma}

\section{Pole behavior and branch continuity}
\label{app:continuity-of-eigenbranches}

The Feshbach effective Hamiltonian
\begin{equation}
    H_{\mathrm{eff}}(\lambda)
    =
    H_{PP}
    -H_{PQ}(H_{QQ}-\lambda I_Q)^{-1}H_{QP}
\end{equation}
is undefined at the poles
$\lambda\in\operatorname{spec}(H_{QQ})$. Nevertheless, only the part
of the reference space that couples to the corresponding eigenspace
of $H_{QQ}$ is singular. The remaining finite eigenbranches have
well-defined one-sided limits and can be matched continuously across
the pole. This section makes that statement precise; the simple-pole
case is discussed in Ref.~\cite{kalantzis_spectral_2016}.

Let $\chi_1,\ldots,\chi_M$ be the distinct eigenvalues of $H_{QQ}$,
and let $\Pi_j^{(Q)}$ be the spectral projector associated with
$\chi_j$. Then
\begin{equation}
    (H_{QQ}-\lambda I_Q)^{-1}
    =
    \sum_{j=1}^{M}
    \frac{\Pi_j^{(Q)}}{\chi_j-\lambda},
    \qquad
    \lambda\notin\operatorname{spec}(H_{QQ}).
\end{equation}
Define the positive-semidefinite coupling operators on $P\mathcal H$,
\begin{equation}
    A_j
    :=
    H_{PQ}\Pi_j^{(Q)}H_{QP}
    \succeq 0.
    \label{eq:pole-coupling-operator}
\end{equation}
The effective Hamiltonian can therefore be written as
\begin{equation}
    H_{\mathrm{eff}}(\lambda)
    =
    H_{PP}
    -\sum_{j=1}^{M}\frac{A_j}{\chi_j-\lambda}.
    \label{eq:Heff-pole-expansion}
\end{equation}
The rank of $A_j$, rather than the degeneracy of $\chi_j$ itself,
determines how many effective-Hamiltonian eigenvalues diverge at that
pole.

Fix a pole $\chi_\ell$ and set
\begin{equation}
    A:=A_\ell,
    \qquad
    r:=\operatorname{rank}(A),
    \qquad
    \mathcal E:=\operatorname{ran}(A),
    \qquad
    \mathcal K:=\ker(A)=\mathcal E^\perp.
\end{equation}
Let $P_{\mathcal E}$ and $P_{\mathcal K}$ denote the orthogonal
projectors onto these subspaces. Separating the singular term in
Eq.~\eqref{eq:Heff-pole-expansion}, define
\begin{equation}
    B_\ell(\lambda)
    :=
    H_{PP}
    -\sum_{\substack{j=1\\j\neq\ell}}^{M}
    \frac{A_j}{\chi_j-\lambda},
    \qquad
    H_{\mathrm{eff}}(\lambda)
    =
    B_\ell(\lambda)-\frac{A}{\chi_\ell-\lambda}.
    \label{eq:regular-singular-decomposition}
\end{equation}
The operator $B_\ell(\lambda)$ is analytic in a neighborhood of
$\chi_\ell$. The finite limiting operator is its compression to the
uncoupled subspace $\mathcal K$,
\begin{equation}
    \overline H_\ell
    :=
    P_{\mathcal K}B_\ell(\chi_\ell)P_{\mathcal K}
    \big|_{\mathcal K}.
    \label{eq:finite-limiting-operator}
\end{equation}
Write its eigenvalues in increasing order as
\begin{equation}
    \nu_1\le\cdots\le\nu_{d-r}.
\end{equation}
We also write
\begin{equation}
    \xi_1^{\uparrow}(\lambda)
    \le\cdots\le
    \xi_d^{\uparrow}(\lambda)
\end{equation}
for the increasingly ordered eigenvalues of
$H_{\mathrm{eff}}(\lambda)$.

\begin{theorem}[Eigenbranch behavior at a resolvent pole]
\label{thm:branch-continuity}
With the notation above, as $\lambda\to\chi_\ell^-$,
\begin{equation}
    \lim_{\lambda\to\chi_\ell^-}
    \xi_i^{\uparrow}(\lambda)
    =
    \begin{cases}
        -\infty,
        & 1\le i\le r,\\[2pt]
        \nu_{i-r},
        & r<i\le d,
    \end{cases}
    \label{eq:left-pole-limits}
\end{equation}
and, as $\lambda\to\chi_\ell^+$,
\begin{equation}
    \lim_{\lambda\to\chi_\ell^+}
    \xi_i^{\uparrow}(\lambda)
    =
    \begin{cases}
        \nu_i,
        & 1\le i\le d-r,\\[2pt]
        +\infty,
        & d-r<i\le d.
    \end{cases}
    \label{eq:right-pole-limits}
\end{equation}
Thus exactly $r$ eigenvalues diverge at the pole, while the remaining
$d-r$ eigenvalues have matching one-sided limits,
\begin{equation}
    \lim_{\lambda\to\chi_\ell^-}
    \xi_{r+k}^{\uparrow}(\lambda)
    =
    \nu_k
    =
    \lim_{\lambda\to\chi_\ell^+}
    \xi_k^{\uparrow}(\lambda),
    \qquad
    1\le k\le d-r.
    \label{eq:finite-branch-matching}
\end{equation}
\end{theorem}

\begin{proof}
Let
\begin{equation}
    \alpha(\lambda):=\frac{1}{\chi_\ell-\lambda}.
\end{equation}
Since $B_\ell(\lambda)$ is bounded near $\chi_\ell$, there is a
constant $C$ such that $\|B_\ell(\lambda)\|\le C$ in a sufficiently
small punctured neighborhood of the pole. Let
$\sigma_1\ge\cdots\ge\sigma_r>0$ be the nonzero eigenvalues of $A$.
For $\lambda<\chi_\ell$, one has $\alpha(\lambda)\to+\infty$, and
the increasingly ordered eigenvalues of $-\alpha(\lambda)A$ are
\begin{equation}
    -\alpha\sigma_1,
    \ldots,
    -\alpha\sigma_r,
    \underbrace{0,\ldots,0}_{d-r}.
\end{equation}
Weyl's perturbation inequality therefore gives
\begin{align}
    \left|\xi_i^{\uparrow}(\lambda)
    +\alpha(\lambda)\sigma_i\right|
    &\le C,
    &&1\le i\le r,
    \nonumber\\
    \left|\xi_i^{\uparrow}(\lambda)\right|
    &\le C,
    &&r<i\le d.
    \label{eq:weyl-pole-bound}
\end{align}
Hence the first $r$ eigenvalues diverge to $-\infty$, whereas the
remaining $d-r$ eigenvalues stay bounded. When
$\lambda>\chi_\ell$, $-\alpha(\lambda)\to+\infty$; the same argument
shows that the largest $r$ eigenvalues diverge to $+\infty$, while
the other $d-r$ remain bounded.

It remains to identify the bounded limits. Relative to the orthogonal decomposition
$P\mathcal H=\mathcal E\oplus\mathcal K$, write
\begin{equation}
    A=
    \begin{bmatrix}
        A_{\mathcal E} & 0\\
        0 & 0
    \end{bmatrix},
    \qquad
    B_\ell(\lambda)=
    \begin{bmatrix}
        B_{\mathcal{EE}}(\lambda)
        & B_{\mathcal{EK}}(\lambda)\\
        B_{\mathcal{KE}}(\lambda)
        & B_{\mathcal{KK}}(\lambda)
    \end{bmatrix},
    \label{eq:pole-block-decomposition}
\end{equation}
where $A_{\mathcal E}\succ0$. For $z$ in a fixed bounded set and
$|\alpha(\lambda)|$ sufficiently large,
\[
    B_{\mathcal{EE}}(\lambda)
    -\alpha(\lambda)A_{\mathcal E}
    -zI_{\mathcal E}
\]
is invertible, with inverse of norm
$\mathcal O(|\alpha|^{-1})$. The Schur-complement identity then gives
\begin{align}
    &\det\bigl(H_{\mathrm{eff}}(\lambda)-zI_P\bigr)
    \nonumber\\
    &\quad=
    \det\!\left(
        B_{\mathcal{EE}}(\lambda)
        -\alpha(\lambda)A_{\mathcal E}
        -zI_{\mathcal E}
    \right)
    \det\!\bigl(\mathcal S_\lambda(z)\bigr),
    \label{eq:pole-schur-factorization}
\end{align}
where
\begin{align}
    \mathcal S_\lambda(z)
    :=\;&
    B_{\mathcal{KK}}(\lambda)-zI_{\mathcal K}
    \nonumber\\
    &-
    B_{\mathcal{KE}}(\lambda)
    \left(
        B_{\mathcal{EE}}(\lambda)
        -\alpha(\lambda)A_{\mathcal E}
        -zI_{\mathcal E}
    \right)^{-1}
    B_{\mathcal{EK}}(\lambda).
    \label{eq:finite-schur-complement}
\end{align}
As $\lambda\to\chi_\ell^\pm$, the second term on the right-hand side
of Eq.~\eqref{eq:finite-schur-complement} vanishes in norm, while
$B_{\mathcal{KK}}(\lambda)\to\overline H_\ell$. Therefore,
\begin{equation}
    \mathcal S_\lambda(z)
    \longrightarrow
    \overline H_\ell-zI_{\mathcal K}
\end{equation}
uniformly for $z$ in compact sets. Since the first determinant in
Eq.~\eqref{eq:pole-schur-factorization} has no bounded zeros for
sufficiently large $|\alpha|$, the $d-r$ bounded eigenvalues are the
zeros of $\det(\mathcal S_\lambda(z))$. Continuity of the roots of a
polynomial with respect to its coefficients implies that these zeros
converge, including multiplicity, to the eigenvalues
$\nu_1,\ldots,\nu_{d-r}$ of $\overline H_\ell$. Ordering the real
zeros gives Eqs.~\eqref{eq:left-pole-limits} and
\eqref{eq:right-pole-limits}.
\end{proof}

Equation~\eqref{eq:finite-branch-matching} is the precise sense in
which the finite branches continue across a pole: their ordered labels
shift by $r$, because $r$ branches disappear to one infinity and
re-enter from the other. If a limiting eigenvalue $\nu_k$ is simple, the corresponding finite
branch is locally real analytic after this relabeling. Indeed, set
$t=\chi_\ell-\lambda$ and write
\[
    B_{\mathcal{EE}}(t)
    :=
    B_{\mathcal{EE}}(\chi_\ell-t).
\]
The inverse in Eq.~\eqref{eq:finite-schur-complement} can then be
written as
\begin{equation}
    \left(
        B_{\mathcal{EE}}(t)
        -\frac{A_{\mathcal E}}{t}
        -zI_{\mathcal E}
    \right)^{-1}
    =
    t\left(
        t[B_{\mathcal{EE}}(t)-zI_{\mathcal E}]
        -A_{\mathcal E}
    \right)^{-1},
\end{equation}
which is analytic at $t=0$. At a degenerate limiting eigenvalue, an
individual basis inside the finite bundle is not canonical; the
associated spectral projector is the basis-independent object.
Finally, if $A_\ell=0$, then $r=0$ and the apparent pole is removable
from $H_{\mathrm{eff}}(\lambda)$.

\section{Proof of the eigenstate and subspace fidelity theorems}
\label{app:fidelity_proof}

This appendix proves Theorem~\ref{thm:prep-fidelity}. The main
ingredient is a direct residual identity for the approximate wave
operator, which makes the cancellation in the conjugated eigenproblem
explicit without separately estimating the diagonal residual block
$R_{PP}$.

\subsection{Residual identity for the approximate wave operator}

Fix a spectral parameter $\lambda$ in the safe region. For any
$|\phi\rangle\in P\mathcal H$, the restriction of the approximate
wave operator to $P\mathcal H$ is
\begin{equation}
    \widetilde\Omega(\lambda)|\phi\rangle
    =
    \begin{bmatrix}
        |\phi\rangle\\[2pt]
        -f(H_{QQ}-\lambda I_Q)H_{QP}|\phi\rangle
    \end{bmatrix}.
\end{equation}
Direct block multiplication gives
\begin{equation}
\label{eq:lifted-residual-identity}
    (H-\lambda I)\widetilde\Omega(\lambda)|\phi\rangle
    =
    \begin{bmatrix}
        \bigl(
            \widetilde H_{\mathrm{eff}}(\lambda)
            -\lambda I_P
        \bigr)|\phi\rangle\\[2pt]
        R_{QP}(\lambda)|\phi\rangle
    \end{bmatrix},
\end{equation}
where
\begin{equation}
\label{eq:RQP-definition-fidelity}
    R_{QP}(\lambda)
    \coloneqq
    \Bigl[
        I_Q
        -(H_{QQ}-\lambda I_Q)
        f(H_{QQ}-\lambda I_Q)
    \Bigr]H_{QP}.
\end{equation}
Since $f(H_{QQ}-\lambda I_Q)$ is a polynomial in
$H_{QQ}-\lambda I_Q$, the two operators commute. Therefore,
\begin{align}
    \|R_{QP}(\lambda)\|
    &\le
    \|H_{QQ}-\lambda I_Q\|
    \left\|
        (H_{QQ}-\lambda I_Q)^{-1}
        -f(H_{QQ}-\lambda I_Q)
    \right\|
    \|H_{QP}\|
    \nonumber\\[2pt]
    &\le
    \alpha_\lambda\|H_{QP}\|\varepsilon_f
    =
    \mathcal O\!\left(
        \alpha\|H_{QP}\|\varepsilon_f
    \right),
    \label{eq:R_QP_bound}
\end{align}
where $\alpha_\lambda=\mathcal O(\alpha)$ on the queried energy
interval. Equation~\eqref{eq:lifted-residual-identity} shows that the
full-space residual is controlled by the reduced residual and the
QSVT inverse error.

\subsection{Non-degenerate eigenstate}

Let $\hat\lambda_k$ and the normalized vector
$|\hat\phi_k\rangle\in P\mathcal H$ be returned by the
eigenvalue-estimation routine, with
\begin{equation}
    \hat H_{\mathrm{eff}}(\hat\lambda_k)|\hat\phi_k\rangle
    =
    \hat\xi_k(\hat\lambda_k)|\hat\phi_k\rangle,
    \qquad
    |\hat\xi_k(\hat\lambda_k)-\hat\lambda_k|
    \le
    \varepsilon_{\mathrm{app}}.
\end{equation}
Define the reduced residual
\begin{equation}
    |r_{P,k}\rangle
    \coloneqq
    \bigl(
        \widetilde H_{\mathrm{eff}}(\hat\lambda_k)
        -\hat\lambda_k I_P
    \bigr)|\hat\phi_k\rangle.
\end{equation}
Using the effective-Hamiltonian estimation error, we obtain
\begin{align}
    \|r_{P,k}\|
    &\le
    \left\|
        \widetilde H_{\mathrm{eff}}(\hat\lambda_k)
        -
        \hat H_{\mathrm{eff}}(\hat\lambda_k)
    \right\|
    +
    |\hat\xi_k(\hat\lambda_k)-\hat\lambda_k|
    \nonumber\\[2pt]
    &\le
    \varepsilon_{\mathrm{est}}
    +
    \varepsilon_{\mathrm{app}}.
    \label{eq:reduced-residual-nondegenerate}
\end{align}

Let
\begin{equation}
    \eta_k
    \coloneqq
    \left\|
        \widetilde\Omega(\hat\lambda_k)
        |\hat\phi_k\rangle
    \right\|^{-2}
    =
    \mathcal O(p_k),
    \label{eq:eta-definition}
\end{equation}
and define the normalized lifted state
\begin{equation}
    |\widetilde\Psi_k\rangle
    \coloneqq
    \sqrt{\eta_k}\,
    \widetilde\Omega(\hat\lambda_k)|\hat\phi_k\rangle.
\end{equation}
Applying Eq.~\eqref{eq:lifted-residual-identity} at
$\lambda=\hat\lambda_k$ gives
\begin{align}
    \left\|
        (H-\hat\lambda_k I)|\widetilde\Psi_k\rangle
    \right\|^2
    &=
    \eta_k
    \left(
        \|r_{P,k}\|^2
        +
        \|R_{QP}(\hat\lambda_k)
        |\hat\phi_k\rangle\|^2
    \right)=
    \mathcal O\!\left(
        p_k\varepsilon^2
    \right),
    \label{eq:full-residual-nondegenerate}
\end{align}
where $\varepsilon$ is the total residual budget in
Eq.~\eqref{eq:total_error_budget}.

Expand
$|\widetilde\Psi_k\rangle=\sum_j c_j|\Psi_j\rangle$
in an eigenbasis of $H$. Let $\lambda_k$ be the target eigenvalue
and let $\Delta$ denote its separation, at the working energy
$\hat\lambda_k$, from the complementary spectrum. Then
\begin{align}
    \left\|
        (H-\hat\lambda_k I)|\widetilde\Psi_k\rangle
    \right\|^2
    &=
    \sum_j
    |c_j|^2(\lambda_j-\hat\lambda_k)^2 \ge
    \Delta^2
    \sum_{j\neq k}|c_j|^2.
\end{align}
Since
\begin{equation}
    1-F(\Psi_k,\widetilde\Psi_k)
    =
    \sum_{j\neq k}|c_j|^2,
\end{equation}
Eq.~\eqref{eq:full-residual-nondegenerate} implies
\begin{equation}
    1-F(\Psi_k,\widetilde\Psi_k)
    =
    \mathcal O\!\left(
        \frac{p_k\varepsilon^2}{\Delta^2}
    \right),
\end{equation}
which proves Eq.~\eqref{eq:fidelity_nondeg}. If $\Delta$ is defined
using the exact eigenvalue $\lambda_k$, the same conclusion follows
whenever $|\hat\lambda_k-\lambda_k|$ is at most a fixed fraction of
that gap.

\subsection{Target spectral subspace}

Let $\mathcal S$ be an $m$-dimensional target spectral subspace and
let $\hat\lambda$ be the common working energy used for the selected
branch bundle. Collect the corresponding orthonormal $P$-space
eigenvectors into
\begin{equation}
    \Phi
    =
    \bigl[
        |\hat\phi_1\rangle,\ldots,|\hat\phi_m\rangle
    \bigr],
    \qquad
    \Phi^\dagger\Phi=I_m.
\end{equation}
If
\begin{equation}
    \hat H_{\mathrm{eff}}(\hat\lambda)\Phi
    =
    \Phi\hat\Xi,
    \qquad
    \|\hat\Xi-\hat\lambda I_m\|
    \le
    \varepsilon_{\mathrm{app}},
\end{equation}
then the reduced-residual matrix
\begin{equation}
    R_P
    \coloneqq
    \bigl(
        \widetilde H_{\mathrm{eff}}(\hat\lambda)
        -\hat\lambda I_P
    \bigr)\Phi
\end{equation}
satisfies
\begin{equation}
    \|R_P\|
    \le
    \varepsilon_{\mathrm{est}}
    +
    \varepsilon_{\mathrm{app}}.
    \label{eq:bundle-RP-bound}
\end{equation}

For each column, define
\begin{equation}
    \eta_k
    \coloneqq
    \left\|
        \widetilde\Omega(\hat\lambda)
        |\hat\phi_k\rangle
    \right\|^{-2}
    =
    \mathcal O(p_k),
    \qquad
    D_\eta
    \coloneqq
    \operatorname{diag}
    \bigl(
        \sqrt{\eta_1},\ldots,\sqrt{\eta_m}
    \bigr).
\end{equation}
The individually normalized, but generally nonorthogonal, lifted
columns are
\begin{equation}
    \widetilde B
    \coloneqq
    \widetilde\Omega(\hat\lambda)\Phi D_\eta
    =
    \bigl[
        |\widetilde\Psi_1\rangle,\ldots,
        |\widetilde\Psi_m\rangle
    \bigr].
    \label{eq:normalized-lifted-matrix}
\end{equation}
Let
\begin{equation}
    G
    \coloneqq
    \widetilde B^\dagger\widetilde B
\end{equation}
be their Gram matrix. Assuming $G\succ0$, the matrix
\begin{equation}
    B
    \coloneqq
    \widetilde B G^{-1/2}
\end{equation}
has orthonormal columns and spans the same prepared subspace.

Applying Eq.~\eqref{eq:lifted-residual-identity} columnwise yields
\begin{equation}
    R
    \coloneqq
    (H-\hat\lambda I)\widetilde B
    =
    \begin{bmatrix}
        R_P\\[2pt]
        R_{QP}(\hat\lambda)\Phi
    \end{bmatrix}
    D_\eta.
    \label{eq:collective-residual-factorization}
\end{equation}
The operator norm of the first factor obeys
\begin{align}
    \left\|
        \begin{bmatrix}
            R_P\\
            R_{QP}(\hat\lambda)\Phi
        \end{bmatrix}
    \right\|^2
    &=
    \left\|
        R_P^\dagger R_P
        +
        \Phi^\dagger
        R_{QP}(\hat\lambda)^\dagger
        R_{QP}(\hat\lambda)
        \Phi
    \right\|
    \nonumber\\[2pt]
    &\le
    \|R_P\|^2
    +
    \|R_{QP}(\hat\lambda)\|^2
    \le
    \varepsilon^2.
    \label{eq:collective-factor-bound}
\end{align}
Moreover,
\begin{equation}
    \|D_\eta\|^2
    =
    \max_k\eta_k
    =
    \mathcal O(p_{\max}).
\end{equation}
Consequently,
\begin{equation}
    \|R\|^2
    =
    \mathcal O\!\left(
        p_{\max}\varepsilon^2
    \right).
    \label{eq:collective-residual-operator-bound}
\end{equation}
For the Frobenius norm,
Eq.~\eqref{eq:collective-residual-factorization} gives
\begin{align}
    \|R\|_F^2
    &=
    \sum_{k=1}^m
    \eta_k
    \left(
        \|R_Pe_k\|^2
        +
        \|R_{QP}(\hat\lambda)
        |\hat\phi_k\rangle\|^2
    \right)
    \nonumber\\[2pt]
    &=
    \mathcal O\!\left(
        \varepsilon^2\sum_{k=1}^m p_k
    \right)
    =
    \mathcal O\!\left(
        mp_{\mathrm{avg}}\varepsilon^2
    \right),
    \label{eq:collective-residual-Frobenius-bound}
\end{align}
where
\begin{equation}
    p_{\max}
    \coloneqq
    \max_k p_k,
    \qquad
    p_{\mathrm{avg}}
    \coloneqq
    \frac{1}{m}\sum_{k=1}^m p_k.
\end{equation}
Equation~\eqref{eq:collective-residual-operator-bound} controls the
residual of every normalized linear combination of the lifted
columns and therefore does not introduce an additional factor of
$m$ in the worst-case bound.

Set
\begin{equation}
    E
    \coloneqq
    R G^{-1/2}.
\end{equation}
Then
\begin{equation}
    (H-\hat\lambda I)B=E,
    \qquad
    B^\dagger B=I_m,
    \label{eq:orthonormal_residual_subspace}
\end{equation}
and
\begin{align}
    \|E\|^2
    &=
    \mathcal O\!\left(
        \|G^{-1}\|
        p_{\max}\varepsilon^2
    \right),
    \label{eq:E-operator-bound}
    \\
    \|E\|_F^2
    &=
    \mathcal O\!\left(
        \|G^{-1}\|
        mp_{\mathrm{avg}}\varepsilon^2
    \right).
    \label{eq:E-Frobenius-bound}
\end{align}

Let $A$ be an orthonormal basis matrix for $\mathcal S$, and let
$A_\perp$ be an orthonormal eigenbasis matrix for
$\mathcal S^\perp$, so that
\begin{equation}
    HA_\perp=A_\perp\Lambda_\perp.
\end{equation}
Let
\begin{equation}
    \Delta
    \coloneqq
    \operatorname{dist}\!\left(
        \hat\lambda,
        \operatorname{spec}\!\left(
            H\big|_{\mathcal S^\perp}
        \right)
    \right)
    >0.
\end{equation}
Left-multiplying
Eq.~\eqref{eq:orthonormal_residual_subspace}
by $A_\perp^\dagger$ gives
\begin{equation}
    (\Lambda_\perp-\hat\lambda I_{N-m})
    A_\perp^\dagger B
    =
    A_\perp^\dagger E.
\end{equation}
Therefore,
\begin{equation}
    \|A_\perp^\dagger B\|
    \le
    \frac{\|E\|}{\Delta},
    \qquad
    \|A_\perp^\dagger B\|_F
    \le
    \frac{\|E\|_F}{\Delta}.
    \label{eq:BperpA_bounds}
\end{equation}
The singular values of $A_\perp^\dagger B$ are the sines of the
principal angles between the target and prepared subspaces. Hence
\begin{align}
    1-F_{\min}
    &=
    \|A_\perp^\dagger B\|^2,
    \label{eq:Fmin}
    \\
    1-F_{\mathrm{avg}}
    &=
    \frac{1}{m}
    \|A_\perp^\dagger B\|_F^2.
    \label{eq:Favg}
\end{align}
Combining
Eqs.~\eqref{eq:E-operator-bound}--\eqref{eq:Favg}
yields
\begin{align}
    1-F_{\min}
    &=
    \mathcal O\!\left(
        \frac{
            \|G^{-1}\|
            p_{\max}\varepsilon^2
        }{\Delta^2}
    \right),
    \\
    1-F_{\mathrm{avg}}
    &=
    \mathcal O\!\left(
        \frac{
            \|G^{-1}\|
            p_{\mathrm{avg}}\varepsilon^2
        }{\Delta^2}
    \right),
\end{align}
which proves the subspace-fidelity bounds in
Theorem~\ref{thm:prep-fidelity}.

\clearpage
\bibliographystyle{quantum}
\bibliography{references}

\end{document}